\documentclass[aps,prx,twocolumn,secnumarabic,nobalancelastpage,amsmath,amssymb,nofootinbib,floatfix]{revtex4-2}

\usepackage[dvipsnames]{xcolor}
\usepackage[colorlinks=true,linkcolor=MidnightBlue,citecolor=blue!70!black,filecolor=green,urlcolor=PineGreen]{hyperref}
\usepackage{graphicx}
\usepackage{tikz}
\usetikzlibrary{%
    decorations.pathreplacing,%
    decorations.pathmorphing,%
    arrows,%
    patterns
}
\usepackage{amsmath,amssymb}
\usepackage{mathrsfs}
\usepackage{dsfont}

\def\be{\begin{equation}}
\def\ee{\end{equation}}
\newcommand{\ensavg}[1]{\left\langle #1 \right\rangle}
\newcommand{\tr}[1]{\mathrm{tr} \left( #1 \right)}
\newcommand{\ket}[1]{| #1 \rangle}

\newcommand{\kett}[1]{| \,#1\, \rangle\kern-0.2em\rangle}
\newcommand{\braa}[1]{\langle\kern-0.2em\langle \,#1\, |}
\newcommand{\bra}[1]{\langle #1 |}

\newcommand{\ip}[2]{\langle #1 | #2 \rangle}
\newcommand{\op}[2]{\ket{#1}\bra{#2}}
\renewcommand{\eqref}[1]{\textsc{eq.}~(\ref{#1})}
\newcommand{\secref}[1]{\S\,\ref{#1}}

\newcommand{\limit}[2]{\underset{#1\,\rightarrow\,#2}{\mathrm{lim}}\;}

\begin{document}

\definecolor{amber}{RGB}{255,191,0}
\definecolor{satinsheengold}{rgb}{0.8, 0.63, 0.21}
\definecolor{caribbeangreen}{rgb}{0.0, 0.8, 0.6}
\definecolor{chromeyellow}{rgb}{1.0, 0.65, 0.0}
\definecolor{deepfuchsia}{rgb}{0.76, 0.33, 0.76}
\definecolor{coquelicot}{rgb}{1.0, 0.22, 0.0}
\definecolor{crimsonglory}{rgb}{0.75, 0.0, 0.2}
\definecolor{deeppink}{rgb}{1.0, 0.08, 0.58}
\definecolor{electricviolet}{rgb}{0.56, 0.0, 1.0}
\definecolor{electricgreen}{rgb}{0.0, 1.0, 0.0}
\definecolor{mint}{rgb}{0.24, 0.71, 0.54}
\definecolor{dodgerblue}{rgb}{0.12, 0.56, 1.0}
\definecolor{lincolngreen}{rgb}{0.11, 0.35, 0.02}
\definecolor{persianblue}{rgb}{0.11, 0.22, 0.73}
\definecolor{patriarch}{rgb}{0.5, 0.0, 0.5}
\definecolor{charcoal}{rgb}{0.21, 0.27, 0.31}
\definecolor{darkblue}{rgb}{0.0, 0.0, 0.55}
\definecolor{darkred}{rgb}{0.55, 0.0, 0.0}
\definecolor{ceruleanblue}{rgb}{0.16, 0.32, 0.75}
\definecolor{darkspringgreen}{rgb}{0.09, 0.45, 0.27}
\definecolor{deeplilac}{rgb}{0.6, 0.33, 0.73}
\definecolor{goldenyellow}{rgb}{1.0, 0.87, 0.0}
\definecolor{lightgray}{rgb}{0.8 0.8 0.8}
\definecolor{oceanboatblue}{rgb}{0.0, 0.47, 0.75}
\definecolor{lightseagreen}{rgb}{0.13, 0.7, 0.67}
\definecolor{aqua}{rgb}{0.0, 1.0, 1.0}
\definecolor{mikadoyellow}{rgb}{1.0, 0.77, 0.05}
\definecolor{richcarmine}{rgb}{0.84, 0.0, 0.25}
\definecolor{xkcd_purple}{RGB}{126, 30, 156}
\definecolor{xkcd_lilac}{RGB}{206, 162, 253}
\definecolor{xkcd_sky}{RGB}{117, 187, 253}
\definecolor{xkcd_cobalt}{RGB}{3, 10, 167}
\definecolor{xkcd_pine}{RGB}{10, 72, 30}
\definecolor{xkcd_teal}{RGB}{2, 147, 134}
\definecolor{xkcd_aquamarine}{RGB}{4, 216, 178}
\definecolor{xkcd_chartreuse}{RGB}{193, 248, 10}
\definecolor{xkcd_goldenrod}{RGB}{250, 194, 5}
\definecolor{xkcd_pumpkin}{RGB}{225, 119, 1}
\definecolor{xkcd_brightred}{RGB}{255, 0, 13}
\definecolor{xkcd_deeppink}{RGB}{203, 1, 98}
\definecolor{xkcd_wine}{RGB}{128, 1, 63}
\definecolor{xkcd_charcoal}{RGB}{52, 56, 55}
\definecolor{xkcd_greypurple}{RGB}{130, 109, 140}

\title{Noise--Canceling Quantum Feedback: non-Hermitian Dynamics \\ with Applications to State Preparation and Magic State Distillation}

\author{Tathagata Karmakar} \email{tatha@berkeley.edu} \thanks{these two authors contributed equally.}
\affiliation{Department of Chemistry, University of California, Berkeley, CA 94720, USA}
\affiliation{Berkeley Center for Quantum Information and Computation, Berkeley, CA 94720, USA}
\author{Philippe Lewalle} \email{plewalle@berkeley.edu} \thanks{these two authors contributed equally.}
\affiliation{Department of Chemistry, University of California, Berkeley, CA 94720, USA}
\affiliation{Berkeley Center for Quantum Information and Computation, Berkeley, CA 94720, USA}
\author{Yipei Zhang}
\affiliation{Department of Chemistry, University of California, Berkeley, CA 94720, USA}
\affiliation{Berkeley Center for Quantum Information and Computation, Berkeley, CA 94720, USA}
\author{K. Birgitta Whaley} 
\affiliation{Department of Chemistry, University of California, Berkeley, CA 94720, USA}
\affiliation{Berkeley Center for Quantum Information and Computation, Berkeley, CA 94720, USA}
\date{\today}

\begin{abstract}
Time--continuous quantum measurement allows for the tracking of a quantum system in real time via sequences of short, and individually weak, measurement intervals.
Such measurements are necessarily invasive, imparting backaction to the system, and allowing the observer to update their state estimate based on stochastic measurement outcomes.  
Feedback control then involves real--time interventions by an observer, conditioned on the time-continuous measurement signal that they receive. 
We here consider diffusive quantum trajectories, and focus on the ``noise--canceling'' subset of feedback protocols that aim to minimize the degree of stochasticity in the dynamics. 
We derive such a class of feedback operations, showing that under the idealized assumptions of pure states, unit measurement efficiency, and zero time--delay in implementing feedback operations, perfectly noise--canceling feedback \emph{always} exists. 
We consider the resulting noise--canceled dynamics generated by an effective non-Hermitian Hamiltonian; while non-Hermitian Hamiltonians from continuous monitoring generally describe rare dynamics (accessible by costly post--selection), the use of noise--canceling feedback here leads to non-Hermitian dynamics that occur deterministically. 
We demonstrate this via examples of entangled state preparation and stabilization.
We then illustrate the potential for the application of noise--cancellation to boost success rates in magic state distillation protocols. 
We show that adding feedback based on noise--cancellation into a time--continuous 5-to-1 distillation protocol leads to higher probabilities of successful distillation across a range of input errors, and extends the threshold on input errors for which the protocol is effective.  Our results highlight the efficacy of noise--canceling feedback--aided protocols for quantum state preparation and stabilization tasks.
\end{abstract}
     
\maketitle

\section{Introduction \label{sec-intro}}

After several decades of rapid progress, time--continuous quantum monitoring is both theoretically \cite{Hudson_Parthasarathy, BookCarmichael, BookWiseman, BookBarchielli, BookJacobs, Chantasri_2021, BookJordan} and experimentally well--developed \cite{BookJordan, Gambetta2008, Clerk_RMP_2010, Murch2013, ShayLeigh2016, Ficheux2018, Gourgy_Martin_Review_2020, Blais_CQED}.
Such monitoring is a pre-requisite for measurement--based feedback, which has also been developed in general terms \cite{PhysRevLett.70.548, PhysRevA.49.2133, PhysRevA.49.1350, Jacobs_Shabani_2008,Gough_2012, Zhang2017} and applied to numerous settings \cite{Tanaka_2012, martin2015deterministic, martin2017optimal, Minev2018, Leigh_Phase, zhang2020locally, martin2019single, chen2020quantum, jiang2020optimality, sundaresan2023demonstrating, Mirrahimi_2007, amini2011stability, Ticozzi_2013, Benoist_2017, Cardona_2018, liang2019exponential, Cardona_2020, amini2023exponential, liang2023exploring, LIANG2024, Ticozzi_2008}. 
Continuous feedback control is of interest e.g.~for continuous quantum error correction \cite{Ahn_2002, Ahn_2003, Ahn_2004, Sarovar_2004, vanhandel2005optimal, Oreshkov_2007, Mascarenhas_2010, Atalaya_CQEC, Mohseninia2020alwaysquantumerror, Atalaya_2021_CQEC, Convy2022logarithmicbayesian, Livingston_2022, Convy_2022}, wherein errors are detected and corrected in close to real time based on always--on continuous monitoring, instead of alternating between rounds of projective measurement and corrective operations. 

Another research area experiencing rapid progress concerns the use of non-Hermitian dyamical generators \cite{Bender_1998, Mostafazadeh_1, *Mostafazadeh_2, *Mostafazadeh_3, Heiss_2004, Heiss_2012, ElGanainy_2018, Ozdemir_2019, Miri_2019, Ashida_2020}.  
Non-Hermitian Hamiltonians (nHH) arise naturally in the context of many open (non-conservative) systems, including those subject to continuous monitoring. 
These objects offer a distinct analytical toolbox for describing open and monitored quantum systems; in particular, the complex eigenvalues and eigenvectors of the NH Hamiltonian \cite{Dembowski_2004, Zhong2018} lead to gain/loss effects that can break principles of adiabatic following \cite{Kvitsinsky_1991, Nenciu_1992, Berry_2011, Berry_2011_b, Leclerc_2013, Viennot_2014, Milburn2015} and manifest chiral behavior \cite{Doppler2016, Hassan2017, Zhong2018, Pick2019, Wang_2019, Zhong:19, Holler2020, Zhong_2021} based on the underlying topology of the complex eigenspectrum. 
That eigenspectrum is largely structured around exceptional points (EPs, degeneracies of both the eigenvalues and eigenvectors), which have been shown to have consequences for quantum sensing \cite{Wiersig_2016, *Wiersig:20, Lau_2018, Cook_2025}, and are central to many approaches aiming to control such systems \cite{Ribiero_2021, abbasi2021topological, Guria_2024, chavva2025}. 

One realization of quantum nHH dynamics arises from post--selecting an open quantum system's evolution on realizations in which quantum jumps did not occur \cite{Wu_2019, Naghiloo_2019_EP, Chen2021, Chen2022, abbasi2021topological}. 
This is formally indistinguishable from a situation in which the emission of quanta is continuously monitored, with a post-selection implemented on trajectories lacking emission \cite{Minganti_2019, Minganti_2020, FlorTeach2019, Lewalle_Pontryagin_2022}. 
Non-Hermitian dynamics can also be applied to model diffusive monitoring \cite{Minganti_2022}. 
While nHH dynamics realized by post-selection on no emission of quanta are now well--established, the necessity of a viable post-selection protocol is a limiting factor in accessing those dynamics: In particular, longer--duration nHH dynamics tend to be rare events, making their post--selection cost relatively high. 

In this work, we apply quantum feedback control to generate a new form of nHH dynamics that is diffusively conditioned rather than post-selected on lack of quantum (usually photon) emission.
While ideal feedback (with high--efficiency measurements and negligible processing delay time to implement feedback operators) is difficult to achieve in practice, we show here that the diffusively conditioned approach has a major advantage. This is that when successfully performed, such feedback can lead to deterministically--occurring nHH dynamics, completely circumventing the post--selection issue present in other quantum realizations. 
Here we focus on the particular form of feedback that generates such deterministic dynamics from stochastic (diffusive) quantum trajectories.  In this situation, for the dynamics to become deterministic, the feedback must necessarily remove the stochasticity from the backaction introduced by the measurement itself. 
We call such feedback \emph{noise--canceling} \cite{zhang2020locally}, and one of the main contributions of this manuscript is a constructive proof that under ideal conditions (unit--efficiency measurements, acting on pure states, and delay--free feedback) a noise--canceling quantum feedback (NCQF) operation \emph{always} exists. 
In fact, the presence of free control parameters in the derivations ensures that we are guaranteed not only one NCQF solution, but the existence of an infinite family of them under ideal conditions.
We give several examples of noise--canceling feedback for the preparation of highly entangled quantum states.

While NCQF solutions always exist for pure states, they do not necessarily exist for all combinations of mixed states and measurement channels. 
In this work, we also derive the conditions that must be met to guarantee the existence of NCQF solutions for mixed states. 
These conditions are an essential stepping stone towards analyzing the utility of NCQF in situations where the general quantum operations may be subject to errors. 
We illustrate the construction of NCQF protocols for mixed states by applying these conditions to the problem of state purification and to the problem of magic state distillation (MSD).

Magic states are single-qubit states that help realize non-Clifford gates through gate teleportation \cite{Bravyi_Kitaev_MSD}. Such states should be prepared with very high fidelities since in a quantum circuit, thousands of non-Clifford gates might be needed \cite{litinski_magic_2019}. 
MSD, introduced by \textcite{Bravyi_Kitaev_MSD}, is the process of preparing high--fidelity magic states from several copies of lower fidelity magic states through Clifford operations and measurements. 
MSD involves projecting the imperfect magic states to the codespace of a stabilizer code. 
If the initial fidelity is above a threshold, upon successful projection, the output multiqubit state can be decoded to obtain magic states with higher fidelities. 
Several protocols for magic state distillation have been proposed   \cite{Meier2012MagicstateDW, PhysRevA.86.052329, PhysRevA.87.022328, PhysRevA.87.042305, PhysRevA.95.022316, Gidney2019efficientmagicstate, Liu_2023, PhysRevA.105.022602, PhysRevLett.111.090505, PhysRevX.2.041021}.   Logical 5-to-1 MSD has been realized with neutral atoms  \cite{rodriguez2024experimentaldemonstrationlogicalmagic}. 
Due to inherent postselection, MSD is expensive and contributes to a significant resource overhead in realizing fault-tolerant quantum computing. 
In this work, we explore whether NCQF can alleviate the postselection cost associated with MSD. 
We consider 5-to-1  distillation based on the $[[5,1,3]]$ code, using continuous measurements of the stabilizers (rather than discrete rounds of projectors), in the spirit of continuous quantum error correction \cite{Ahn_2002, Ahn_2003, Ahn_2004, Sarovar_2004, vanhandel2005optimal, Oreshkov_2007, Mascarenhas_2010, Atalaya_CQEC, Mohseninia2020alwaysquantumerror, Atalaya_2021_CQEC, Convy2022logarithmicbayesian, Livingston_2022, Convy_2022}. 
On its own, such continuous measurements has a negligible impact on the performance of MSD.
However, this unlocks the ability to perform feedback ``as the measurement unfolds'' rather than merely after it is completed, and we will show that NCQF (even imperfect NCQF) can take advantage of this, suppressing unwanted stochastic outcomes in the process:
We see improvement in post-selection probabilities on logical magic states, as well as output fidelities, across a range of input fidelities. 

We address these points as follows: 
We lay out the main mathematical model and conditions for NCQF in \secref{sec-NCQF}. 
We then explore three examples of entangling measurements boosted by NCQF in \secref{sec-nHH_examples}. 
This serves a dual purpose of i) allowing us to demonstrate several concrete examples of noise--canceled non-Hermitian Hamiltonian (NC-nHH) dynamics, while ii) showcasing several ways in which ideal NCQF offers a route towards the deterministic preparation of highly entangled states. 
\secref{sec_Purify} serves a transitional role, illustrating how rapid state purification is connected to the existence of perfectly noise--canceling feedback solutions on mixed states. 
Finally, in \secref{sec-MSD} we pivot to a different type of example, demonstrating the application of NCQF to MSD. 
Our conclusions and perspective for further development of NCQF are summarized in \secref{sec-conclude}. 
A number of mathematical derivations supporting and elaborating on the main text appear in the Appendices. 

\section{Noise--Canceling Feedback \label{sec-NCQF}}

We here summarize noise-canceling feedback.
Full derivations, generalizing beyond these assumptions, appear in Appendix \ref{sec-NCQF-fullderive}. 
We first define the problem, and then show that for pure states a completely noise--canceling solution always exists. 
We characterize the dynamics under that solution in terms of a novel non-Hermitian effective Hamiltonian. 

\subsection{Dynamics from Continuous Quantum Monitoring and Instantaneous Feedback}

The It\^{o} Stochastic Master Equation (SME) for diffusive quantum trajectories typically reads
\be \label{ITO_SME} \begin{split}
d\hat{\rho} = & \left(\hat{L}\,\hat{\rho}\,\hat{L}^\dag - \tfrac{1}{2}\,\hat{\rho}\,\hat{L}^\dag\hat{L} - \tfrac{1}{2}\,\hat{L}^\dag\hat{L}\,\hat{\rho}\right)\,dt \\
& + \sqrt{\eta} \left(\hat{L}\,\hat{\rho} + \hat{\rho}\,\hat{L}^\dag - \hat{\rho}\,\tr{\hat{L}\,\hat{\rho} + \hat{\rho}\,\hat{L}^\dag} \right)\,dW
\end{split} \ee
where $\hat{L}$ is Lindblad jump operator, monitored with efficiency $\eta$, and $dW$ is a diffusive Wiener process \cite{BookBarchielli, BookWiseman, BookJacobs, BookJordan}. 
The term on the first line describes (Markovian) Lindblad dissipation, while the second line describes the stochastic effect of monitoring on the state evolution, when that evolution is conditioned on the measurement record (outcomes, or readout).
We may understand the readout 
\be \label{ITO_RO}
r\,dt = \sqrt{\eta}\,\tr{\hat{L}\,\hat{\rho} + \hat{\rho}\,\hat{L}^\dag}\,dt + dW = \sqrt{\eta}\,\mathsf{s}\,dt + dW
\ee
as a sum of expected signal $\mathsf{s} = \langle \hat{L} + \hat{L}^\dag \rangle$ plus noise $dW$. 

Instantaneous coherent feedback can be added in the It\^{o} picture by considering $\hat{\mathcal{U}}\,(\hat{\rho} + d\hat{\rho})\,\hat{\mathcal{U}}^\dag$, where $\hat{\mathcal{U}} = e^{-i\,\hat{dh}}$ is an infinitesimal increment of the unitary evolution driven by open--loop $\hat{\Omega}$ controls and noise--proportional feedback $\hat{\omega}$, as per $\hat{dh} = \hat{\Omega}\,dt + \hat{\omega}\,dW$ \cite{martin2015deterministic, martin2017optimal, Zhang2017, zhang2020locally, Cardona_2018, Cardona_2020}. 
The resulting evolution reads 
\begin{subequations} \label{ito-fb-singlemeas} \be \begin{split} 
d\hat{\rho} &= \left\lbrace i\left[\hat{\rho},\hat{\Omega}\right] + i\left[\widehat{\mathsf{b}}_L[\hat{\rho}],\hat{\omega}\right] + \widehat{\mathsf{a}}_L[\hat{\rho}] + \widehat{\mathsf{a}}_\omega[\hat{\rho}]   \right\rbrace dt \\ & \quad\quad+ \left\lbrace \widehat{\mathsf{b}}_L[\hat{\rho}] + i\left[\hat{\rho},\hat{\omega}\right] \right\rbrace dW \\
&= \widehat{\mathcal{A}}[\hat{\rho}]\,dt + \widehat{\mathcal{B}}[\hat{\rho}]\,dW,
\end{split}
\label{eq:SME_wfb}\ee
where we have defined the superoperators 
\be 
\widehat{\mathsf{a}}_\bullet[\hat{\rho}] = \hat{\bullet}\,\hat{\rho}\,\hat{\bullet}^\dag -\tfrac{1}{2}\,\hat{\bullet}^\dag\hat{\bullet}\,\hat{\rho} - \tfrac{1}{2}\,\hat{\rho}\,\hat{\bullet}^\dag\hat{\bullet}, \quad\text{and}
\ee \be 
\widehat{\mathsf{b}}_\bullet[\hat{\rho}] = \sqrt{\eta}\left( \hat{\bullet}\,\hat{\rho} + \hat{\rho}\,\hat{\bullet}^\dag - \hat{\rho}\,\tr{\hat{\bullet}\,\hat{\rho} + \hat{\rho}\,\hat{\bullet}^\dag }\right).
\ee \end{subequations}

For a pure state $\ket{\psi}$, this may equivalently be expressed as a Stochastic Schr\"{o}dinger Equation (SSE) 
\be \label{CSSE_full} \begin{split}
d\ket{\psi} =& -i \left\lbrace\hat{\Omega} + \hat{\omega}\,\hat{L} - \tfrac{i}{2}\left(\hat{L}^\dag\hat{L} + \hat{\omega}^2 \right) + \tfrac{i}{2}\,\hat{\mathcal{J}} \right\rbrace\ket{\psi}\,dt \\ &+\left(\hat{L} - i\,\hat{\omega} - \tfrac{1}{2}\,\mathsf{s}\,\hat{\mathds{1}} \right)\ket{\psi}\,dW
\end{split} \ee 
where we have defined $\hat{\mathcal{J}} = \mathsf{s}\,(\hat{L}+i\,\hat{\omega} -\tfrac{1}{4}\,\mathsf{s}\,\hat{\mathds{1}} )$, and the first line is analogous to $\widehat{\mathcal{A}}$ in \eqref{ito-fb-singlemeas}, and the second line to $\widehat{\mathcal{B}}$ in \eqref{ito-fb-singlemeas}. 

\subsection{Perfect Noise Cancellation: Solution and Dynamics}

\begin{figure}
\includegraphics[width = \columnwidth]{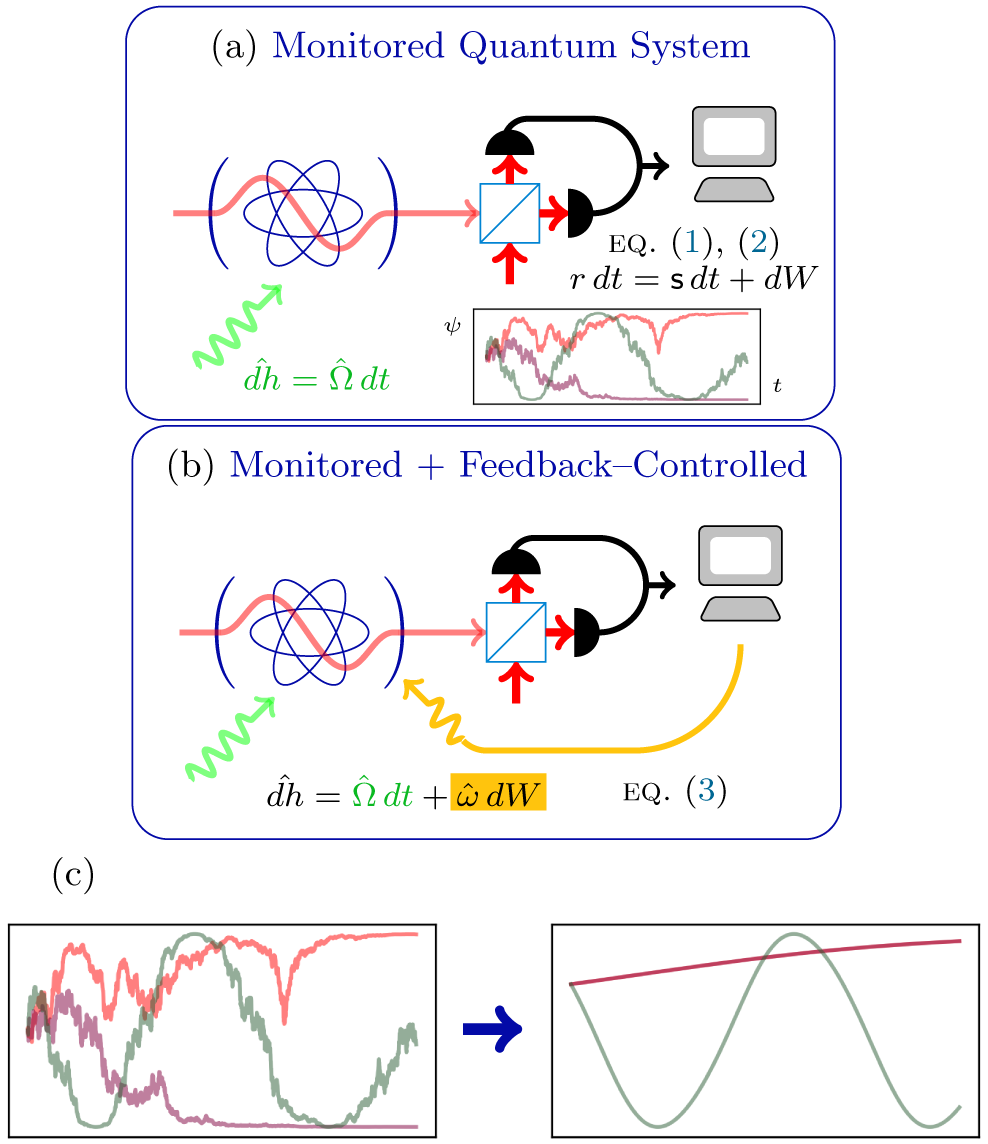}
\caption{
Cartoon illustration of noise-canceling quantum feedback (NCQF). 
A monitored quantum system is depicted in panel (a): the system (cartoon atom) is probed and then measured (red optics), leading to a classical measurement record $r$ \eqref{ITO_RO}. Open loop controls (green optics) may or may not be present. 
The result is that the user has access to stochastic quantum trajectories, i.e.~dynamics conditioned on the stochastic measurement record, illustrated qualitatively in the inset. Such dynamics are solutions to \eqref{ITO_SME}. 
In panel (b), the system is supplemented with feedback of the form \eqref{ito-fb-singlemeas}, meaning that we now consider the possibility of adding coherent feedback proportional on the measurement noise, and potentially dependent on the real--time conditional state of the system.  
In panel (c) we illustrate the aim of \emph{noise--canceling} feedback specifically: We aim to cancel all stochasticity out of the conditional dynamics (left), leaving behind dynamics that, while still influenced by the dissipation and control protocol, are \emph{completely deterministic} and smooth (right). 
One of the main results of this manuscript is to demonstrate that it is \emph{always} possible to generate a NCQF protocol of the type illustrated given pure states and instantaneous feedback. 
} \label{fig_cartoon_statement}
\end{figure}

\eqref{CSSE_full} describes measurement-- and feedback--controlled dynamics for pure states, and \eqref{ito-fb-singlemeas} describes the same generalized to mixed states. 
These equations both contain terms expressing deterministic (or average) dynamics $\sim dt$, and noisy backaction $\sim dW$. 
Feedback that is \emph{noise--canceling} is feedback that ensures that the noisy backaction terms vanish. 
In other words, complete noise--cancellation implies that we choose $\hat{\omega}$ such that
\begin{subequations} \label{NCC_both} \be 
\left(\hat{L} - i\,\hat{\omega} -\tfrac{1}{2}\,\mathsf{s}\,\hat{\mathds{1}} \right) \ket{\psi} = 0
\ee 
at every instant of our system's evolution. 
The mixed state generalization of the above reads
\be 
\widehat{\mathsf{b}}_L[\hat{\rho}] + i[\hat{\rho},\hat{\omega}] = 0.
\label{eq:NCQF_cond_rho}
\ee 
\end{subequations}
We call \eqref{NCC_both} the Noise--Cancellation Condition (NCC), and
in the event that we satisfy it over a sustained time interval of our evolution, we will have  \emph{deterministically} created the dynamics
\be \label{NC-SSE-SME}
\ket{\dot{\psi}} = -i\,\hat{\mathcal{H}}_{NC}\,\ket{\psi} \quad\leftrightarrow\quad \dot{\rho} = \widehat{\mathcal{A}}[\hat{\rho}].
\ee 
For the pure--state case we have defined the effective non-Hermitian Hamiltonian (nHH), from \eqref{CSSE_full}, as
\be \label{NC_NHH_single}
\hat{\mathcal{H}}_{NC} \equiv \hat{\Omega} + \hat{\omega}\,\hat{L} - \tfrac{i}{2}\left(\hat{L}^\dag\hat{L} + \hat{\omega}^2 \right) + \tfrac{i}{2}\,\mathsf{s}\,\left(\hat{L}+i\,\hat{\omega} -\tfrac{1}{4}\,\mathsf{s}\,\hat{\mathds{1}} \right);
\ee
this operator generates the noise--canceled dynamics. 
We re-iterate that full derivations of all of the objects shown here, including the generalization to multiple measurement and feedback channels, is detailed in Appendix~\ref{sec-NCQF-fullderive}. 
A cartoon summarizing the main setting of this paper, and the qualitative properties of NCQF, appears in Fig.~\ref{fig_cartoon_statement}.

We now state one of the main results of this manuscript: NCQF, as just defined, is \emph{always} possible on a pure state $\ket{\psi}$, given unit--efficiency measurements and instantaneous feedback. 
In particular, for $\hat{\rho} = \ket{\psi}\bra{\psi}$ pure and a rank--$1$ projector, it is easy to verify that
\begin{subequations} \label{basic_NCQF_solution} \be 
\hat{\omega}_0 = i\left([\hat{\rho},\hat{X}] - \hat{Y} \right),
\ee \be 
\text{with}\quad \hat{X} \equiv \tfrac{1}{2}\left( \hat{L} + \hat{L}^\dag \right), \quad \hat{Y} \equiv \tfrac{1}{2}\left( \hat{L} - \hat{L}^\dag \right),
\ee \end{subequations}
offers a control solution that satisfies \eqref{NCC_both}. 
This solution works by canceling the stochasticity from transitions between the current state--space $\hat{\Pi} = \ket{\psi}\bra{\psi}$, and the orthogonal space $\hat{\Pi}^\perp = \hat{\mathds{1}} - \hat{\Pi}$; this means that further feedback terms that act purely in $\hat{\Pi}$ or $\hat{\Pi}^\perp$ can be added to the above solution without jeopardizing its fundamental noise cancellation properties. 
The use of these free parameters is discussed at length in \secref{sec_Restricted_Controls}, and Appendices \ref{app_omega_dynamics} and \ref{app_Free_params}.

\subsection{Properties of Noise--Canceled Dynamics and effective non-Hermitian Hamiltonian\label{sec-Meas-ES}}

Let us consider some properties of the dynamics under noise--cancellation (which apply generally, for the basic solution \eqref{basic_NCQF_solution} but also for other noise--canceling solutions).
Note that perfect noise cancellation is possible for impure $\hat{\rho}$ only if the diagonal elements of $\hat{X}$, projected onto the subspace where $\hat{\rho}$ has support, are all equal to $\tfrac{1}{2}\,\mathsf{s}$. See Appendix \ref{app_omega_dynamics}\,\secref{sec_NC-mixed} for proof and details.

Suppose we find ourselves at a fixed point of the measurement--only dynamics, i.e.~we find ourselves at a state $\hat{\rho}_\circ$ where $d\hat{\rho} = i[\hat{\rho}_\circ,\hat{\Omega}]dt + \widehat{\mathsf{a}}_L[\hat{\rho}_\circ]dt + \widehat{\mathsf{b}}_L[\hat{\rho}_\circ]dW = 0$. 
The noise cancellation condition $\widehat{\mathcal{B}}[\hat{\rho}_\circ] = 0$ then reduces simply to $[\hat{\rho}_\circ,\hat{\omega}] = 0$, and the deterministic drift then also simplifies to $\widehat{\mathcal{A}}[\hat{\rho}_\circ] = \widehat{\mathsf{a}}_\omega[\hat{\rho}_\circ] = 0$, because the Lindbladian $\widehat{\mathsf{a}}_\bullet[\hat{\circ}]$ necessarily vanishes for any pair of operators for which $[\hat{\bullet},\hat{\circ}] = 0$. 
We thereby confirm an intuitive result: All fixed points of the dynamics without feedback are necessarily preserved by the addition of NCQF. 
This is a fundamental way in which NCQF preserves structural features of the measurement dynamics, with wide ramifications for this class of feedback protocols. 

Consider for instance, eigenstates of a measurement channel $\hat{L}$: These are necessarily fixed points of the measurement dynamics \eqref{ITO_SME} when $\hat{\Omega} = 0$, and we may thereby understand that the pure--state NCQF dynamics will tend to asymptotically stabilize to a measurement eigenstate (see Appendix \ref{app_omega_dynamics}\,\secref{sec-nHH_stability}), if no open--loop controls or free feedback parameters are used to explicitly orient the dynamics to some other purpose. 

Noise--canceling dynamics are almost exclusively state--dependent. One implication of this is that
while the effective Schr\"{o}dinger equation \eqref{NC-SSE-SME} may appear to suggest that the noise--canceled dynamics are linear in $\ket{\psi}$, this is generally not the case, because $\hat{\omega}$ (and therefore $\hat{\mathcal{H}}_{NC}$ itself), often ends up depending implicitly on $\ket{\psi}$ in order to satisfy \eqref{NCC_both}. 
In particular, it is evident that our basic solution \eqref{basic_NCQF_solution} $\hat{\omega}_0$ does generically depend on $\hat{\rho} = \ket{\psi}\bra{\psi}$, except in two special cases: We may have state--independent noise cancellation if we engineer an anti-Hermitian Lindblad operator (see Refs.~\cite{Szigeti_2014, Saiphet_2021} for detailed discussion of such ``no--knowledge'' measurements), or if $\hat{\rho}$ happens to be an eigenstate of $\hat{X}$, such that the commutator $[\hat{\rho},\hat{X}]$ vanishes. 
Outside of these special cases however, perfect noise cancellation generically \emph{requires} real--time knowledge of the measurement--conditioned dynamics. 
While we will find it useful below to consider the complex eigenspectrum of $\hat{\mathcal{H}}_{NC}$, we acknowledge that this is a potentially bizarre thing to do: $\hat{\mathcal{H}}_{NC}$ is defined with respect to a specific state, and care should be taken when abstracting the operator $\hat{\mathcal{H}}_{NC}$ away from that context. 

The NCC \eqref{NCC_both} can be expressed as an eigenvalue equation for the ``noise operator'' $\hat{\Xi} \equiv \hat{L} - i\,\hat{\omega} = \hat{X} - i\,\tilde{\omega}$ for $\tilde{\omega} \equiv \hat{\omega} + i\,\hat{Y}$, as per
\be \label{NCC_Xi_main}
\hat{\Xi}\ket{\psi} = \left(\tfrac{1}{2}\,\mathsf{s} - i\,\omega_\psi \right)\ket{\psi}, 
\ee
i.e.~the feedback $\hat{\omega}$ within $\hat{\Xi}$ is chosen such that the current state $\ket{\psi}$ is an eigenvector  of $\hat{\Xi}$, with eigenvalue $\Re[\lambda] =\tfrac{1}{2}\,\mathsf{s}$, to guarantee noise cancellation. 
\textcite{zhang2020locally} mention the existence of a global phase parameter in this context; we have here written this as an optional imaginary part of the eigenvalue. 
We say that this is optional because the term arises as a free parameter in our feedback operation $\hat{\omega}$ (see Appendix \ref{app_omega_dynamics}, and \eqref{NCC_phase} in particular, to see how this comes about).
Specifically, this phase emerges in the pure--state case if we add a free component to $\hat{\omega}$ proportional to $\hat{\Pi} = \ket{\psi}\bra{\psi}$, i.e.~we may add $\omega_\psi\,\hat{\Pi}$ to a noise--canceling solution such as \eqref{basic_NCQF_solution}, without impeding its noise--cancellation properties.
We may re-write the nHH \eqref{NC_NHH_single} as
\be \label{HNC_Xi} \begin{split} 
\hat{\mathcal{H}}_{NC} = &~\hat{\Omega} + i\,\mathsf{s}\,\hat{Y} - \tfrac{i}{2}\left( \hat{Y}\,\hat{\Xi} + \hat{\Xi}^\dag\,\hat{Y} \right) 
\\&+ \tfrac{i}{4}\left\lbrace \hat{\Xi}^2 - \hat{\Xi}^{\dag^2} - 2\,\hat{\Xi}^\dag\hat{\Xi} + 2\mathsf{s}\left( \hat{\Xi}^\dag - \tfrac{1}{4}\,\mathsf{s}\,\hat{\mathds{1}}\right) \right\rbrace
\\=&~\hat{\Omega} + i\,\mathsf{s}\,\hat{Y} - \tfrac{1}{2}\,\mathsf{s}\,\tilde{\omega} - \tfrac{i}{2}\left( \hat{Y}\,\hat{\Xi} + \hat{\Xi}^\dag\,\hat{Y} \right) 
\\&+\tilde{\omega}\,\hat{X} - \tfrac{i}{2}\left(\Delta\hat{X}^2 + \tilde{\omega}^2 \right) ~~\text{with}~\Delta\hat{X} = \hat{X} - \tfrac{1}{2}\,\mathsf{s}\,\hat{\mathds{1}},
\end{split}\ee
where the first form is strictly in terms of the noise operator, and the second splits $\hat{\mathcal{H}}_{NC}$ into its Hermitian and anti-Hermitian parts on each line (with the lone exception of the term $\tilde{\omega}\,\hat{X}$). 
It is helpful to write $\hat{\mathcal{H}}_{NC}$ in terms of $\hat{\Xi}$ because \eqref{NCC_Xi_main} indicates that the current state is necessarily an eigenstate of $\hat{\Xi}$ when the noise is canceled.
See \eqref{noise_sub-forms} for further details about the derivation of \eqref{HNC_Xi} from \eqref{NC_NHH_single}.

Based on the discussion of the nHH and fixed points above, we may suppose that the most--stable eigenstate of $-i\hat{\mathcal{H}}_{NC}$ in the long--time limit will also be a measurement eigenstate, and therefore be the asymptotic terminus of the dynamics (absent open--loop controls, or use of free parameters explicitly to the contrary).
Precise comments to this effect may be found in Appendix \ref{app_omega_dynamics}\,\secref{sec-nHH_stability}.
When $\hat{\mathcal{H}}_{NC}$ acts on the intended and current state $\ket{\psi}$ such that noise is canceled, we may now readily find that its expectation value evaluates to
\be \label{HNC_xi-proj} \begin{split}
-i\bra{\psi}\hat{\mathcal{H}}_{NC}\ket{\psi} & = \ensavg{\tfrac{1}{2}\,\mathsf{s}\,\hat{Y}-i\,\hat{\Omega}}-\tfrac{1}{2}\,\omega_\psi^2. 
\end{split}\ee
Such an expression is particularly helpful in the long--time limit, when the current state often converges onto an eigenstate of the nHH (such that the expectation value above is also the eigenvalue).

In the event of a mixed initial state, the terminal point may be either a single pure measurement eigenstate, or a mixture of several of them, depending on how much the intervening NCQF dynamics are able purify the initial state. 
The importance of these points will emerge repeatedly in the concrete examples we develop below. 
In summary, while NCQF can be tailored with free feedback parameters to some extent, we view NCQF's ``default'' behavior as follows: This type of feedback does \emph{not fundamentally alter} the dynamics created by measurements and open loop control that have been put in place; rather, adding NCQF to a system \emph{enhances what is already there}, by eliminating stochasticity from the dynamics while leaving those dynamics' broad structural features in place. 

\subsection{NCQF with Restricted Controls \label{sec_Restricted_Controls}}

We have remarked on the existence of free parameters within NCQF that allow for feedback operators to be tailored, while maintaining perfect noise cancellation.
There are two potential strategies to fruitfully use the feedback parameters: One is to construct $\hat{\omega}_0$ as per \eqref{basic_NCQF_solution}, and then modify the feedback by adding specific free parameters (we discuss this approach in Appendix \ref{app_omega_dynamics}$\,$\secref{sec-NCC_Free}; see \eqref{NCQF_form_general} in particular). 
The alternative, which we explore in this section, is a direct test of whether a given form of $\hat{\omega}$ may satisfy the NCC.
To accomplish this, we re-formulate the NCC in terms of noise magnitude, which may be zeroed or minimized, and can be straightforwardly expressed in terms of flexible control constraints.

\subsubsection{Noise Magnitude}

Let us modestly adjust the notation of \eqref{ito-fb-singlemeas}, such that the noise term in dynamics with continuous monitoring and instantaneous feedback may be expressed
\begin{subequations} 
\be
    \hat{\mathcal{B}}=\hat{\mathcal{B}}^\dagger = \sqrt{\eta}\,\hat{K}-i\left[\hat{\omega},\hat{\rho}\right], \quad\text{with}
\ee\be
\hat{K}=\left\{\Delta\hat{X},\hat{\rho}\right\}+\left[\hat{Y},\hat{\rho}\right].
\ee \end{subequations}
We then define the ``noise magnitude'' $\mathcal{N}$ similar to a squared trace--norm of $\hat{\mathcal{B}}$, i.e.~we define
\begin{subequations} \label{noise-quad} \be \label{noise-mag}
\mathcal{N} = \tr{\hat{\mathcal{B}}^\dag\hat{\mathcal{B}}}
\ee
such that $\mathcal{N}\,dt$ may be understood as giving a measure of the scale of the noise entering our dynamics at any given moment. 
This quantity is real and non-negative by construction.
We may write $\mathcal{N}  = \mathcal{N}_a + \mathcal{N}_b + \mathcal{N}_c$ for 
\be 
\mathcal{N}_a = \tr{[\hat{\omega},\hat{\rho}][\hat{\rho},\hat{\omega}]} = \tr{[\hat{\rho},\hat{\omega}]^\dag[\hat{\rho},\hat{\omega}]},
\ee \be 
\mathcal{N}_b = 2i\sqrt{\eta}\,\tr{\hat{\omega}\left[\hat{K},\hat{\rho}\right]}, 
\ee \be 
\mathcal{N}_c = \eta\, \tr{\hat{K}^2}. 
\ee \end{subequations}
We have divided these terms according their having (a) a quadratic dependence on the feedback control $\hat{\omega}$, (b) a linear dependence on $\hat{\omega}$, and (c) no dependence on $\hat{\omega}$. 
This implies that $\mathcal{N}_c$ is, on its own, the noise magnitude in the absence of any feedback. 
Notice that $\mathcal{N}_a$ is also strictly non-negative, and only vanishes when the controls satisfy $[\hat{\omega},\hat{\rho}] = 0$, such that the controls are not impactful. 
Moreover, $\mathcal{N}_b$ is necessarily real, since everything inside that term's trace is anti-Hermitian. 

\subsubsection{Noise Minimization with Restricted Controls}

We may now utilize \eqref{noise-quad} to formulate a condition to i) test whether or not a given form of the controls may be used to achieve noise cancellation, or ii) minimize the noise with a given control that cannot completely cancel the noise. 
Suppose that we have controls $\hat{\omega} = f\,\hat{\Theta}$, where $\hat{\Theta}$ is the form of the control operator we wish to test (which should be Hermitian and traceless), and where $f$ is a real scalar to be determined. 
We evaluate \eqref{noise-quad} for such a control, obtaining
\begin{subequations} \label{f-quad-single-general} \begin{align}
\mathcal{N}(f) &= f^2\,a + f\,b + c \quad\text{for} \\
a &= \mathcal{N}_a |_{\hat{\omega} = \hat{\Theta}} = \tr{[\hat{\Theta},\hat{\rho}][\hat{\rho},\hat{\Theta}]}, \\
b &= \mathcal{N}_b |_{\hat{\omega} = \hat{\Theta}} = 
2i\,\sqrt{\eta} \,\tr{\hat{\Theta}\left[\hat{K},\hat{\rho}\right]},\\
c &= \mathcal{N}_c.
\end{align} \end{subequations}
The solution to $\mathcal{N}(f) = 0$ is of course
\be \label{quad_sol}
f = \frac{-b \pm \sqrt{b^2-4\,a\,c}}{2\,a}.
\ee
If a real solution \eqref{quad_sol} exists, then perfect cancellation of the noise generated by monitoring $\hat{\rho}$ with $\hat{L}$ and efficiency $\eta$ is possible using the feedback $\hat{\omega} = f\,\hat{\Theta}$. 
For this to be the case, the discriminant of the quadratic equation (which will always be real) must be positive, i.e.~there is a perfectly noise--canceling solution under the control constraint $\hat{\omega} = f\,\hat{\Theta}$ iff
\be 
b^2 - 4\,a\,c \geq 0. 
\ee
Since $\mathcal{N}(f) \geq 0$, we may immediately understand that $b^2 - 4\,a\,c \leq 0$ by construction, such that in practice we will either find a unique root when $b^2 - 4\,a\,c = 0$, or no root at all.
In the event that no root exists, we may still however \emph{minimize} the noise (since $a\geq0$). 
The noise magnitude is minimized by choosing the minimum value of $\mathcal{N}(f)$, i.e.~the control 
\be \label{fmin-single}
f^\star = -\frac{b}{2\,a} \quad\rightarrow\quad \mathcal{N}(f^\star) = c - \frac{b^2}{4\,a}
\ee
minimizes the noise, reducing the noise magnitude (from $c$, which is the noise magnitude without feedback) by an amount $b^2/4a$. 
Perfect noise cancellation occurs when it so happens that $\mathcal{N}(f^\star) = 0$. 
Note that the minimum value obtainable depends on the choice of $\hat{\Theta}$. 
Some control choices are better than the others, and the primary purpose of this formulation is to test the effectiveness of candidate controls $\hat{\Theta}$ that are physically easy or desirable to implement. 
We also clarify that there is no contradiction between the unique optimum we find here (under control constraints), and the existence of an infinite family of solutions mentioned elsewhere (when any coherent feedback is allowed).

Generalizations of these results to controls of the form $\hat{\omega} = \sum_j f_j\,\hat{\Theta}_j$, and special cases of direct use in the examples considered in later sections, appear in Appendix \ref{app_Free_params}. 
We stress that all of the calculations above may be used to establish the existence of, and then calculate, a NC feedback $\hat{\omega}$ \emph{at a particular state}, and therefore at a particular time in a system's evolution. 
It will be interesting to see whether the existence of a perfectly noise--canceling solution at a certain instant in time guarantees the same throughout the evolution, for a consistent choice of $\hat{\Theta}$.

\section{NCQF for entangled state preparation \label{sec-nHH_examples}}

We here reconsider a few known examples of noise-canceling quantum feedback from the literature, from the perspective of the non-Hermitian Hamiltonian \eqref{NC_NHH_single}. 
As we mentioned in the introduction, non-Hermitian Liouvillians and effective Hamiltonians have garnered growing research interest in quantum information science recently \cite{Ashida_2020}. 
In the previous section, we have demonstrated a method by which to \emph{deterministically} obtain an effective non-Hermitian Hamiltonain (nHH), avoiding the post--selection cost that is usually associated with realizing such objects in the quantum setting. 
We consequently adapt some of the ``usual'' analysis strategies of the nHH literature in this setting, and apply them to NCQF situations where \eqref{NC_NHH_single} applies.
Specifically, we will consider the entanglement of two qubits by measurement, first via continuous half--parity monitoring \cite{martin2015deterministic, martin2017optimal, zhang2020locally, Elouard2019singleshotenergetic}, and then via joint fluorescence measurement \cite{martin2019single, Lewalle:21, Lewalle_2020_limitcycle}. 
In both examples, entanglement can be deterministically generated via NCQF that maintains the dynamics perpetually within a subspace where the expected signal is zero ($\mathsf{s} = 0$) for all measurements. 
While these characteristics are shared with optimal solutions known from prior literature, \eqref{basic_NCQF_solution} does lead to slightly different controls leading to slightly different dynamics compared to those derived previously \cite{zhang2020locally, martin2019single}. 
We have mentioned the existence of free parameters and multiple solutions above, and this discussion will be made concrete by comparing the solutions \eqref{basic_NCQF_solution} and/or \eqref{fmin-single} to those from the prior literature.
We will furthermore discuss the behavior of the nHH \eqref{NC_NHH_single} along the entanglement--generating NCQF dynamics, and discuss stabilization of the entanglement in terms of the nHH. 

\subsection{Bell States via Half--Parity NCQF \label{sec:half-par}}

\begin{figure}
\includegraphics[width= \columnwidth]{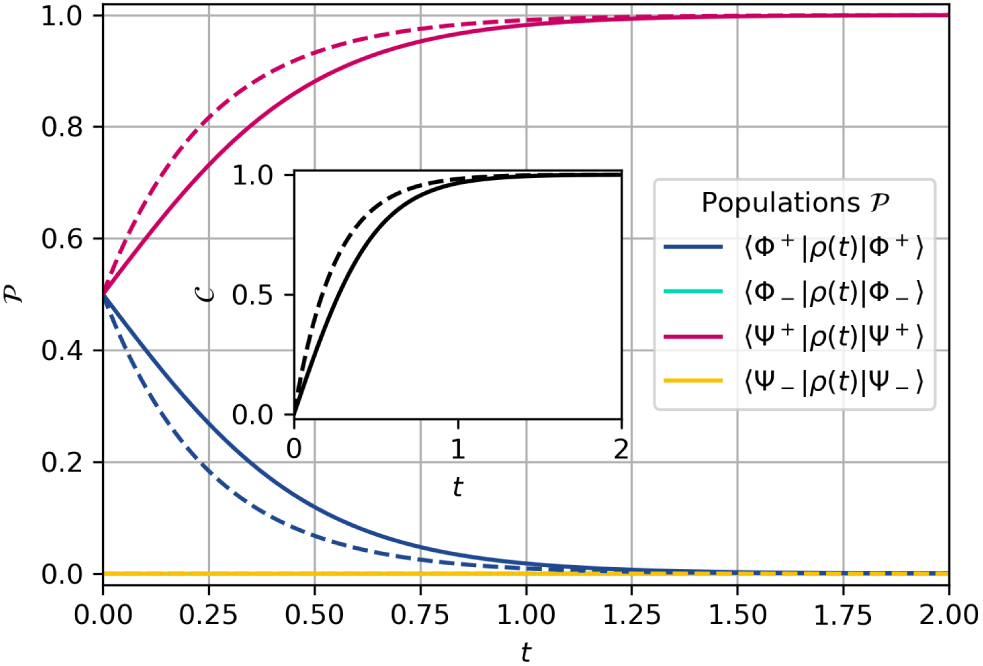}
\caption{
We illustrate the dynamics arising from applying NCQF to the half--parity measurement \eqref{L_hp}. 
Solid lines show the dynamics arising from the feedback rule \eqref{basic_NCQF_solution}, while the dashed lines show dynamics arising from a locally--optimal feedback rule \cite{martin2015deterministic, martin2017optimal, zhang2020locally}; \emph{both} sets of controls are perfectly noise canceling. 
Populations are plotted in the Bell basis in the main figure, with $\ket{\Phi^+_-} \sim \ket{ee} \pm \ket{gg}$ and $\ket{\Psi^+_-} \sim \ket{eg} \pm \ket{ge}$.  
Concurrence $\mathcal{C}$ \cite{Wooters_1998} is plotted in the inset. 
Time, on the horizontal axes, is shown in units of the inverse measurement rate $\Gamma^{-1}$ [see \eqref{L_hp}]. 
The initial state is the separable state $\left(\ket{e}/\sqrt{2} + \ket{g}/\sqrt{2} \right)^{\otimes 2}$. 
We see that half--parity monitoring supplemented by NCQF leads to deterministic (and asymptotically stable) preparation of the Bell state $\ket{\Psi^+}$, with dynamics that stay entirely within the $\mathsf{s}_{hp} = 0$ subspace (this is true for both feedback rules shown). 
We see that a simple solution \eqref{basic_NCQF_solution}, derived from the premise of noise cancellation \eqref{NCC_both}, performs almost as well as a previously known optimal solution (which happens to also be noise canceling). 
} \label{fig:half-par-dynamics}
\end{figure}

The two--qubit half--parity measurement is characterized by the Lindblad operator 
\be \label{L_hp}
\hat{L}_{hp} = \sqrt{\Gamma}\,\left(\hat{\mathds{1}}\otimes \hat{\sigma}_z + \hat{\sigma}_z \otimes \hat{\mathds{1}} \right) = \sqrt{\Gamma}\left(\begin{array}{cccc}  
2 & 0 & 0 & 0 \\ 0 & 0 & 0 & 0 \\ 0 & 0 & 0 & 0 \\ 0 & 0 & 0 & -2
\end{array}\right),
\ee
which is effectively a measurement of the excitation number of a two--qubit system. 
We begin with this system because this measurement is known to be entangling, and moreover the optimal local feedback based on diffusive monitoring of \eqref{L_hp} is known to be noise canceling \cite{martin2015deterministic, martin2017optimal, zhang2020locally}. 
We compare the dynamics under the locally--optimal feedback rule $\hat{\omega}^\star_{hp} = f(\hat{\rho})(\hat{\sigma}_y \otimes \hat{\mathds{1}} + \hat{\mathds{1}} \otimes \hat{\sigma}_y)$ \cite{martin2015deterministic, martin2017optimal, zhang2020locally} against our NCQF rule \eqref{basic_NCQF_solution} in Fig.~\ref{fig:half-par-dynamics}. 
It is clear the the solutions are \emph{not} identical, which highlights the fact that NCQF solutions are not unique in general. 
However, the solutions are quite similar: They both deterministically transform the separable state $\left(\ket{e}/\sqrt{2} + \ket{g}/\sqrt{2} \right)^{\otimes 2}$ into the Bell state $\ket{\Psi^+}$, and asymptotically stabilize that entangled state. 
This is possible because the stabilized Bell state is in the nullspace of the measurement \eqref{L_hp}, such that the half--parity measurement can naturally Zeno--stabilize that Bell state \cite{Lewalle_CDJP} provided the dynamics can be guided there. 
We do find that the locally--optimal solution of \cite{martin2015deterministic, martin2017optimal, zhang2020locally} can be re-derived by using \eqref{quad_sol} and \eqref{f-quad-single-general} in conjunction with $\hat{\Theta} = \hat{\sigma}_y^{(1)} + \hat{\sigma}_y^{(2)}$. 
It equivalently re-emerges as the unique perfectly noise--canceling solution to \eqref{f-quad-multi-general} given $\hat{\Theta}_1 = \hat{\sigma}_y^{(1)}$ and $\hat{\Theta}_2 = \hat{\sigma}_y^{(2)}$.

\begin{figure}
\includegraphics[width = \columnwidth]{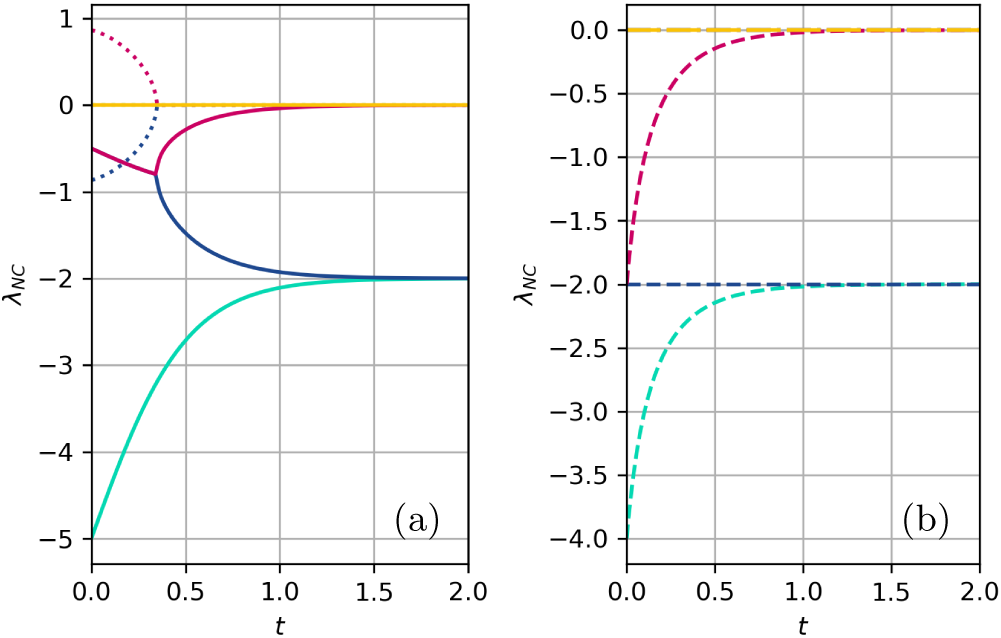}
\caption{
We plot the eigenvalues $\lambda$ of the dynamical matrix $-i\,\hat{\mathcal{H}}_{NC}$, as a function of time, for the control protocols generating the dynamics shown in Fig.~\ref{fig:half-par-dynamics}. 
Eigenvalues are shown in units of the measurement rate $\Gamma$. 
Panel (a) shows the eigenvalues associated with the NCQF rule \eqref{basic_NCQF_solution} (shown in solid lines in Fig.~\ref{fig:half-par-dynamics}), while panel (b) shows the eigenvalues associated with a locally--optimal feedback rule \cite{martin2015deterministic, martin2017optimal, zhang2020locally} (shown in dashed lines in Fig.~\ref{fig:half-par-dynamics}). 
Real parts of the eigenvalues, denoting damping of the associated right eigenstate, are shown as solid (a) or dashed (b) lines. 
The imaginary parts are shown as dotted (a) or dash--dotted (b) lines. 
In both cases, the eigenstates of $-i\,\hat{\mathcal{H}}$ as $t\rightarrow \infty$ are the standard Bell states; colors are therefore assigned to match the associated populations plotted in Fig.~\ref{fig:half-par-dynamics}. 
} \label{fig:half-par-spectra}
\end{figure}

The eigenspectrum of the effective nHH \eqref{NC_NHH_single} evaluated along the dynamics of Fig.~\ref{fig:half-par-dynamics}, shown in Fig.~\ref{fig:half-par-spectra}, offers some explanation of how the dynamics are guided to that Bell state. 
The real parts of the eigenvalues of $-i\,\hat{\mathcal{H}}_{NC}$ can be interpreted as gain/loss rates on the populations of their associated eigenstates \cite{Ashida_2020, Nenciu_1992}, while the imaginary parts dictate relative phases or oscillation frequencies (like the eigenenergies of a typical Hermitian $\hat{H}$). 
The locally--optimal feedback $\omega^\star_{hp}$ and \eqref{basic_NCQF_solution} here both lead to spectra of $-i\,\hat{\mathcal{H}}_{NC}$ that asymptotically stabilize to the standard Bell basis, with the spectrum 
\be 
\limit{t}{\infty} -i\,\hat{\mathcal{H}}_{NC}^{hp} \quad\rightarrow\quad \bigg\lbrace \begin{array}{lc}
\lambda = 0 ~~& \ket{\lambda} = \ket{\Psi^{+}_-} \\ 
\lambda = -2\Gamma ~~ & \ket{\lambda} = \ket{\Phi^{+}_-}
\end{array} .
\ee
As shown in Fig.~\ref{fig:half-par-spectra}, the initial state $\ket{\Psi^+} + \ket{\Phi^+}$ is transformed to $\ket{\Psi^+}$, because $\ket{\Phi^+}$ is lossy compared to the stabilized state, and therefore effectively cedes its population to $\ket{\Psi^+}$ over time. 
The NCQF rule \eqref{basic_NCQF_solution} leads to loss rates that are equal for some time before they split, while the locally--optimal rule \cite{martin2015deterministic, martin2017optimal, zhang2020locally} generates an anit-Hermitian $\hat{\mathcal{H}}_{NC}$ with gain/loss rates between the relevant states that split right away. 
The latter reaches the target state $\ket{\Psi^+}$ faster by virtue of separating $\ket{\Psi^+}$ and $\ket{\Phi^+}$ (splitting their relative gain/loss rates) more efficiently, but the end result is quite similar regardless. 

This behavior is an example of that described generically in \secref{sec-Meas-ES}: If $\hat{\Omega}$ is not used, asymptotic convergence towards a measurement eigenspace is the expected long--time behavior under perfect NCQF.
We have here used a measurement whose eigenspaces correspond to Bell states, such that preparation of a superposition of measurement eigenstates leads the feedback to ``pick'' one of the eigenspaces as the ``more stable'' one, and drive deterministic convergence towards that eigenspaces. 
We do not explicitly analyze the full--parity measurement here for brevity, but it is not difficult to apply the analysis above and the appropriate literature \cite{Williams_2008, martin2015deterministic, martin2017optimal, zhang2020locally} and find that the qualitative aspects of the half--parity measurement carry over to the full--parity measurement as well. 
See \secref{sec:GHZ} for continued discussion. 

\begin{figure}
\includegraphics[width = \columnwidth]{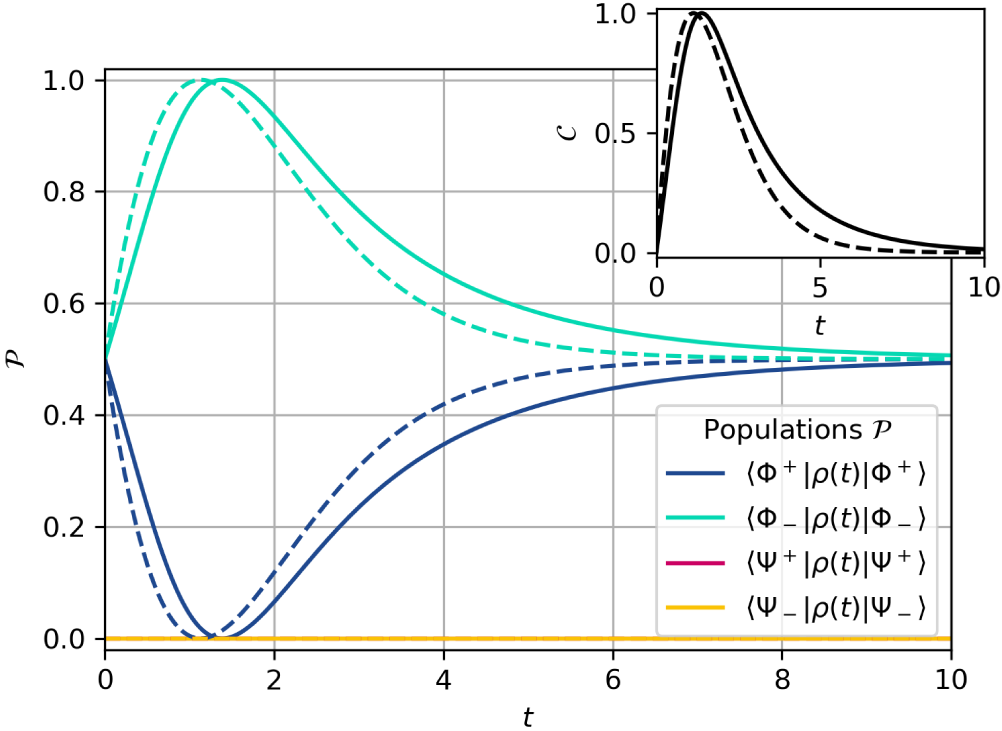}
\caption{
We illustrate the dynamics arising from applying NCQF to the joint fluorescence measurements \eqref{L_flor} \cite{Lewalle:21}. 
Solid lines show the dynamics arising from the feedback rule \eqref{basic_NCQF_solution}, while the dashed lines show dynamics arising from a locally--optimal feedback rule \cite{martin2019single, Lewalle_2020_limitcycle}; \emph{both} sets of controls are perfectly noise canceling. 
Populations are plotted in the Bell basis in the main figure, with $\ket{\Phi^+_-} \sim \ket{ee} \pm \ket{gg}$ and $\ket{\Psi^+_-} \sim \ket{eg} \pm \ket{ge}$.  
Concurrence $\mathcal{C}$ \cite{Wooters_1998} is plotted in the inset. 
Time, on the horizontal axes, is shown in units inverse to the qubits' spontaneous emission rate $\gamma^{-1}$ \eqref{L_flor}. 
The initial state is the separable state $\ket{ee}$. 
We see that monitoring supplemented by NCQF leads to deterministic preparation of the Bell state $\ket{\Psi^-}$ at a specific point in time, with dynamics that stay entirely within the $\mathsf{s}_{f1} = 0 = \mathsf{s}_{f2}$ subspace as the state evolves from $\ket{ee}$ towards $\ket{gg}$. 
Again, we find that the simple solution \eqref{basic_NCQF_solution}, derived from the premise of noise cancellation \eqref{NCC_both}, performs almost as well as a previously known optimal solution (which happens to also be noise canceling).
} \label{fig:flor-dynamics}
\end{figure}

\begin{figure}
\includegraphics[width = \columnwidth]{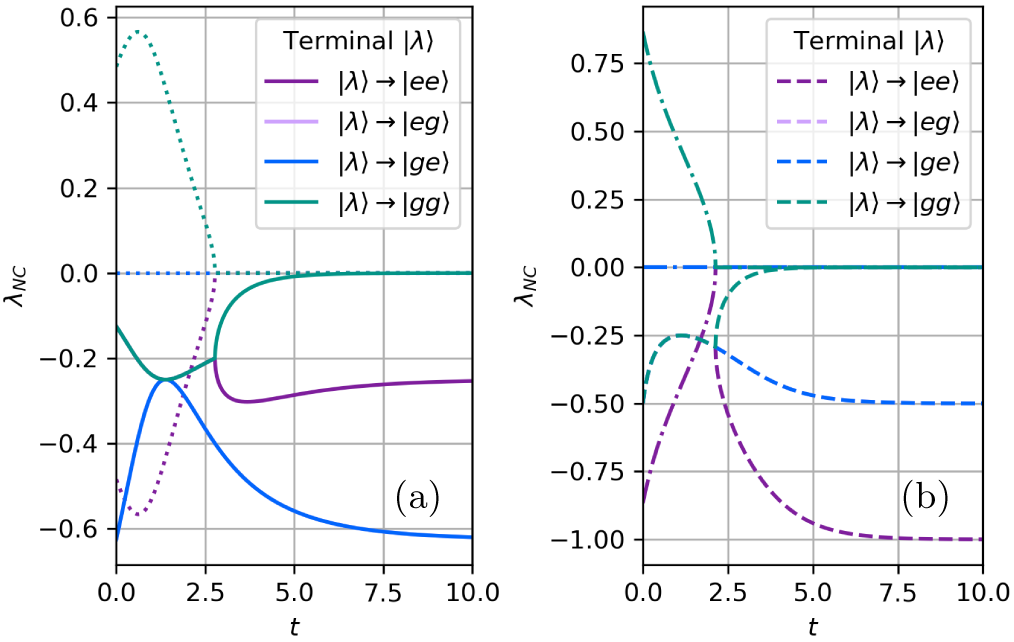}
\caption{
We plot the eigenvalues $\lambda$ of the dynamical matrix $-i\,\hat{\mathcal{H}}_{NC}$, as a function of time, for the control protocols generating the dynamics shown in Fig.~\ref{fig:flor-dynamics}. 
Eigenvalues are shown in units of the decay rate $\gamma$. 
Panel (a) shows the eigenvalues associated with the NCQF rule \eqref{basic_NCQF_solution} (shown in solid lines in Fig.~\ref{fig:flor-dynamics}), while panel (b) shows the eigenvalues associated with a locally--optimal feedback rule \cite{martin2019single, Lewalle_2020_limitcycle, zhang2020locally} (shown in dashed lines in Fig.~\ref{fig:flor-dynamics}). 
Real parts of the eigenvalues, denoting damping of the associated right eigenstate, are shown as solid (a) or dashed (b) lines. 
The imaginary parts are shown as dotted (a) or dash--dotted (b) lines. 
In both cases, the eigenstates of $-i\,\hat{\mathcal{H}}$ as $t\rightarrow \infty$ are the separable energy eigenstates, with the relative dampings supporting relaxation to the steady state $\ket{gg}$. 
Both protocols shown match their respective $\ket{eg}$ and $\ket{ge}$ damping rates. 
The locally--optimal solution (b) \cite{martin2019single, zhang2020locally} exhibits larger phase shifts between $\ket{ee}$ and $\ket{gg}$ in the initial evolution, allowing it to prepare $\ket{\Phi_-}$ more quickly than the basic NCQF of \eqref{basic_NCQF_solution}.
} \label{fig:flor-spectra}
\end{figure}

\subsection{Bell States via Fluorescence NCQF \label{sec:flor}}

Joint fluorescence measurement is characterized by the Lindblad operators
\be \label{L_flor}
\hat{L}_{f1} = \sqrt{\tfrac{\gamma}{2}}\left(\hat{\sigma}_{-}^{(1)} + \hat{\sigma}_{-}^{(2)} \right), \quad
\hat{L}_{f2} = i\,\sqrt{\tfrac{\gamma}{2}}\left( \hat{\sigma}_{-}^{(2)} - \hat{\sigma}_{-}^{(1)} \right),
\ee
where $\hat{\sigma}_{-}^{(1)} = \hat{\sigma}_- \otimes \hat{\mathds{1}}$ and $\hat{\sigma}_{-}^{(2)} = \hat{\mathds{1}} \otimes \hat{\sigma}_-$. 
The phases between \eqref{L_flor} are chosen to maximize the extent to which the measurements generate entanglement between the monitored emitters. 
Detailed discussion of the physical device implementing these measurements can be found in \cite{Lewalle:21} and references therein. 

Once again, we compare the locally--optimal feedback policy \cite{martin2019single, zhang2020locally}
\be 
\hat{\omega}_{f1}^\star = g(\hat{\rho})(\hat{\sigma}_y \otimes \hat{\mathds{1}} + \hat{\mathds{1}} \otimes \hat{\sigma}_y), ~
\hat{\omega}_{f2}^\star = g(\hat{\rho})(\hat{\sigma}_x \otimes \hat{\mathds{1}} - \hat{\mathds{1}} \otimes \hat{\sigma}_x),
\ee
to the application of \eqref{basic_NCQF_solution} to each measurement channel. 
The solutions are again similar to one another, but not identical, as illustrated in Fig.~\ref{fig:flor-dynamics}. 
In contrast with the previous example, we here have a situation wherein a Bell state is \emph{not} a measurement eigenstate.
Instead the joint kernel of the measurement \eqref{L_flor} is $\ket{gg}$, and Figs.~\ref{fig:flor-dynamics} and \ref{fig:flor-spectra} illustrate that this is the asymptotically--stable state, consistent with our remarks in \secref{sec-Meas-ES}. 
A Bell state is still deterministically created via both our basic NCQF protocol \eqref{basic_NCQF_solution}, and the locally--optimal protocol \cite{martin2019single}, but it is now something that is created at a specific time in the dynamics, rather than something which is asymptotically stabilized. 
As in the prior example, the locally--optimal feedback solutions can be recovered by using \eqref{f-quad-single-general} and \eqref{quad_sol}, or as the unique solution to \eqref{f-quad-multi-general} using local $x$ and $y$ rotations.

\begin{figure}
\includegraphics[width = \columnwidth]{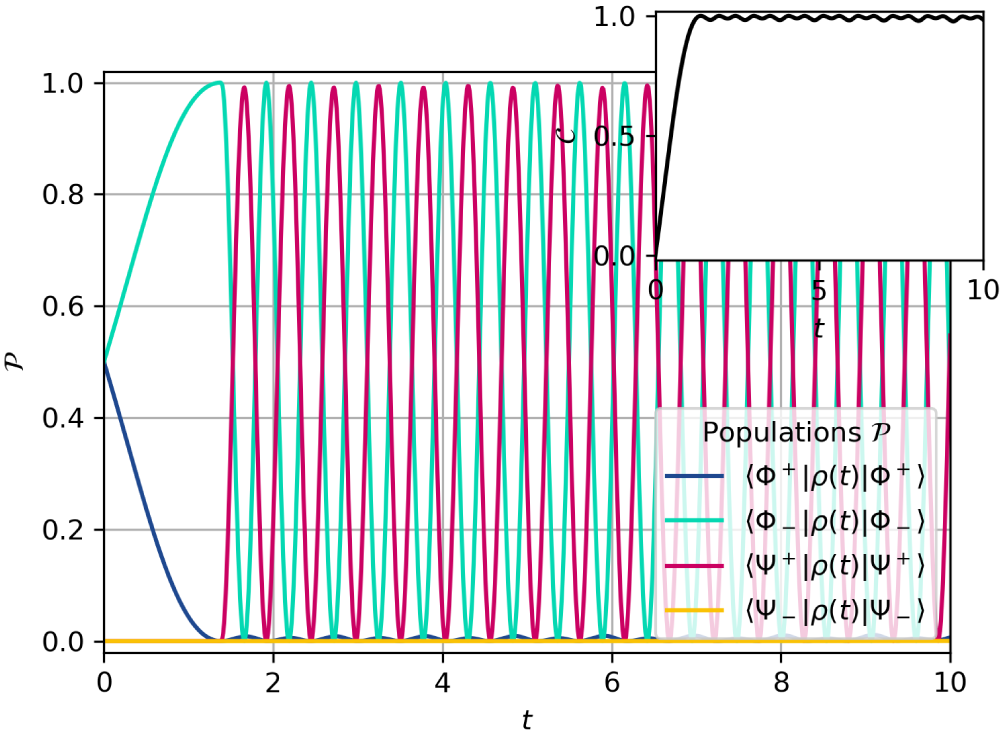}
\caption{
We repeat the NCQF example of Fig.~\ref{fig:flor-dynamics}, using the rule \eqref{basic_NCQF_solution}, but this time add a separable and simple open--loop control $\hat{\Omega}\propto \hat{\sigma}_y^{(1)} + \hat{\sigma}_y^{(2)}$ once the dynamics reach the Bell state $\ket{\Phi^-}$. 
This addition, inspired by the effectiveness of the bang--bang version of this control for entanglement stabilization \cite{Lewalle_2020_limitcycle, Sun_2010}, allows for similar results here. 
In particular, we are now able to drive deterministic flipping between two Bell states in the noise--canceled dynamics, such that the concurrence (inset) becomes trapped in a small limit cycle near $\mathcal{C} = 1$. 
As in \cite{Lewalle_2020_limitcycle}, faster local driving of the qubits can shrink the rapid concurrence oscillations arbitrarily close to $\mathcal{C} = 1$. 
We here have an example where the measurement and NCQF dynamics do \emph{not} naturally stabilize a target state that is of interest to us; however, we are still able to stabilize a quantity of interest by utilizing the open--loop degrees of control freedom $\hat{\Omega}$, which can always be tuned freely atop any NCQF rule that satisfies \eqref{NCC_both}. 
}\label{fig:flor_stabilized}
\end{figure}

Stabilizing the concurrence is however still possible if we utilize the open--loop control $\hat{\Omega}$. 
We may draw inspiration from prior literature \cite{Lewalle_2020_limitcycle, Sun_2010}, where it is found that applying fast local qubit rotations leads to entanglement stabilization under similar joint--fluorescence measurements. 
We illustrate in Fig.~\ref{fig:flor_stabilized} that this remains true when basic NCQF \eqref{basic_NCQF_solution} is supplemented with fast local rotations $\hat{\Omega} \propto \hat{\sigma}_y^{(1)} + \hat{\sigma}_y^{(2)}$, that are turned on at the moment the bare NCQF dynamics strike the Bell state $\ket{\Phi_-}$: 
Such a drive traps the dynamics in an oscillation between two Bell states, with concurrence nearly $1$ over the entire oscillatory cycle. 
Thus, this example demonstrates that the open--loop degree of control freedom can be utilized to stabilize a quantity of interest, even though the desired property is \emph{not} held by stable measurement eigenstates towards which the system converges when $\hat{\Omega} = 0$. 

\subsection{Four--Qubit GHZ State via NCQF \label{sec:GHZ}}

\begin{figure}
    \centering
    \includegraphics[width = \columnwidth]{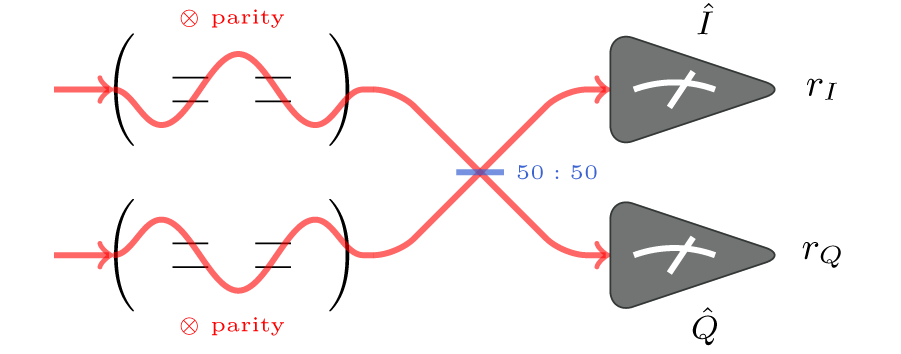}
    \caption{
    We illustrate a model in which parity probes are cascaded into a Hong--Ou--Mandel (HOM) / Bell State Measurement (BSM) device that monitors quadratures of the mixed field modes. 
    This is effectively a combination of the devices employed in making two--qubit Bell states described in earlier examples: If the output of a single cavity containing two qubits were monitored on its own, we recover parity monitoring as described in \secref{sec:half-par}. 
    If instead each cavity shown here contained a single qubit instead of two, we recover the model of \secref{sec:flor}.  
    We show in \secref{sec:GHZ} that NCQF can enable deterministic creation of Greenberger--Horne--Zeilinger (GHZ) states in the four--qubit combined device illustrated here.
    Specifically, full--parity signals that are mixed and monitored in quadratures $90^\circ$ apart, supplemented by basic NCQF \eqref{basic_NCQF_solution}, deterministically transforms $(\tfrac{1}{\sqrt{2}}\ket{e} + \tfrac{1}{\sqrt{2}}\ket{g})^{\otimes 4}$ into $\tfrac{1}{\sqrt{2}}\ket{eeee} - \tfrac{1}{\sqrt{2}}\ket{gggg}$ (up to single--qubit rotations). 
    The quadrature monitoring may be implemented via e.g.~optical homodyne detection, or the corresponding phase--sensitive amplification protocols (where the this cartoon is more suggestive of the latter, and Fig.~\ref{fig_cartoon_statement} is more suggestive of the former).
    }\label{fig:GHZ_setup}
\end{figure}

We now consider a combination of the previous two schemes, in order to form one that is (to our knowledge) novel. 
Suppose that an optical mode couples to a pair of qubits so as to serve as a pointer degree of freedom for a parity measurement; instead of amplifying/measuring this right away, we now imagine that the optical mode is mixed with another that was similarly coupled to a different qubit pair, and then joint quadrature measurements are performed. 
The optical readout is like the one we employed for the fluorescence measurement (and closely related to the Einstein--Podolsky--Rosen / EPR measurement \cite{EPR_1935, Lewalle:21}), but is now coupled in such a way as to mediate the monitoring of four qubits instead of two. 
See Fig.~\ref{fig:GHZ_setup}. 
The EPR measurement of quadratures $\hat{I} \sim \hat{a}+\hat{a}^\dag$ and $\hat{Q} \sim \hat{a} - \hat{a}^\dag$ of the mixed field can be realized via two optical homodyne detection devices \cite{Tyc2004, Reid_2009}, or the corresponding amplification hardware \cite{Reid_1989, Flurin_2012, Silveri_2016}. 

\begin{figure*}
\centering
\includegraphics[width = \textwidth]{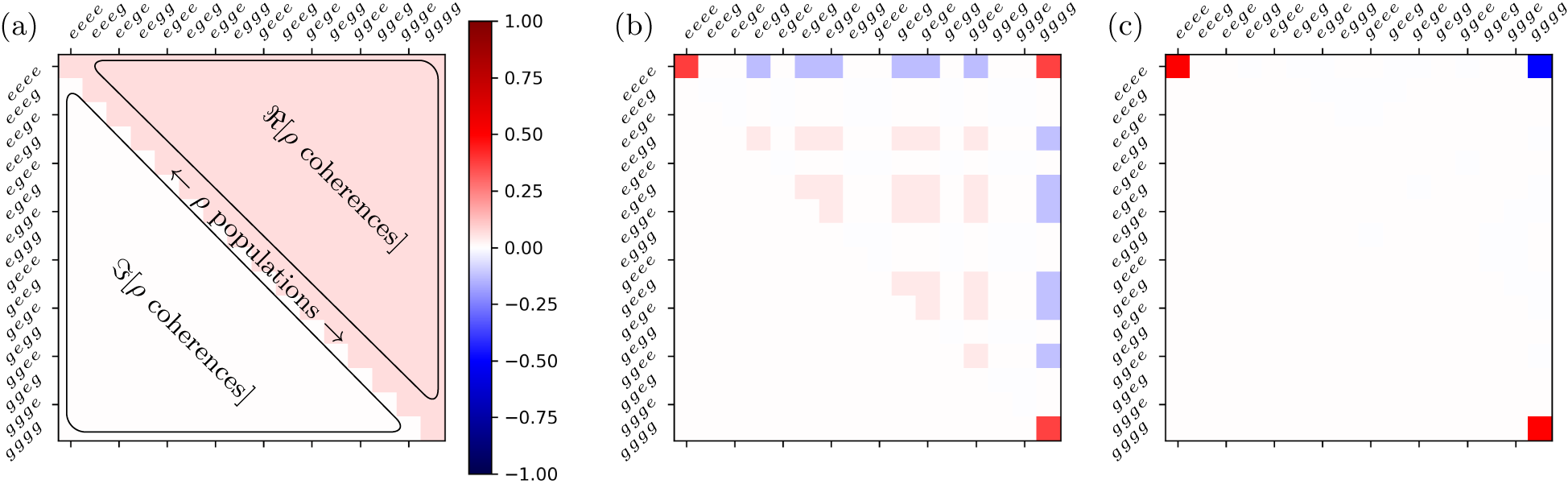}
\caption{
Panels here illustrate four--qubit states (density matrices), with populations down the diagonal, the real part of coherences in the upper triangle, and imaginary parts of coherences in the lower triangle. White is used for zero values, with deeper shades of red and blue used for positive and negative values, respectively (see colorbar, which is the same for all panels).
In (a) we illustrate the separable equal--superposition state $\propto (\ket{e} + \ket{g})^{\otimes 4}$ in this representation for reference. 
The state shown in (a) is used as the initial state for the NCQF protocols illustrated in panels (b) and (c), and based on the device of Fig.~\ref{fig:GHZ_setup}.  
Panel (b) illustrates the final state after joint half--parity monitoring \eqref{L_hp_IQ} boosted by NCQF \eqref{basic_NCQF_solution}, while panel (c) illustrates the final state after joint full--parity monitoring \eqref{L_fp_IQ} boosted by NCQF \eqref{basic_NCQF_solution}. 
Here the ``final'' state is plotted after the protocol has run for a duration $T = 10\,\Gamma^{-1}$. 
In panels (b) and (c), all qubits are conjugated by a Hadamard gate after the evolution, as per $\hat{\rho}_\mathrm{plot} = \hat{\mathsf{H}}_4\,\hat{\rho}_f\,\hat{\mathsf{H}}_4^\dag$ for $\hat{\mathsf{H}}_4 = (1/4)(\hat{\sigma}_x + \hat{\sigma}_z)^{\otimes 4}$ and $\hat{\rho}_f$ the output of the NCQF boosted dynamics; this is a local unitary operation and therefore cannot affect the entanglement of the four qubit system, and is applied here because it makes the entanglement structure of final states easier to understand visually. 
We see that the four--qubit parity protocols deterministically generate highly entangled states \eqref{joint_par_finstates} in the long--time limit, including a four--qubit GHZ state \eqref{4Q_GHZm} in the full--parity case [see panel (c)]. 
}\label{fig:GHZ}
\end{figure*}

Recall that half--parity monitoring of two qubits may be modeled by the Lindblad operator $\hat{L}_{hp} = \sqrt{\Gamma}\,\left(\hat{\mathds{1}}\otimes \hat{\sigma}_z + \hat{\sigma}_z \otimes \hat{\mathds{1}} \right)$ \eqref{L_hp}, and that full parity monitoring is given by $\hat{L}_{fp} = \sqrt{\Gamma}\,\hat{\sigma}_z \otimes \hat{\sigma}_z$ \cite{Williams_2008, martin2015deterministic, martin2017optimal, zhang2020locally}. 
Mixing two half--parity signals and then performing the EPR quadrature measurement just described leads to 
\begin{subequations} \label{L_hp_IQ}\be 
\hat{L}_{hp}^{(I)} = \tfrac{1}{\sqrt{2}}\left(\hat{L}_{hp}\otimes\hat{\mathds{1}} + \hat{\mathds{1}} \otimes \hat{L}_{hp}\right),
\ee \be 
\hat{L}_{hp}^{(Q)} = \tfrac{i}{\sqrt{2}}\left(\hat{L}_{hp}\otimes\hat{\mathds{1}} - \hat{\mathds{1}} \otimes \hat{L}_{hp}\right),
\ee \end{subequations}
and application of the same process to full--parity coupling instead leads to
\begin{subequations} \label{L_fp_IQ} \be 
\hat{L}_{fp}^{(I)} = \tfrac{1}{\sqrt{2}}\left(\hat{L}_{fp}\otimes\hat{\mathds{1}} + \hat{\mathds{1}} \otimes \hat{L}_{fp}\right),
\ee \be 
\hat{L}_{fp}^{(Q)} = \tfrac{i}{\sqrt{2}}\left(\hat{L}_{fp}\otimes\hat{\mathds{1}} - \hat{\mathds{1}} \otimes \hat{L}_{fp}\right).
\ee \end{subequations}
We show in Fig.~\ref{fig:GHZ} that highly entangled states can be generated by applying these measurements to the initial separable state $\propto (\ket{e} + \ket{g})^{\otimes 4}$, using NCQF \eqref{basic_NCQF_solution} to render the dynamics deterministic. 
The entanglement structure is clearer if we apply a Hadamard gate to all four qubits at the end of the NCQF protocol, where such an operation is given by
\be \label{hadamard_4}
\hat{\mathsf{H}}_4 = \tfrac{1}{4}\left(\hat{\sigma}_x + \hat{\sigma}_z \right)^{\otimes 4}. 
\ee
This gate is local (such that it cannot affect the entanglement between the four qubits), Hermitian, and unitary. 
Then if $\ket{\psi_f}$ is the final output of each NCQF protocol, the half--parity protocol gives 
\begin{subequations} \label{joint_par_finstates} \be \begin{split}
\hat{\mathsf{H}}_4 \ket{\psi_f^{(hp)}} =& \sqrt{\tfrac{3}{8}}\left(\ket{eeee} + \ket{gggg}\right) \\& - \tfrac{1}{\sqrt{24}}\left(\ket{eegg} + \ket{ggee} + \ket{\psi^+}\otimes\ket{\psi^+}\right),
\end{split} \ee 
where we have shorthanded the un-normalized Bell state $\ket{\psi^+} = \ket{eg} + \ket{ge}$,
and the full parity protocol generates a Greenberger--Horne--Zeilinger (GHZ) state \cite{GHZ_1989, GHZ_1990, Horodecki_Entangle_RMP, Verstraete_2002}
\be \label{4Q_GHZm}
\hat{\mathsf{H}}_4\ket{\psi_f^{(fp)}}  = \tfrac{1}{\sqrt{2}}\ket{eeee} - \tfrac{1}{\sqrt{2}}\ket{gggg}. 
\ee \end{subequations}
These states are generated exactly in the long--time limit, and are already generated with extremely high fidelities within a few $\Gamma\,T$.
Generation of these states after $T = 10\,\Gamma^{-1}$ is visualized in Fig.~\ref{fig:GHZ}.

\subsection{Discussion \label{sec_Entangle_sub-discuss}}

Above we have considered three distinct setups in which the probabilistic creation of entanglement via continuous measurement is turned into a deterministic process via the addition of NCQF. 
In \secref{sec:half-par} and \secref{sec:flor} we re-considered physical situations where locally--optimal and noise--canceling feedbacks were already known \cite{martin2015deterministic, martin2017optimal, martin2019single, zhang2020locally, Lewalle_2020_limitcycle}; we find that in both cases qualitatively similar behavior may be obtained by demanding noise cancellation via the simple expression \eqref{basic_NCQF_solution}. 
That solution is not unique, and the existence of previously--known and distinct solutions illustrates that free parameters allowed within the NCQF framework can in fact be used constructively to fine--tune NCQF dynamics under constraints. 
In \secref{sec:GHZ} we elaborated on the previously known schemes, illustrating that four--qubit GHZ states may be deterministically prepared by invoking the same principles and process. 
While there have been previous feedback protocols that prepare GHZ states \cite{zhang2020locally}, the one proposed here is (to our knowledge) the first to do so via completely deterministic dynamics in a measurement--driven approach.
These findings collectively highlight that NCQF has a lot of potential in supporting entanglement generation and stabilization tasks in monitored quantum systems. 

To this end, we would like to highlight a common theme between all three of the schemes above that suggests a route towards further progress. 
\emph{Every} NCQF--based entanglement generation scheme above leads to dynamics that take place entirely within zero--signal ($\mathsf{s} = 0$) subspaces.
Recall that the expected signal of a continuous measurement is $\mathsf{s} = \langle \hat{L} + \hat{L}^\dag \rangle$, with the readout given by $r\,dt = \sqrt{\eta}\,\mathsf{s}\,dt + dW$.
In order to better understand what it means for our dynamics to stay in a $\mathsf{s} = 0$ subspace, let us define signal operator $\hat{\mathsf{s}} = \hat{L} + \hat{L}^\dag$ such that $\mathsf{s}  = \langle \hat{\mathsf{s}} \rangle$, and let $\ket{\psi_\mathsf{s}^{(n)}}$ be the $n^\mathrm{th}$ eigenstate of the signal operator, with eigenvalue $\lambda^{(n)}_\mathsf{s}$. 
We may write the current state in the signal eigenbasis as per $\ket{\psi} = \sum_n c_n\,\ket{\psi_\mathsf{s}^{(n)}}$, and then see that
\be \label{s0_condition} \begin{split}
\ensavg{\hat{\mathsf{s}}} &= \sum_{m,n} \bra{\psi_\mathsf{s}^{(m)}} c_m^\ast\,c_n\,\lambda_\mathsf{s}^{(n)} \ket{\psi_\mathsf{s}^{(n)}} = \sum_n |c_n|^2\,\lambda_\mathsf{s}^{(n)}.
\end{split} \ee
This implies that in order to have $\mathsf{s} = 0$, the state must be either i) in the nullspace of $\hat{\mathsf{s}}$, and/or ii) have populations ``balanced'' across the eigenspaces of $\hat{\mathsf{s}}$, so that positive and negative signs on the eigenvalues combine in such a way as to cancel populations out of the sum. 
For the purposes of this discussion, it is helpful to adopt a convention in which measurement operators $\hat{L}$ are normalized so as to be traceless, i.e.~the spectrum of $\hat{\mathsf{s}}$ should be shifted so that $\sum_n \lambda_\mathsf{s}^{(n)} = 0$. 
While we do not show that this is necessarily the only approach to take, it is intuitive that dynamics residing within zero--signal subspaces are especially good at generating balanced superpositions of different signal eigenstates, which tend to be interesting entangled states.

\section{Rapid Purification Protocols as NCQF on Mixed States \label{sec_Purify}}

While it is always possible to satisfy the pure--state NCC \eqref{NCC_Xi_main}, e.g.~via \eqref{basic_NCQF_solution}, it is not always possible to satisfy the mixed--state version of \eqref{NCC_both} for every $\hat{\rho}$.
Eigenspaces of measurement channels, or sub-spaces complementary to a measurement (e.g.~sub-spaces where $\mathsf{s} = 0$ for traceless measurement channels) are among those where perfect NCQF is however sometimes possible even for impure states (see Appendix \ref{app_omega_dynamics} \secref{sec_NC-mixed} for details). 
Here we provide a simple (single--qubit) example of NCQF for mixed states, where the qubit is kept in the complementary subspace of the measurement observable. 
This turns out to be exactly the well-known rapid purification protocol for a single qubit, which results in the reduction of the purification time by a factor of 2 compared to the situation without feedback \cite{rapid_purification}.

We are interested in purifying a single-qubit state by measuring a fixed Pauli observable, i.e.~$\hat{L} = \sqrt{\Gamma}\,\hat{\sigma}_z$. 
Without loss of generality (WLOG), we can always assume the state is in the $xz$-plane of the Bloch sphere, so we have 
\be 
\hat{\rho} = \tfrac{1}{2}\left(\hat{\mathds{1}} + R\,(\cos\theta\,\hat{\sigma}_z + \sin\theta\,\hat{\sigma}_x)\right)
\ee 
for the Bloch radius $0\leq R \leq 1$. 
Appendix \ref{app_omega_dynamics} \secref{sec_NC-mixed}, and the NCC \eqref{mixed-state-NCC-exist} in particular, demonstrates that perfect noise cancellation is possible iff 
\be \label{mixed_NCC_maintext}
\bra{\lambda_n}\hat{X}\ket{\lambda_n} = \tfrac{1}{2}\,\mathsf{s} \quad\forall~n~\text{with}~\ket{\lambda_n}\notin\mathrm{ker}(\hat{\rho}),
\ee
where $\ket{\lambda_n}$ are the pure eigenstates of $\hat{\rho}$, and all those with $\lambda_n > 0$ are included. 
Direct calculation shows that the eigenvalues of $\hat{\rho}$ here are 
\begin{equation}
    \lambda_1 = \frac{1+R}{2}, \quad\quad\lambda_2 = \frac{1-R}{2},
\end{equation}
with eigenstates 
\begin{equation}
    |\lambda_1\rangle = \cos\left(\frac{\theta}{2}\right)|0\rangle + \sin\left(\frac{\theta}{2}\right)|1\rangle ,
\end{equation}
and 
\begin{equation} 
    |\lambda_2\rangle = -\sin\left(\frac{\theta}{2}\right)|0\rangle + \cos\left(\frac{\theta}{2}\right)|1\rangle.
\end{equation}
The state $\hat{\rho} = \lambda_1\ket{\lambda_1}\bra{\lambda_1} + \lambda_2\ket{\lambda_2}\bra{\lambda_2}$ has purity $\tr{\hat{\rho}^2} = (1+R^2)/2$.
Working in the $\hat{\rho}$--eigenbasis, we find $X_{11} = \langle\lambda_1|\hat{X}|\lambda_1\rangle = \sqrt{\Gamma}\cos\theta$ and $X_{22} = -\sqrt{\Gamma}\cos\theta$. \eqref{mixed_NCC_maintext} is then equivalent to 
\begin{equation}
    \cos\theta = R\, \cos\theta, \quad -\cos\theta = R\, \cos\theta.
\end{equation}
where the signal is $\mathsf{s} = 2\sqrt{\Gamma}\,R\,\cos\theta$.
When $R \neq 1$ these conditions can be true only if $\cos\theta = 0$, which means that perfect noise cancellation is possible along the axis of the Bloch sphere perpendicular to the measurement axis (i.e.~states in the $z = 0$ Bloch--plane admit a perfectly noise--canceling solution if we monitor along $\hat{\sigma}_z$).
Notice that when $R =1$, we have a single equation instead of two, and there is no constraint.
We may then note that this is i) another instance in which a $\mathsf{s} = 0$ subspace takes on special significance, and ii) exactly the condition imposed in a well--known rapid purification protocol \cite{rapid_purification}. 
The latter demands that the relationship between the state and measurement axis should be controlled to maximize the rate at which the observer gains classical information about the quantum state \cite{BookBarchielli, Philippe_Thesis, Cylke_2022, Lewalle_CDJP}; feedback is used to maintain such a situation throughout the dynamics \cite{rapid_purification, Wiseman_2006, Wiseman_2008, Combes_2008, Combes_2010, Combes_2011, Combes_2011b, Ralph_2011, Teo_2014}. 

Let us now work in the required subspace, defined by $\cos\theta = 0$; we will demonstrate that a feedback of the form \begin{equation}
    \hat{\omega}=\hat{\Pi}\,\hat{A}\,\hat{\Pi}+i\left[\hat{\Pi},\hat{X}\right] - i\,\hat{Y}+\hat{\Pi}^\perp\hat{B}\hat{\Pi}^\perp,
    \label{NCQF_ABform_main}
\end{equation} that satisfies \eqref{mixed_NCC_maintext} maximizes the purification rate, and reproduces the results of \cite{rapid_purification}. (See \eqref{NCQF_form_general} for a significant elaboration on the origins and uses of such an expression; this is \eqref{basic_NCQF_solution} plus some free terms based on matrices $\hat{A}$ and $\hat{B}$ to be determined.)
Here $\hat{\Pi} = \sum_n \ket{\lambda_n}\bra{\lambda_n}$ for all $\ket{\lambda_n}$ for which $\lambda_n \neq 0$, with $\hat{\rho} = \sum_n \lambda_n \ket{\lambda_n}\bra{\lambda_n}$.
We may have the state along the $x$-axis by assuming the NCC is satisfied in the preceding time, i.e.~$\hat{\rho} = ( \hat{\mathds{1}} + R\,\hat{\sigma}_x)/2$. 
Direct calculation gives $X_{12} = \langle\lambda_1|\hat{X}|\lambda_2\rangle = -\sqrt{\Gamma} = X_{21}$. 
Using the process detailed in Appendix \ref{app_omega_dynamics}\,\secref{sec_NC-mixed} (see \eqref{a_mnvals} in particular) to derive a feedback satisfying the NCC, we find that the off--diagonals of $\hat{A}$ must be such that $\hat{A} = \sqrt{\Gamma}\,\hat{\sigma}_y/R$, where diagonal elements may be added freely. 
We choose $A_{11} = A_{22} = 0$, finding
\begin{equation}\label{rapid_purification_NCQF}
    \hat{\omega} = \hat{A} = \frac{\sqrt{\Gamma}\,\hat{\sigma}_y}{R}.
\end{equation}
This makes sense physically, as measurement of $\hat{\sigma}_z$ will disturb the state $\hat{\rho} = (\hat{\mathds{1}} + R\,\hat{\sigma}_x)/2$ and make it drift towards $|e\rangle$ or $|g\rangle$ stochastically in the $xz$-plane. 
The feedback unitary then brings it back to $x$-axis by applying a rotation around $y$-axis. 
However, as expected, this NCQF is not unique: one may further rotate the state around $z$-axis by an arbitrary angle, in which case we will have non-zero diagonal elements $A_{11}$ and $A_{22}$.
Note that the purification with NCQF is consistent with \eqref{eq:lambda_m_evol}, where we see that largest eigenvalue of the density matrix $\hat{\rho}$ always increases under complete noise cancellation (when all the free parameters are set to zero). 

We have shown that \eqref{rapid_purification_NCQF} is a noise--canceling feedback for qubit states prepared along the $x$--axis of the Bloch sphere (for any purity). 
Now we show why this feedback is equivalent to the optimal purification result in this same system. 
We again assume that $\hat{\rho} = ( \hat{\mathds{1}} + R\,\hat{\sigma}_x)/2$ holds WLOG.
We then search for $f \in \mathbb{R}$ in the control $\hat{\omega}/\sqrt{\Gamma} = f\,\hat{\sigma}_y$ to maximize the rate of purification.
Translating \eqref{ito-fb-singlemeas} into a dynamical system of Bloch coordinates, we have 
\begin{subequations}\begin{equation}\label{bloch_eqx}
    dx = 2\Gamma\,(2\,f- x - f^2 \,x)dt,
\end{equation}
\begin{equation}\label{bloch_eqz}
    dz = 2\sqrt{\Gamma}(1-f\, x)dW,
\end{equation}\end{subequations}
via $\hat{\rho} + d\hat{\rho} = (\hat{\mathds{1}} + (x+dx)\hat{\sigma}_x + dz\, \hat{\sigma}_z)/2$. 
We have $x = R$ as $\hat{\rho}$ is initially along the $x$-axis. 
Obviously, in order to keep the state perpendicular to the $z$-axis (i.e.~following the principles of maximal information extraction of e.g.~\cite{rapid_purification, Combes_2011}, and to cancel the noise), one should choose $f = 1/R$.
This recovers \eqref{rapid_purification_NCQF}. 
However, this solution also extremizes $\dot{R}$, satisfying $\partial_f \dot{R} = 0$, thereby confirming the optimality of NCQF in single qubit purification tasks. 
One can understand this coincidence in the following way: i) the state being perpendicular to the measurement axis  ensures both the NCC \eqref{mixed-state-NCC-exist} and the optimality condition for qubit purification, and ii) the dynamics of $z$ \eqref{bloch_eqz} are purely stochastic and stochasticity also only comes into play in \eqref{bloch_eqz}, therefore eliminating $z$ dynamics via \eqref{rapid_purification_NCQF} not only achieves noise-canceling, but also freezes the $z$ coordinate at $z=0$, ensuring an optimal purification rate.
While this mathematical coincidence is not guaranteed in larger systems, many of the same principles continue to apply \cite{Combes_2011}.

\section{NCQF for Magic State Distillation \label{sec-MSD}}
So far, we have illustrated that noise--canceling feedback is a concept from which one may develop many examples of entanglement generation and stabilization, as well as protocols to rapidly purify mixed states. 
We now develop an example that is qualitatively different, and (to our knowledge), new to the literature on time--continuous feedback and time--continuous error correction: Magic State Distillation (MSD) via continuous measurement may be boosted via NCQF. 
This will provide an example of NCQF boosting the performance of a stabilizer--based Quantum Error Correction (QEC) code. 

\subsection{Review of MSD}

Magic states are pure single qubit states that help us realize non-Clifford gates through gate teleportation \cite{Bravyi_Kitaev_MSD}. In MSD, we start with several copies of faulty magic states. Distillation  involves projecting those imperfect magic states to the codespace of a stabilizer code. 
If the initial fidelity is above a threshold,  successful projection leads to a mixed state of logical magic states. The final multiqubit state is then decoded to obtain a single qubit magic state.
The distillation procedure results in  a quadratic error suppression in the fidelity \cite{Bravyi_Kitaev_MSD, PhysRevA.95.022316}. In the following, we consider 5-to-1 distillation of a magic state based on the $[[5,1,3]]$ code. First,  we briefly review the distillation procedure. We intend to prepare the state
\begin{equation}
    \ket{F_0}=\cos\beta\ket{0}+e^{i\pi/4}\sin\beta\ket{1}, \quad \cos 2\beta=\frac{1}{\sqrt{3}},
\end{equation}
which can help realize a $\pi/6$
 gate. We start with an initial noisy state
\begin{equation}
    \hat{\rho} =(1-\epsilon_{\textrm{in}})\ket{F_0}\bra{F_0}+\epsilon_{\textrm{in}}\ket{F_1}\bra{F_1},
    \label{rho_1q_noisy}
\end{equation}
where $\ket{F_1}= \hat{\sigma}_y\,\hat{\mathsf{H}}\,\ket{F_0}$ and $\hat{\mathsf{H}}$ denotes the Hadamard gate. 
The noisy state in \eqref{rho_1q_noisy} can be prepared by applying the dephasing transformation
\begin{equation}
\hat{F} = \frac{e^{i\pi/4}}{\sqrt{2}}\begin{pmatrix}
1  & 1\\
i & -i
\end{pmatrix}, 
\end{equation}
and $\hat{F}^\dagger$, and the identity $\hat{\mathds{1}}$, to an input state with $1/3$ probability each. 
Note, in the original notation of Ref.~\cite{Bravyi_Kitaev_MSD}, these states were denoted as $\ket{T}$ states. For distillation, we start with five copies of the initial noisy state \eqref{rho_1q_noisy}. In other words, $\hat{\rho}_{\textrm{in}}=\hat{\rho}^{\otimes 5}$. The stabilizers for the five qubit code are \cite{PhysRevLett.77.198}
\begin{equation} \label{eq_stabilizers}
    \begin{array}{ccccccccccc}
        \hat{S}_1 & = & \hat{\sigma}_x & \otimes & \hat{\sigma}_z & \otimes & \hat{\sigma}_z & \otimes & \hat{\sigma}_x & \otimes & \hat{\mathds{1}}, \\
         \hat{S}_2 & = & \hat{\mathds{1}} & \otimes & \hat{\sigma}_x & \otimes & \hat{\sigma}_z & \otimes & \hat{\sigma}_z & \otimes & \hat{\sigma}_x,\\
         \hat{S}_3 & = & \hat{\sigma}_x & \otimes & \hat{\mathds{1}} & \otimes & \hat{\sigma}_x & \otimes & \hat{\sigma}_z & \otimes & \hat{\sigma}_z, \\
         \hat{S}_4 & = & \hat{\sigma}_z & \otimes & \hat{\sigma}_x & \otimes & \hat{\mathds{1}} & \otimes & \hat{\sigma}_x & \otimes & \hat{\sigma}_z.
    \end{array}
\end{equation}
Following a successful  $(+1,+1,+1,+1) $ syndrome, the state is a mixture of the logical  magic states 
\begin{equation}\begin{split}
    \hat{\rho}_s=&\left[\frac{\epsilon_{\textrm{in}}^5+5\epsilon_{\textrm{in}}^2(1-\epsilon_{\textrm{in}})^3}{6}\right]\ket{F_0^L}\bra{F_0^L} \\ &+\left[\frac{(1-\epsilon_{\textrm{in}})^5+5\epsilon_{\textrm{in}}^3(1-\epsilon_{\textrm{in}})^2}{6}\right]\ket{F_1^L}\bra{F_1^L},
    \label{rho_s}
\end{split}\end{equation}
with
\begin{equation}\begin{split}
    \ket{F_0^L}=\sqrt{6}\hat{\Pi}\ket{F_1}\otimes\ket{F_1}\otimes \ket{F_1}\otimes\ket{F_1}\otimes\ket{F_1}, \\ \ket{F_1^L}=\sqrt{6}\hat{\Pi}\ket{F_0}\otimes\ket{F_0}\otimes \ket{F_0}\otimes\ket{F_0}\otimes\ket{F_0},
\end{split}\end{equation}
where 
\begin{equation}
    \hat{\Pi}=\frac{1}{16}\prod_{j=1}^4(I+S_j),
\end{equation}
is the projector into the codespace. 
The probability of successful syndrome is given by
\begin{equation}
    p_s=\frac{\epsilon_{\textrm{in}}^5+5\epsilon_{\textrm{in}}^2(1-\epsilon_{\textrm{in}})^3+5\epsilon_{\textrm{in}}^3(1-\epsilon_{\textrm{in}})^2+(1-\epsilon_{\textrm{in}})^5}{6}, \label{eq:ps_analytical}
\end{equation}
such that MSD has an inherent post-selection cost. 
We have $p_s \leq 1/6$, with the best value $p_s = 1/6$ achieved in the limit $\epsilon_\mathrm{in} \rightarrow 0$.
After the successful stabilizer measurements, decoding and swap are applied to distill $\ket{F_0}$. 
To increase the distillation yield, $\hat{\rho}_s$ should be as close to $\ket{F_1^L}$ as possible. 
The final single qubit state obtained is 
\begin{equation}
    \hat{\rho}_{\textrm{out}}=(1-\epsilon_{\textrm{out}})\ket{F_0}\bra{F_0}+\epsilon_{\textrm{out}}\ket{F_1}\bra{F_1},
\end{equation}
with 
\begin{equation}
    \epsilon_{\textrm{out}}=\frac{\epsilon_{\textrm{in}}^5+5\epsilon_{\textrm{in}}^2(1-\epsilon_{\textrm{in}})^3}{\epsilon_{\textrm{in}}^5+5\epsilon_{\textrm{in}}^2(1-\epsilon_{\textrm{in}})^3+5\epsilon_{\textrm{in}}^3(1-\epsilon_{\textrm{in}})^2+(1-\epsilon_{\textrm{in}})^5}.
    \label{eps_out_analytical}
\end{equation}
If $\epsilon_{\textrm{in}}<\epsilon_{\textrm{thr}}=\frac{1}{2}\left(1-\sqrt{\frac{3}{7}}\right)\approx0.173$, the distillation leads to an increase in the output fidelity. 
Thus, the threshold fidelity for MSD to be effective is $1-\epsilon_{\textrm{thr}}$. 

\subsection{MSD Boosted by NCQF}

We will now see how noise canceling feedback can augment the distillation process. 
We consider weak continuous measurements of the stabilizers with perfect detectors. 
This means that we consider continuous measurement of four channels $\hat{L}_j = \sqrt{\Gamma}\,\hat{S}_j$, where $\hat{S}_j$ are the stabilizers  defined in \eqref{eq_stabilizers}. 
With such continuous stabilizer monitoring, and the possibility of instantaneous feedback, we then have dynamics generated by a sum over stabilizer channels of terms like those in \eqref{ITO_SME} through \eqref{ito-fb-singlemeas}.

Here, our feedback unitary is of the form $\hat{\mathcal{U}}_{\textrm{fb}}=e^{-i\sum_{j=1}^4\hat{\omega}_jdW_j}$. 
In general, it is not possible to find completely noise--canceling feedback for mixed state dynamics (see Appendix \ref{app_omega_dynamics}\,\secref{sec_NC-mixed} for detailed comments). Mixed--state dynamics are an integral part of MSD.
While perfect noise cancellation will not be possible in this example, we can find feedback that cancels a significant amount of the noise with \eqref{NCQF_form_general} and \eqref{a_mnvals}.
\begin{figure}
    \centering
    \includegraphics[width=\linewidth]{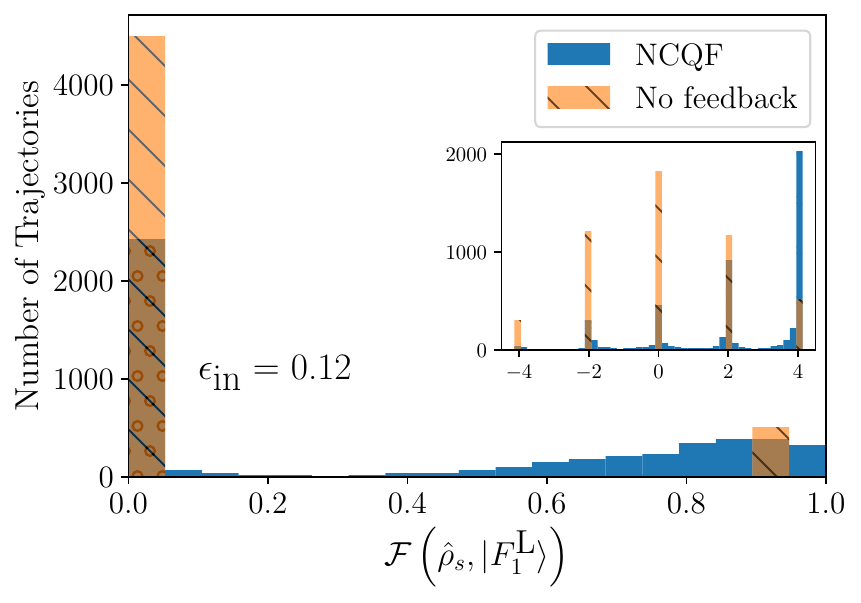}
    \caption{In this plot, we show the distribution of the fidelities of the logical magic state $\ket{F_1^\textrm{L}}$ in the final state ($t=5.0$ in the units of inverse measurement rate) for an initial infidelity of $\epsilon_{\textrm{in}}=0.12$. The orange (blue) histogram with diagonal (circular) hatches shows the fidelities without (with) NCQF for 5000 simulated trajectories. In agreement with \cite{Bravyi_Kitaev_MSD}, around 10\% of the trajectories lead to a successful distillation without feedback. In the presence of the noise-canceling feedback,  we see a significant decrease in the fraction of failed postselection. Additionally, instead of a single final fidelity, we find a distribution of values with a peak near 0.916.     The inset shows the distribution of the sum of the expectation values of the stabilizers at the final time without (orange hatched) and with (blue) noise-canceling feedback.   With noise-canceling feedback, we see a spread around each possible integer value. Also, a higher number of trajectories end up in the success space corresponding to  $\sum_j\left\langle{S_j}\right\rangle=4$.}
    \label{fig:msd_histo}
\end{figure}

\begin{figure}
    \centering
    \includegraphics[width=\linewidth]{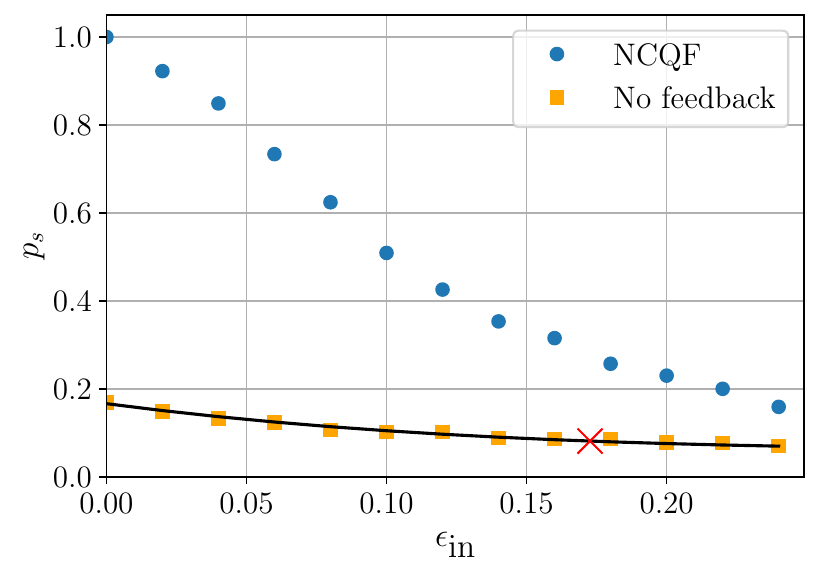}
\caption{This plot shows the successful postselection probability as a function of the initial infidelity for trajectories with (blue dots) and  without (orange squares) noise canceling feedback. Here, we define successful postselection probability as the fraction of trajectories with a total stabilizer expectation value $\sum_j\left\langle S_j\right \rangle>3.9$. The black line shows the analytical probability \eqref{eq:ps_analytical} from Ref.~\cite{Bravyi_Kitaev_MSD}. The results without feedback are consistent with the analytical expression. The distillation threshold is marked as a red cross. With noise-canceling feedback, we see a 3-4 times increase in the successful postselection probability. Also, if the input is a perfect $\ket{F_0}$ state with $\epsilon_{\textrm{in}}=0$, the noise canceling feedback prepares $\ket{F_1^\textrm{L}}$ deterministically. 
}
    \label{fig:msd_probs}
\end{figure}

\begin{figure}
    \centering
    \includegraphics[width=\linewidth]{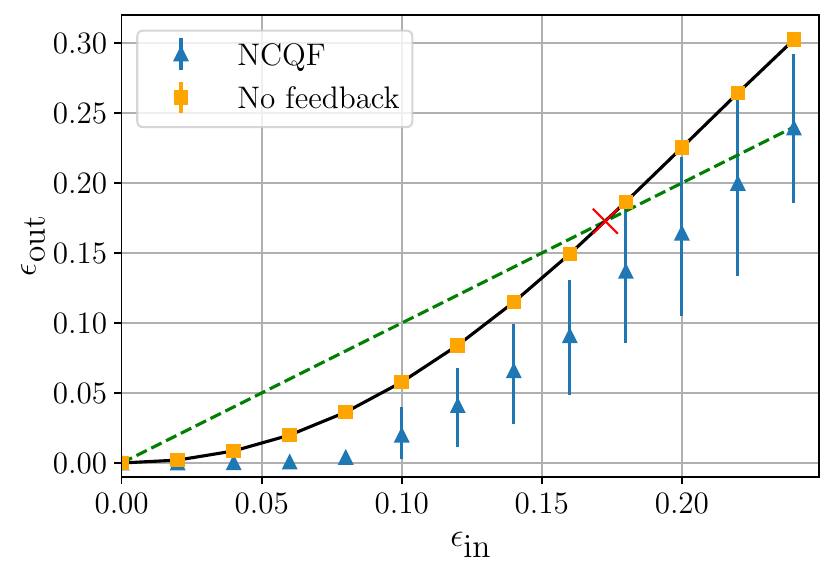}
\caption{In this plot, we show the output infidelity as a function of input infidelity with (blue triangles) and without (orange squares) noise canceling feedback. The error bars correspond to 16 and 84 percentile values. The green dashed  line corresponds $\epsilon_{\textrm{out}} = \epsilon_{\textrm{in}}$. The distillation process is beneficial only when its performance stays below the green line. Here $\epsilon_{\textrm{out}}$ without noise canceling feedback is calculated for successful postselections. The black line shows the analytical expression for $\epsilon_{\textrm{out}}$ from \eqref{eps_out_analytical}. Since in the presence of noise-canceling feedback the number of successful postselections increases significantly (Fig.~\ref{fig:msd_probs}), we only select $N_{\textrm{ps}}(\epsilon_{\textrm{in}})$ trajectories with the lowest $\epsilon_{\textrm{out}}$. Here, $N_{\textrm{ps}}(\epsilon_{\textrm{in}})$ denotes the number of postselected trajectories under no feedback for a given $\epsilon_{\textrm{in}}$. The threshold is marked as a red cross and is the intersection point of the green dashed and black solid lines. We see noise noise-canceling feedback helps us improve the distilled state fidelity. Also, the threshold with noise canceling feedback is lowered to $\epsilon_{\textrm{in}}\sim .24$, which improves substantially on the threshold $\epsilon_\mathrm{thr} \approx 0.173$ described near \eqref{eps_out_analytical}. This is below the highest magic state   fidelity $\simeq .789$ that can be achieved from only Clifford operations before distillation (see Ref.~\cite{Bravyi_Kitaev_MSD}).
}\label{fig:msd_epsvseps}
\end{figure}

Fig.~\ref{fig:msd_histo} shows the histogram of 5000 trajectories with and without noise-canceling feedback. Here, the initial infidelity  is $\epsilon_{\textrm{in}}=0.12$. In the presence of noise-canceling feedback, we see a significant increase in the fraction of successful postselection. Such behavior is clarified in Fig.~\ref{fig:msd_probs}. There, we see a significant increase in the postselection probability in the presence of noise-canceling feedback.  Also, if the initial infidelity $\epsilon_{\textrm{in}}=0.0$, the noise canceling feedback deterministically prepares $\ket{F_1^\textrm{L}}$, consistent with our analysis for pure state noise cancellation. 
This is in contrast with the measurement-only case, where $p_s \rightarrow 1/6$ \eqref{eq:ps_analytical} as $\epsilon_\mathrm{in} \rightarrow 0$.
So, while there is no reason to perform MSD if the input state is already perfect, we here see that as we are approaching very good magic states, NCQF effectively reduces the MSD runtime overhead six-fold, by eliminating the post--selection cost.

Fig.~\ref{fig:msd_epsvseps} shows the output infidelity as a function of input infidelity for trajectories with and without noise-canceling feedback. As Fig.~\ref{fig:msd_histo} and Fig.~\ref{fig:msd_probs} suggest, with noise-canceling feedback, successful postselection leads to a larger number of trajectories with varying output fidelities. To make a valid comparison, we select  $N_{\textrm{ps}}(\epsilon_{\textrm{in}})$ trajectories with the lowest $\epsilon_{\textrm{out}}$, where $N_{\textrm{ps}}(\epsilon_{\textrm{in}})$ denotes the number of postselected trajectories under no feedback for a given $\epsilon_{\textrm{in}}$.
We see a clear improvement of the output fidelities under noise canceling feedback. We also see the threshold effectively getting lowered beyond $\epsilon_0\sim0.211$, achievable through Clifford operations on single qubit states \cite{Bravyi_Kitaev_MSD}. 

The influence of noise-canceling feedback for MSD can be understood from \eqref{eq:lambda_m_evol}, which shows the growth of different populations under perfect noise cancellation. Although \eqref{eq:lambda_m_evol} has been derived with the perfect noise-cancelation assumption, it can still be insightful for analyzing the MSD results.  The equation shows that the largest eigenvalue of the state $\hat{\rho}$ grows under noise-canceling feedback. Thus, during the distillation, the NCQF increases $\ket{F_1^L}$ fidelity, which corresponds to the dominant population in the five--qubit state. This behavior materializes in  Fig.~\ref{fig:msd_histo} and Fig.~\ref{fig:msd_epsvseps}.

As mentioned before, noise-canceling feedback leads to a number of trajectories avoiding unsuccessful postselection. 
These trajectories achieve different fidelities at the final evolution time of $t=5$ (in units of the inverse measurement rate $\Gamma^{-1}$). We expect, as we increase the integration time, most of these trajectories under successful postselection will end up with $\epsilon_{\textrm{out}}$, as predicted by \eqref{eq:lambda_m_evol}. Due to the expensive nature of simulations,  we leave such explorations for future work. A drawback of this approach is the need to incorporate non-local and non-Clifford feedback unitaries. 
However, as pointed out throughout the article, noise-canceling feedback allows free parameters, and we have explicitly shown in \secref{sec_Restricted_Controls} and Appendix \ref{app_Free_params} how to solve a noise minimization problem for restricted controller forms.
We believe it will be interesting and worthwhile to consider the effectiveness of the noise--minimizing feedback under simplified control assumptions in future work.

In summary, noise-canceling feedback can help with magic state distillation in two ways. First, by boosting the postselection probabilities, such feedback lowers the number of unsuccessful runs.  Second, for successful postselections, the feedback increases the yields, thus reducing the need for successive distillation stages. These two aspects can help a lot with decreasing resource overhead with magic state distillation \cite{litinski_magic_2019} and also might be useful for postselection-heavy schemes such  magic state cultivation \cite{gidney2024magicstatecultivationgrowing}.   

\section{Concluding Discussion \label{sec-conclude}}
 
Several examples of feedback on diffusive quantum trajectories that exhibit complete noise cancellation, resulting in deterministic processes, have been discovered in the literature (see \textcite{zhang2020locally} and references therein). 
This property has largely arisen by happenstance, in the course of research seeking to optimize a feedback for other tasks.
In the present paper, we have performed the first (to our knowledge) exploration of noise--canceling feedback protocols as a distinct category of operations in their own right. 
We have been able to establish some foundational results for NCQF in the course of this investigation. 
In particular, we have found that i) complete noise cancellation is \emph{always} possible for pure states, and ideal measurements and feedback operations, 
ii) there are generically some free parameters in the feedback, such that perfect NCQF is not just possible, but encompasses a family of controls rather than a single control solution. 
(In particular, there are an infinite number of feedback operators that completely cancel the stochastic noise if any exist at all.)
Moreover, iii) ideal NCQF dynamics may be characterized in terms of an effective (state--dependent) non-Hermitian Hamiltonian. 
Going beyond pure states, we have found that iv) complete cancellation of noise is possible for specific combinations of mixed states and measurement operators, but is not generically possible for all mixed states given particular measurement channels. 
We have characterized the conditions for such noise cancellation on mixed states to exist, and v) provided alternative protocols for noise minimization when the control is constrained. 
Furthermore, vi) we have demonstrated that feedback derived from a noise--cancellation principle can boost the performance of tasks of interest to the quantum information science and technology community, even in cases where complete cancellation of the noise is not possible. 

We have concretely illustrated these broad points i) -- v) by developing a series of examples. 
We have been able to reproduce known examples from the feedback literature, and apply continuous feedback to wholly new areas, for both pure and mixed states. 
Specifically, in \secref{sec-nHH_examples} we have demonstrated several instances in which ideal NCQF is beneficial in state preparation tasks. 
Here, probabilistic preparation of entangled states by measurement is rendered deterministic by adding NCQF into a continuous measurement process. 
We have used our formalism to reproduce several Bell--state preparation results from the literature \cite{martin2015deterministic, martin2017optimal, zhang2020locally, martin2019single, Lewalle_2020_limitcycle}, and demonstrated the effectiveness of NCQF for larger four--partite entangled state preparation. 
With regards to the utility of NCQF for tasks with mixed states, we have reproduced a known example of rapid purification by feedback (which exhibits perfect noise cancellation despite the impure states; see \secref{sec_Purify}), and then demonstrated in \secref{sec-MSD} that NCQF can greatly enhance the success probability of Magic State Distillation (MSD), even though noise is not perfectly canceled throughout such a task. 

MSD is a key primitive for performing universal quantum computation as magic states help us realize non-Clifford gates \cite{Bravyi_Kitaev_MSD}. Noise-canceling feedback provides an avenue to reduce the cost associated with  postselection.   
Such findings are promising since postselection can pose a significant bottleneck in realizing such a procedure on near term quantum hardware \cite{litinski_magic_2019, rodriguez2024experimentaldemonstrationlogicalmagic, gidney2024magicstatecultivationgrowing}.
Additionally, our success in boosting the performance of MSD using NCQF suggests further lines of investigation to establish when and how NCQF may be a beneficial tool for continuous quantum error correction \cite{Mohseninia2020alwaysquantumerror, Atalaya_CQEC, Atalaya_2021_CQEC, Convy_2022, Convy2022logarithmicbayesian, Livingston_2022} more broadly. 
We have described protocols for (optimal) noise minimization under restricted controls in \secref{sec_Restricted_Controls}; these are expected to be useful tools for future work on MSD boosted by NCQF, as well as in other settings.
Investigating the robustness of the NCQF boosted MSD in the presence of noise in the drive and time delay in the feedback will be crucial to quantify the performance of such protocols in experiments.

Looking forward, it is clear that our present investigation of noise \emph{canceling} feedback considers a special case of a broader problem of noise \emph{minimizing} feedback, which might be fruitfully considered in future work. 
While we have here proven the existence of NCQF under idealized conditions, we expect that under realistic laboratory conditions noise minimization, rather that strict cancellation, will be the more relevant and attainable goal in the near term. 
This is due to finite measurement efficiencies $\eta < 1$, which generate mixed states, as well as delay / latency in performing feedback operations (the latter of which we have not considered in this manuscript). 

The first results presented here suggest that NCQF, or generalizations thereof, may be a strong candidate methodologies for further research. 
Based on our investigations so far, we expect such strategies to be generically good at boosting the success probability of non-deterministic processes in the absence of open--loop controls, and that there is a great deal more to learn about how open--loop controls may be used to modify those dynamics.

\par \emph{Acknowledgements --- } This work has been supported by AFOSR MURI Grant No.~FA9550-21-1-0202. 
PL was additionally partially supported by US-Israel BSF Grant no.~2020166.
Support is also acknowledged from the U.S.~Department of Energy, Office of Science, National Quantum Information Science Research Centers, Quantum Systems Accelerator. 
We are grateful to Alain Sarlette for stimulating discussions about some further applications of noise--canceling feedback. 
We also thank Kater Murch, Julian Wolf, Olive Eilbott, and Shay Hacohen-Gourgy for insightful discussions.
We are grateful to Chen Zhao and Hengyun Zhou for insights into some aspects of MSD. 
We thank Kelley McDonald,  Bert de Jong, Christopher Paciorek, Mark Yashar, and Avijit Shee for their help with HPC. 
PL is grateful to the UMass Lowell department of Physics \& Applied Physics for their hospitality during part of this manuscript’s preparation.


\appendix

\section{General Description of \\ Instantaneous Quantum Feedback \label{sec-NCQF-fullderive}}

We here provide full detail of derivations that are abbreviated in the main text. 
This appendix is intended to be relatively self--contained, using notations consistent with those in the main text. 

Some notations common to the derivations below are laid out here. 
Kraus \cite{Kraus_1983} operators $\hat{\mathcal{M}}(r)$ are elements of a positive operator--valued measure (POVM), i.e.
\be 
\sum_n\int_{-\infty}^{\infty} dr_c\,\hat{\mathcal{M}}_{c,n}^\dag\hat{\mathcal{M}}_{c,n} = \hat{\mathds{1}} 
\ee 
holds for each measurement channel $c$, where losses indexed by $n$ are needed only if the measurements have efficiency $\eta_c < 1$. 
(In this manuscript we mostly consider the ideal case $\eta_c = 1$). 
The readout $r_c$ for each channel can be expressed as
\begin{subequations} \be 
r_c\,dt = \sqrt{\eta_c}\,\mathsf{s}_c\,dt + dW_c,
\ee
which is the sum of an expected signal
\be 
\mathsf{s}_c = \ensavg{\hat{\mathsf{s}}_c} = \left\langle \psi \left| \hat{L}_c + \hat{L}_c^\dag \right| \psi \right\rangle = \tr{\hat{L}_c\,\hat{\rho} + \hat{\rho}\,\hat{L}_c^\dag},
\ee \end{subequations}
and noise. 
Wiener noise $dW(t)$ is Gaussian white noise, obeying $\ensavg{dW_j(t)\,dW_n(t')} = \delta_{jn}\,\delta(t-t')$ and It\^{o}'s lemma $dW^2 = dt$. 
Our Kraus operators $\hat{\mathcal{M}}_c$ are related to their Lindblad operators $\hat{L}_c$ via the typical short--time (infinitesimal $dt$) It\^{o} expansion \cite{BookWiseman, Wiseman1996, Chantasri_2021, Rouchon_2015, Wonglakhon_2024}
\begin{subequations} \label{Kraus-expand} \begin{align}
\label{Kraus-expand-0}\hat{\mathcal{M}}_{c,0} &\approx \mathcal{N}\,e^{-r_c^2\,dt/4} \left\lbrace \hat{1} + \hat{\mathfrak{Z}}_c\,dt + \mathcal{O}(dt^2) \right\rbrace \\ &\text{with}\quad \hat{\mathfrak{Z}}_c = \sqrt{\eta_c}\,r_c\,\hat{L}_c - \tfrac{1}{2}\, \hat{L}_c^\dag\hat{L}_c , \quad\& \\ \label{Kraus-expand-1}
\hat{\mathcal{M}}_{c,1} &\approx \mathcal{N}\,e^{-r_c^2\,dt/4} \left\lbrace
\sqrt{1-\eta_c}\,\sqrt{dt}\,\hat{L}_c + \mathcal{O}(dt^\frac{3}{2}) 
\right\rbrace.
\end{align}\end{subequations}
Losses of $n = 0$ or $n = 1$ energy quanta are adequate to characterize the time--continuum limit. 
It will be convenient later to have introduced the superoperator shorthands
\begin{subequations} \be 
\widehat{\mathsf{a}}_\bullet[\hat{\rho}] = \hat{\bullet}\,\hat{\rho}\,\hat{\bullet}^\dag -\tfrac{1}{2}\,\hat{\bullet}^\dag\hat{\bullet}\,\hat{\rho} - \tfrac{1}{2}\,\hat{\rho}\,\hat{\bullet}^\dag\hat{\bullet},
\ee \be 
\widehat{\mathsf{b}}_\bullet[\hat{\rho}] = \sqrt{\eta}\left( \hat{\bullet}\,\hat{\rho} + \hat{\rho}\,\hat{\bullet}^\dag - \hat{\rho}\,\tr{\hat{\bullet}\,\hat{\rho} + \hat{\rho}\,\hat{\bullet}^\dag }\right),
\ee \end{subequations}
that typically act on the density matrix $\hat{\rho}$, and describe the time--continuous dynamics arising from continuous measurement as described below. 
Finally, we introduce coherent dynamics via the unitary 
\begin{subequations}\be \begin{split}
\hat{\mathcal{U}} = e^{-i\,\hat{dh}} &\approx  \hat{\mathds{1}} - i\,d\hat{h} - \tfrac{1}{2}\,d\hat{h}^2 \\
& \approx \hat{\mathds{1}} - i(\hat{\Omega}\,dt + \boldsymbol{\hat{\omega}}\cdot d\mathbf{W}) - \tfrac{1}{2}\, \boldsymbol{\hat{\omega}}^2\,dt
\end{split} \ee 
expanded via It\^{o} calculus, and using the infinitesimal Hamiltonian increment
\be 
\hat{dh} = \hat{\Omega}\,dt + \boldsymbol{\hat{\omega}}\cdot d\mathbf{W} = \hat{\Omega}\,dt + \sum_c \hat{\omega}_c\,dW_c
\ee \end{subequations}
with both open--loop $\hat{\Omega}$ (deterministic) and feedback $\boldsymbol{\hat{\omega}}$ (proportional to the noise) terms. 
We here consider coherent control, such that $\hat{\Omega} = \hat{\Omega}^\dag$ and $\hat{\omega}_c = \hat{\omega}_c^\dag~\forall~c$.

\subsection{Pure States via Stochastic Schr\"{o}dinger Equation \label{sec-CSSE}}

We here derive a description of quantum feedback and the noise--cancellation condition ideal measurement ($\eta = 1$) and feedback operations acting on pure states.
We first consider a single measurement and feedback channel for simplicity. 
We draw the notations above together by considering the state update due to measurement and feedback, written to $O(dt)$ using It\^{o} calculus: 
\be \label{SSE_ito-op}
\ket{\psi(t+dt)} = \frac{\hat{\mathcal{U}}\,\hat{\mathcal{M}}\ket{\psi(t)}}{|\hat{\mathcal{U}}\,\hat{\mathcal{M}}\ket{\psi(t)} |} 
\ee
Notice the leading normalization terms $\mathcal{N}\,e^{-r^2\,dt/4}$ from \eqref{Kraus-expand} cancel out of the numerator and denominator immediately. 
Focusing just on the numerator of \eqref{SSE_ito-op} (i.e.~momentarily neglecting norm--preservation in the state dynamics), we have
\begin{widetext} \begin{subequations}\be\begin{split}
\ket{\psi(t+dt)} &\approx \hat{\mathcal{U}}\left( \hat{\mathds{1}} - \tfrac{1}{2}\,\hat{L}^\dag\hat{L}\,dt + \hat{L}\,\left(\mathsf{s}\,dt + dW\right) \right)\ket{\psi(t)} 
\\ & \approx \left( \hat{\mathds{1}} - i\,\hat{\Omega}\,dt - \tfrac{1}{2}\,\hat{\omega}^2\,dt - i\,\hat{\omega}\,dW \right) \left( \hat{\mathds{1}} - \tfrac{1}{2}\,\hat{L}^\dag\hat{L}\,dt + \hat{L}\,dW \right)\ket{\psi(t)} 
\\ &= \left\lbrace \hat{\mathds{1}} - \left(i\,\hat{\Omega} + \tfrac{1}{2}\,\hat{\omega}^2 + \tfrac{1}{2}\,\hat{L}^\dag\hat{L} + i\,\hat{\omega}\,\hat{L} - \mathsf{s}\,\hat{L} \right)dt + \left(\hat{L} - i\,\hat{\omega} \right)dW \right\rbrace \ket{\psi(t)} 
\end{split}\ee
such that we have (un-normalized) evolution
\be 
d\ket{\psi} \sim \left(\hat{L} - i\,\hat{\omega} \right) \ket{\psi}\,dW - \left(i\,\hat{\Omega} + \tfrac{1}{2}\,\hat{\omega}^2 + \tfrac{1}{2}\,\hat{L}^\dag\hat{L} + i\,\hat{\omega}\,\hat{L} - \mathsf{s}\,\hat{L} \right)\ket{\psi}\,dt.
\ee \end{subequations}
Notice above that the ordering of the operations is important: We explicitly suppose that the measurement precedes the feedback even in the limit where the feedback occurs ``instantaneously'', because there is physically no way to perform the feedback operation without first having the measurement record available. 
Clearly $(\hat{L}-i\,\hat{\omega})\ket{\psi} = 0$ is the \emph{unnormalized} noise-cancellation condition (NCC). 

Next we explicitly normalize these dynamics. 
We must evaluate
\begin{subequations}\label{SSE_normfactor}\be 
|\hat{\mathcal{U}}\,\hat{\mathcal{M}}\ket{\psi}|^{-1} = \ensavg{\hat{\mathcal{M}}^\dag\,\hat{\mathcal{U}}^\dag\hat{\mathcal{U}}\,\hat{\mathcal{M}}}^{-\tfrac{1}{2}} 
=  \ensavg{\hat{\mathcal{M}}^\dag\hat{\mathcal{M}}}^{-\tfrac{1}{2}} 
\ee
in order to do this. 
Using the expansion in \eqref{SSE_ito-op}, we have
\begin{align}
\ensavg{\hat{\mathcal{M}}^\dag\hat{\mathcal{M}}}^{-\tfrac{1}{2}} 
& = \ensavg{\left(\hat{\mathds{1}} - \tfrac{1}{2}\,\hat{L}^\dag\hat{L}\,dt + \hat{L}^\dag\,(\mathsf{s}\,dt + dW) \right) \left( \hat{\mathds{1}} - \tfrac{1}{2}\,\hat{L}^\dag\hat{L}\,dt + \hat{L}\,(\mathsf{s}\,dt + dW)\right)}^{-\tfrac{1}{2}}  \\
& = \ensavg{\left(\hat{\mathds{1}} +[\hat{L}+\hat{L}^\dag]dW + \mathsf{s}[\hat{L} + \hat{L}^\dag]dt\right)}^{-\tfrac{1}{2}} \\
& = \left(1 + \mathsf{s}\,dW + \mathsf{s}^2\,dt\right)^{-\tfrac{1}{2}} \\
& = 1 - \tfrac{1}{2}\,\mathsf{s}\,dW - \tfrac{1}{8}\,\mathsf{s}^2\,dt,
\end{align}\end{subequations}
where the It\^{o} rule is used throughout in simplifying expressions to $\mathcal{O}(dt)$. 
The normalized dynamics are then obtained by appending the normalization factor onto the un-normalized dynamics (again using the It\^{o} rule), as per
\be \label{CSSE_normalized} \begin{split}
d\ket{\psi} &= \left(1 - \tfrac{1}{2}\,\mathsf{s}\,dW - \tfrac{1}{8}\,\mathsf{s}^2\,dt \right)\left(\left(\hat{L} - i\,\hat{\omega} \right) \ket{\psi}\,dW - \left(i\,\hat{\Omega} + \tfrac{1}{2}\,\hat{\omega}^2 + \tfrac{1}{2}\,\hat{L}^\dag\hat{L} + i\,\hat{\omega}\,\hat{L} - \mathsf{s}\,\hat{L} \right)\ket{\psi}\,dt\right)
\\ & = \left(\hat{L} - i\,\hat{\omega} - \tfrac{1}{2}\,\mathsf{s}\,\hat{\mathds{1}} \right)\ket{\psi}\,dW -i \left(\hat{\Omega} + \hat{\omega}\,\hat{L} - \tfrac{i}{2}\left(\hat{L}^\dag\hat{L} + \hat{\omega}^2 \right) + \tfrac{i}{2}\,\mathsf{s}\left[\hat{L}+i\,\hat{\omega} -\tfrac{1}{4}\,\mathsf{s}\,\hat{\mathds{1}} \right] \right)\ket{\psi}\,dt
\end{split} \ee 
The noise--cancellation condition (NCC) with normalization included therefore reads
\be \label{CSSE_NCcondition}
\left(\hat{L} - i\,\hat{\omega} -\tfrac{1}{2}\,\mathsf{s}\,\hat{\mathds{1}} \right) \ket{\psi} = 0,
\ee 
and is a key result of this derivation. 
Note that it is acceptable for this condition to be met up to a global phase, i.e.~we may re-write the above like an eigenvalue equation $(\hat{L} - i\,\hat{\omega})\ket{\psi} = (\tfrac{1}{2}\,\mathsf{s} + i\,\phi)\ket{\psi}$ \eqref{NCC_Xi_main}, where the real part of the eigenvalue is fixed to be $\mathsf{s}/2$, but the imaginary part $\phi$ is completely free \cite{zhang2020locally}. 
Additional comments appear in Appendix \ref{app_omega_dynamics}\,\secref{sec_SolveNCC_pure}. 

Above we performed our derivations under the assumption that there is a single measurement operator $\hat{L}$ and corresponding feedback $\hat{\omega}$. 
However the extension of these derivations to multiple channels with mutually un-correlated noise is trivial (use $\hat{\mathcal{M}} \rightarrow \prod_c \hat{\mathcal{M}}_c$ each expanded into an operator $\hat{L}_c$ as above and associated with a noise channel $dW_c$, use $\hat{dh} = \hat{\Omega}\,dt + \sum_c \hat{\omega}_c\,dW_c$, and then the time--continuum dynamics are derived exactly as below, with a sum over each measurement $\hat{L}_c$ / feedback $\hat{\omega}_c$ pair). 
When \eqref{CSSE_NCcondition} is satisfied, then the remaining deterministic dynamics in \eqref{CSSE_normalized} may be expressed in terms of a Schr\"{o}dinger equation with effective non-Hermitian Hamiltonian, namely
\be \label{NC_nHH}
\ket{\dot{\psi}} = -i\,\hat{\mathcal{H}}_{NC}\ket{\psi}  \quad\text{with}\quad \hat{\mathcal{H}}_{NC} = \hat{\Omega} + \sum_c \hat{\omega}_c\,\hat{L}_c - \tfrac{i}{2}\left(\hat{L}_c^\dag\hat{L}_c + \hat{\omega}_c^2 \right) + \tfrac{i}{2}\,\mathsf{s}_c\left[\hat{L}_c+i\,\hat{\omega}_c -\tfrac{1}{4}\,\mathsf{s}_c\,\hat{\mathds{1}} \right],
\ee \end{widetext}
including the generalization to multiple measurement and feedback channels. 

\subsection{Mixed States via Stochastic Master Equation \label{sec-CSME}}

Generalization of the above to mixed states can be performed using the It\^{o} Stochastic Master Equation (SME) as a point of departure.
The dynamics due to measurement alone (without feedback) can be expressed \cite{BookBarchielli, BookWiseman, BookJacobs}
\be \label{Ito-SME}
d\hat{\rho} = \widehat{\boldsymbol{\mathsf{a}}}_{L}[\hat{\rho}]\,dt + \widehat{\boldsymbol{\mathsf{b}}}_L[\hat{\rho}]\cdot d\mathbf{W}.
\ee
Here the boldface notation indicates that many measurement channels might be present, such that $\widehat{\boldsymbol{\mathsf{a}}}$ is a sum over many individual $\widehat{\mathsf{a}}$, and $\widehat{\boldsymbol{\mathsf{b}}}\cdot d\mathbf{W}$ denotes a sum over many terms $\widehat{\mathsf{b}}_{L_c}\,dW_c$. 
Feedback can be added to those measurement dynamics via
\be \label{ito-fb-ansatz}
\hat{\rho}(t+dt) = \hat{\mathcal{U}}\left(\hat{\rho} + d\hat{\rho} \right)\hat{\mathcal{U}}^\dag. 
\ee
Expanding with It\^{o} calculus in a similar manner as the pure--state case above leads us to 
\be \begin{split} \label{ito-fb} 
d\hat{\rho} &= \left\lbrace i\left[\hat{\rho},\hat{\Omega}\right] + i\left[\widehat{\boldsymbol{\mathsf{b}}}_L[\hat{\rho}],\hat{\boldsymbol{\omega}}\right] + \widehat{\boldsymbol{\mathsf{a}}}_L[\hat{\rho}] + \widehat{\boldsymbol{\mathsf{a}}}_\omega[\hat{\rho}]   \right\rbrace dt \\ & \quad\quad+ \left\lbrace \widehat{\boldsymbol{\mathsf{b}}}_L[\hat{\rho}] + i\left[\hat{\rho},\hat{\boldsymbol{\omega}}\right] \right\rbrace \cdot d\mathbf{W} \\
&= \widehat{\boldsymbol{\mathcal{A}}}[\hat{\rho}]\,dt + \widehat{\boldsymbol{\mathcal{B}}}[\hat{\rho}]\cdot d\mathbf{W}.
\end{split}\ee
Note that all instances of $\widehat{\boldsymbol{\mathsf{a}}}$ and $\widehat{\boldsymbol{\mathsf{b}}}$ are implicitly summed over measurement / feedback channels $c$. 
Noise--cancellation requires that we set $\widehat{\boldsymbol{\mathcal{B}}}[\hat{\rho}] = 0$, such that 
\be \label{CSME_noise_cancel_condition}
\sqrt{\eta_c} \left( \hat{L}_c\,\hat{\rho} + \hat{\rho}\,\hat{L}_c^\dag - \mathsf{s}_c\,\hat{\rho} \right) + i [\hat{\rho},\hat{\omega}_c] = 0 \quad\forall~c
\ee
is the mixed--state generalization of the NCC \eqref{CSSE_NCcondition}. 
In the event that this noise--cancellation condition is satisfied for each channel $c$, we then have noise--canceled dynamics expressed by 
\be \label{CME_ANC} \begin{split}
\dot{\rho} &= \widehat{\boldsymbol{\mathcal{A}}}[\hat{\rho}]
\\ &= i\left[\hat{\rho},\hat{\Omega}\right] + i\left[\widehat{\boldsymbol{\mathsf{b}}}_L[\hat{\rho}],\hat{\boldsymbol{\omega}}\right] + \widehat{\boldsymbol{\mathsf{a}}}_L[\hat{\rho}] + \widehat{\boldsymbol{\mathsf{a}}}_\omega[\hat{\rho}]
\\ & = i\left[\hat{\rho},\hat{\Omega}\right] + i \sum_c \sqrt{\eta_c}\,\left[\hat{L}_c\,\hat{\rho} + \hat{\rho}\,\hat{L}_c^\dag - \mathsf{s}_c\,\hat{\rho} , \hat{\omega}_c\right] 
\\ & \quad\quad + \hat{L}_c\,\hat{\rho}\,\hat{L}_c^\dag - \tfrac{1}{2}\,\hat{L}_c^\dag\hat{L}_c\,\hat{\rho} - \tfrac{1}{2}\,\hat{\rho}\,\hat{L}_c^\dag\hat{L}_c 
\\ & \quad\quad + \hat{\omega}_c\,\hat{\rho}\,\hat{\omega}_c - \tfrac{1}{2}\,\hat{\omega}_c^2\,\hat{\rho} - \tfrac{1}{2}\,\hat{\rho}\,\hat{\omega}_c^2, 
\end{split} \ee
which generalizes \eqref{NC_nHH}. 
Note that in the event that \eqref{CSSE_NCcondition} and/or \eqref{CSME_noise_cancel_condition} are \emph{not} satisfied, \eqref{CME_ANC} still has physical meaning, in that these dynamics represent the average evolution under measurement and instantaneous feedback.

\section{Noise Canceling Feedback: Solutions and their Dynamics \label{app_omega_dynamics}}

We now consider the construction of feedback controls $\hat{\omega}$ that satisfy \eqref{CSSE_NCcondition} and \eqref{CSME_noise_cancel_condition}. 
This means that we solve the simplest form of the problem we are immediately confronted with when looking to perform NCQF: Given $\hat{L}$ and and real--time estimate of $\ket{\psi}$, what $\hat{\omega}$ can we choose to render the controlled dynamics deterministic rather than stochastic?
Two short explorations will allow us to understand our eventual solution for pure $\ket{\psi}$, as well as the corresponding generalization to mixed states and inefficient monitoring (see \secref{sec-NCC_Decompose}): First, it will be convenient to separate the Hermitian and Anti-Hermitian parts of the NCC, as they behave distinctly. 
Second, we define subspaces containing and orthogonal to the current quantum state.
After writing each of these definitions in turn, it will relatively straightforward to show that the NCC can be satisfied by a feedback that cancels out noisy transitions between the state $\ket{\psi}$ and the space orthogonal to that state. 

In this section we will always perform calculations for a single measurement channel. 
Based on the preceding section, we may appreciate that noise cancellation may be treated separately for each channel with the overall dynamics a sum over those in each channel, such that the generalization from one channel to multiple measurements and feedbacks is quite trivial. 

\subsection{Preparation: Manipulating the Noise--Cancellation Condition and Dynamics}

To begin, it is helpful to decompose the measurement $\hat{L}$ and NCC into its Hermitian and anti-Hermitian parts. 
Let $\hat{L} = \hat{X} + \hat{Y}$ with
\be 
\hat{X} = \tfrac{1}{2}\left(\hat{L} + \hat{L}^\dag \right) = \hat{X}^\dag, \quad \hat{Y} = \tfrac{1}{2}\left(\hat{L} - \hat{L}^\dag\right) = -\hat{Y}^\dag
\ee 
We may immediately observe that only the Hermitian quadrature contributes signal to the measurement, because
\be 
\ensavg{\hat{\mathsf{s}}} = \ensavg{\hat{L}+\hat{L}^\dag} = \ensavg{\hat{X} + \hat{Y} + \hat{X}^\dag + \hat{Y}^\dag} = 2 \ensavg{\hat{X}}
\ee

The noise--cancellation condition \eqref{NCC_both} may be generalized for arbitrary measurement efficiency, reading
\be \label{NCC_eta_Xi}
\hat{\Xi}\,\hat{\rho}+\hat{\rho}\,\hat{\Xi}^\dagger = \sqrt{\eta}\,\mathsf{s} \, \hat{\rho},
\ee 
with the noise operator $\hat{\Xi}$ defined as
\be 
\hat{\Xi}= \sqrt{\eta}\,\hat{L} - i\,\hat{\omega} = \sqrt{\eta}\,\hat{X}-i\,\tilde{\omega} \quad\text{with}\quad \tilde{\omega} = \hat{\omega}+i\sqrt{\eta}\,\hat{Y}.
\ee 
For pure states, 
we may re-write \eqref{CSSE_NCcondition} in similar notation, obtaining
\be \begin{split}
0 & = \left( \sqrt{\eta}\,\hat{X} + \sqrt{\eta}\,\hat{Y} - i\,\hat{\omega} - \tfrac{1}{2}\,\sqrt{\eta}\,\mathsf{s}\,\hat{\mathds{1}} \right) \ket{\psi} \\
 & =
\left(  \sqrt{\eta} \left(\hat{X} - \ensavg{\hat{X}}\,\hat{\mathds{1}} \right) - i\left( \hat{\omega} + i\sqrt{\eta}\,\hat{Y} \right) \right) \ket{\psi} \\ 
& = \left(\sqrt{\eta}\,\Delta \hat{X} - i \,\tilde{\omega} \right) \ket{\psi},
\end{split} \ee 
which is a special case of \eqref{NCC_eta_Xi}. 
Clearly $\tilde{\omega} = \hat{\omega}+i\,\sqrt{\eta}\,\hat{Y}$ is a Hermitian control, indicating that we may effectively absorb the anti-Hermitian part of $\hat{L}$ into $\tilde{\omega}$ without loss of generality. 
We also introduce the shorthand
\be 
\Delta\hat{X} \equiv \hat{X}-\ensavg{\hat{X}} = \tfrac{1}{2}\left( \hat{\mathsf{s}} - \ensavg{\hat{\mathsf{s}}}\right).
\ee 

\subsection{Solving the NCC for Pure States \label{sec_SolveNCC_pure}}

We now consider the case of pure states and perfect measurement efficiency in more detail: 
In that case of perfect measurement efficiency $\eta = 1$, and pure state $\hat{\rho} = \ket{\psi}\bra{\psi}$, it is \emph{always} possible to find a solution to \eqref{nc_rho_space}. 
Consider
\be 
\hat{D}_\Pi = \hat{U}^\dag\hat{\rho}\,\hat{U} \quad\text{for}\quad \hat{U}^\dag\hat{U} = \hat{\mathds{1}} \quad\&\quad \hat{D}_\Pi = \mathrm{diag}\,\boldsymbol{\lambda},
\ee
where the unitary $\hat{U}$ is assembled from the orthonormal eigenvectors of $\hat{\rho}$, and $\hat{\Pi}$ is the diagonal matrix of its real eigenvalues. 
In the event that $\hat{\rho}$ represents a pure state, the density matrix is a rank--$1$ projector, and will consequently have one $\lambda = 1$ and all other $\lambda_j = 0$. 
Thus in the case $\hat{\rho} = \ket{\psi}\bra{\psi} = \hat{\Pi}$ we can, without loss of generality, write
\begin{subequations} \label{diagonalize_NCC_basis} \be 
\hat{D}_\Pi = \left( \begin{array}{cccc}
1 & 0 & 0 & \cdots \\
0 & 0 & 0 & \cdots \\
0 & 0 & 0 & \cdots \\
\vdots & \vdots & \vdots & \ddots
\end{array} \right) \quad\text{with}
\ee \be 
\hat{U} = \left( \ket{\psi} ~~ \ket{\psi^\perp_1} ~~ \ket{\psi^\perp_2} ~~ \cdots \right) \quad\&\quad 
\hat{U}^\dag = \left( \begin{array}{c} 
\bra{\psi} \\ \bra{\psi^\perp_1} \\ \bra{\psi^\perp_2} \\ \vdots
\end{array} \right).
\ee \end{subequations}

To proceed with finding a solution in this special case, let us apply the frame change implied by \eqref{diagonalize_NCC_basis} to the NCC \eqref{NCC_both}, i.e.
\be \begin{split}
\lbrace \hat{U}\hat{\Pi}\hat{U}^\dag &, \Delta\hat{X} \rbrace = i[\tilde{\omega},\hat{U}\hat{\Pi}\hat{U}^\dag] \\&\rightarrow\quad 
\lbrace \hat{\Pi}, \hat{U}^\dag\Delta\hat{X}\hat{U} \rbrace = i[\hat{U}^\dag\tilde{\omega}\hat{U},\hat{\Pi}].
\end{split}\ee
With the density matrix diagonalized, we may now re-write the matrix elements in this expression explicitly:
\be \label{solve-omega-tilde} \begin{split}
&\left( \hat{U}^\dag\Delta\hat{X}\hat{U}\right)_{mn} (\lambda_m+\lambda_n) = i\left( \hat{U}^\dag\tilde{\omega}\hat{U}\right)_{mn} (\lambda_n - \lambda_m) \\&\rightarrow\quad \left( \hat{U}^\dag\tilde{\omega}\hat{U}\right)_{mn} = i\left( \hat{U}^\dag\Delta\hat{X}\hat{U}\right)_{mn} \frac{\lambda_m+\lambda_n}{\lambda_m -\lambda_n} = \Upsilon_{mn}. 
\end{split} \ee 
In this equivalent formulation, elements of $\hat{\Upsilon} \equiv \hat{U}^\dag\tilde{\omega}\hat{U}$ for which $\lambda_m - \lambda_n = 0$ (thereby including all of the diagonals) are free parameters.
Those elements of $\tilde{\omega}$ that are \emph{required} to a take on some non-zero value to satisfy the NCC are those mediating a transition between the $\hat{\Pi}$ (i.e.~the unique $\lambda  = 1$ in the pure--state case) and $\hat{\Pi}^\perp$ (i.e.~$\lambda = 0$ in the pure--state case) spaces.
For the pure state case of immediate interest, that means we must fix elements $\tilde{\omega}_{mn}$ for which $(\lambda_m+\lambda_n)/(\lambda_m-\lambda_n) = \pm 1$ instead of $0$. 

Let us focus on these transition elements between $\hat{\Pi}$ and $\hat{\Pi}^\perp$: For these elements only \emph{one of} $\lambda_m,\,\lambda_n = 1$, with the other zero. 
Consider first $m = 1$, and $n \neq 1$, such that $\lambda_m = 1$ and $\lambda_n = 0$:
\begin{subequations} \be 
\frac{\lambda_m+\lambda_n}{\lambda_m -\lambda_n} = 1 \quad\rightarrow\quad \Upsilon_{1n} = -i\,\bra{\psi}\hat{U}^\dag\,\Delta\hat{X}\,\hat{U}\ket{\psi_{n-1}^\perp}.
\ee
The transposed elements ($m \neq 1$, $n = 1$) similarly read
\be 
\frac{\lambda_m+\lambda_n}{\lambda_m -\lambda_n} = -1 \quad\rightarrow\quad \Upsilon_{m1} = i\,\bra{\psi^\perp_{m-1}}\hat{U}^\dag\,\Delta\hat{X}\,\hat{U}\ket{\psi},
\ee \end{subequations}
which makes explicit that $\hat{\omega}$ has remained Hermitian throughout this construction. 
It follows that for pure state $\hat{\rho} = \ket{\psi}\bra{\psi}$, we can construct the necessary elements of $\hat{\Upsilon} \equiv \hat{U}^\dag\tilde{\omega}\hat{U}$ from given quantities, and then reverse our basis transformation to find
\be \label{omega_upsilon}
\hat{\omega} = \hat{U}\hat{\Upsilon}\hat{U}^\dag -i\,\hat{Y}.
\ee
This can be written in more operationally straightforward form by noting that $\hat{U}$ is itself assembled from orthonormal basis vectors of the $\hat{\Pi}$ and $\hat{\Pi}^\perp$ spaces: We therefore re-iterate \eqref{omega_upsilon} as [recall \eqref{basic_NCQF_solution}]
\be \label{pure_NC_single}
\hat{\omega}_0 = i\left( [\hat{\rho},\hat{X}] - \hat{Y} \right),
\ee
which accounts for the necessary transition elements only.
It is easy to verify that the NCC \eqref{CSSE_NCcondition} and/or \eqref{CSME_noise_cancel_condition} are satisfied by this solution, e.g.
\be \begin{split}
&\left(\hat{L} - i\,\hat{\omega}_0 -\tfrac{1}{2}\,\mathsf{s}\,\hat{\mathds{1}} \right) \ket{\psi} \\ &\quad = \left( \hat{X} + [\hat{\Pi},\hat{X}] - \tfrac{1}{2}\,\mathsf{s}\,\mathds{1} \right) \ket{\psi} \\ &\quad = \hat{X}\ket{\psi} + \ket{\psi}\bra{\psi}\hat{X}\ket{\psi} - \hat{X}\ket{\psi} - \tfrac{1}{2}\,\mathsf{s}\ket{\psi} = 0
\end{split} \ee 
where we have used $\ensavg{\hat{X}} = \tfrac{1}{2}\,\mathsf{s}$. 

\subsection{General Subspace Decomposition \label{sec-NCC_Decompose}}

Let us generalize the eigen-decomposition of $\hat{\rho}$ beyond the pure state case (of $\hat{\rho}$ being a rank-$1$ projector). 
Suppose that $\hat{\rho}$ can be written in the form
\be 
\hat{\rho} = \sum_{m=1}^M \lambda_m(t) \ket{m(t)}\bra{m(t)} = \sum_{m=1}^M \lambda_m\,\hat{\Pi}_m,
\label{rho_diagonal_decomp}
\ee
where $\lambda_m$ are the eigenvalues of $\hat{\rho}$, which may be interpreted as probabilities obeying $\sum_m \lambda_m = 1$ and $0\leq \lambda_m \leq 1~\forall~m$. 
We also have $\hat{\Pi}_m$ the projectors onto the associated eigenstates. 
Here $M$ is the rank of $\hat{\rho}$, and for $\hat{\rho} \in \mathbb{C}^{N \times N}$ we have $\mathrm{dim}(\mathrm{ker}(\hat{\rho})) = N-M$, where $\mathrm{ker}(\hat{\rho})$ denotes the kernel (or nullspace) of $\hat{\rho}$.
Then we may write a basis for $\hat{\rho}$ in terms of 
\be 
\hat{\Pi} = \sum_{m = 1}^M \hat{\Pi}_m \quad\&\quad \hat{\Pi}^\perp = \hat{\mathds{1}} - \hat{\Pi},
\label{projector_eq1}
\ee
such that $\hat{\Pi}^\perp$ is the projector into the nullspace of $\hat{\rho}$, and $\hat{\Pi}$ is the projector into a space where $\hat{\rho}$ has non-zero support.  
These projectors necessarily have the properties
\be
\hat{\Pi}\,\hat{\rho} = \hat{\rho} = \hat{\rho}\,\hat{\Pi} \quad\&\quad
\hat{\Pi}^\perp\hat{\rho} = 0 = \hat{\rho}\,\hat{\Pi}^\perp
\label{projector_eq2}
\ee

Feedback may be written in this basis as
\begin{equation}
    \tilde{\omega}=\hat{\Pi}\,\hat{A}\,\hat{\Pi}+i\sqrt{\eta}\,\left[\hat{\Pi},\hat{X}\right]+\hat{\Pi}^\perp\hat{B}\hat{\Pi}^\perp,
    \label{NCQF_form_general}
\end{equation}
where $\hat{A}$ and $\hat{B}$ are Hermitian matrices to be determined. 
Practically speaking, this form divides the feedback into three terms, whose function (from left to right) is to create dynamics within $\hat{\Pi}$, across the partition separating $\hat{\Pi}$ from $\hat{\Pi}^\perp$, and within the nullspace $\hat{\Pi}^\perp$. 
Substituting this form into \eqref{NCC_eta_Xi}, we obtain
\begin{subequations} \label{NCC_projection}
\be
\hat{\Xi}\,\hat{\rho} + \hat{\rho}\,\hat{\Xi}^\dag = \hat{\Pi} \left( \lbrace \sqrt{\eta}\,\hat{X},\hat{\rho}\rbrace + i[\hat{\rho},\hat{A}] \right) \hat{\Pi} = \sqrt{\eta}\,\mathsf{s}\,\hat{\rho}.
\ee
It is clear that $\hat{B}$ is a completely free matrix, as it specifies elements of $\tilde{\omega}$ that act only within the nullspace of $\hat{\rho}$, and consequently does not appear in the NCC.  
The question of whether or not a noise--canceling feedback can be found is then restricted to the subspace where $\hat{\rho}$ has support, and reduced to ascertaining whether a Hermitian $\hat{A}$ exists to satisfy the restricted NCC. 
In particular, using the notation $\hat{\Pi}\,X\,\hat{\Pi} = \hat{X}_\rho$, $\hat{\Pi}\,\hat{A}\,\hat{\Pi} = \hat{A}_\rho$, and so on, we may search for $\hat{A}$ such that 
\be 
\left(\left\lbrace \sqrt{\eta}\,\hat{X}_{\hat{\rho}},\hat{\rho}\right\rbrace-\sqrt{\eta}\,\mathsf{s}\,\hat{\rho}\right)-i\left[\hat{A}_{\hat{\rho}},\hat{\rho}\right]=0
    \label{nc_rho_space}
\ee \end{subequations}
is satisfied. 
Clearly, when $\hat{\Pi}$ is rank-$1$, \eqref{nc_rho_space} reduces to a scalar equation $2\,X_\rho = 2\,\mathsf{s}$, which is satisfied by \emph{any} choice of $\hat{A}_\rho$. 

\subsection{Noise-canceling feedback beyond pure states \label{sec_NC-mixed}}

We have seen that \eqref{NCC_projection} admits a family of solutions when $\hat{\rho}$ is a pure a state. 
This is not clearly guaranteed for mixed states however, which will generically occur any evolution with $\eta < 1$. 
Let us investigate these cases by seeking to constrain the elements of $\hat{A}$ further.  
We suppose that we may write any operator $\hat{O}$ in $\{\ket{m}\}$ basis ($\hat{\rho}$ eigenbasis, see \eqref{rho_diagonal_decomp}) as per
\begin{equation}
    \hat{O}= \sum_{m,n} O_{mn}\ket{m}\bra{n}.
\end{equation}
Then \eqref{nc_rho_space} can be written as
\begin{subequations} \be \begin{split}
     \sum_{m,n} \Big\{&(\sqrt{\eta}\,X_{mn} -i\,A_{mn}) \lambda_m \\&+ (\sqrt{\eta}\,X_{mn}+i\,A_{mn}) \lambda_n \Big\}\ket{m}\bra{n} \\ =& \sum_{m} \sqrt{\eta}\,\mathsf{s}\,\lambda_m\ket{m}\bra{m}.\label{nc_cond_mn_elements}
\end{split} \ee
For $m = n$ we have
\begin{equation}
    \sqrt{\eta}\,X_{mm}=\tfrac{1}{2}\,\sqrt{\eta}\,\mathsf{s} = \sqrt{\eta} \sum_m \lambda_m \bra{m}\hat{X}\ket{m},
    \label{nc_xmm}
\end{equation}
with the elements $A_{mm}$ left as free parameters. 
For $m\neq n$, we find instead 
\begin{equation}
    \sqrt{\eta}\,X_{mn}(\lambda_m+\lambda_n)+i\,A_{mn}(\lambda_n-\lambda_m)=0.
    \label{xmn_omegamn}
\end{equation} \end{subequations}
Notice that this is effectively just a repeat of \eqref{solve-omega-tilde}. 
For $\lambda_m\neq\lambda_n$ \eqref{xmn_omegamn} puts constraints on the values of $A_{mn}$. 
Specifically, we have
\begin{equation}
A_{mn}=i\sqrt{\eta}\,X_{mn}\frac{\lambda_m+\lambda_n}{\lambda_m-\lambda_n}
    \label{a_mnvals}
\end{equation}
for all $m \neq n$ for which $\lambda_n \neq \lambda_m$. 
In a degenerate subspace (i.e.~for matrix elements where $\lambda_m = \lambda_n$ for $m \neq n$), we may choose the $A_{mn}$ freely, but \emph{only} if 
\be X_{mn}=0.
\label{xmn_0}
\ee 
This last condition turns out to be a non-issue however: Within any $\lambda$--degenerate subspaces ($\lambda_m = \lambda_n$ for $m \neq n$), we are free to choose a basis that also block--diagonalizes any non-zero elements in $\hat{X}$, such that the corresponding $X_{mn}$ off--diagonal elements can always be made zero, avoiding a divergence in \eqref{a_mnvals}. 
Notice that in the pure state case, \eqref{a_mnvals} is essentially a re-writing of \eqref{solve-omega-tilde} in the $\hat{\rho}$--eigenbasis.

In summary then, we may reduce the noise--cancellation problem to the subspace where $\lambda_m > 0$, i.e.~the subspace where $\hat{\rho}$ has non-zero support. 
A perfectly noise--canceling feedback exists \emph{iff} for $m$ with $\lambda_m > 0$ we have 
\be \label{mixed-state-NCC-exist}
X_{nn} = \sum_m \lambda_m\,X_{mm} = \tfrac{1}{2}\mathsf{s} ~\forall~n.
\ee
Notice that for a pure state, we have a single $\lambda_m = 1$, in which case the RHS of this condition contains a single term, and is trivially satisfied. 
For mixed states, the NCC \eqref{nc_rho_space} is \emph{unsatisfiable} specifically in the case where the projection of the Hermitian part of the measurement operator into the $\hat{\rho}$--eigenbasis has different elements down the diagonal, inconsistent with \eqref{mixed-state-NCC-exist}. 
Physically, this condition suggests a few cases where perfect NCQF can be obtained: If $\hat{\rho}$ is i) in a superposition or mixture entirely within a Zeno subspace (or decoherence--free subspace) of the measurement channel of interest, such that the measurement does not induce noise to begin with, or ii) in a superposition or mixture of states complementary to the measurement of interest, then a noise--canceling solution can be found.
These cases can be understood as the extremes where the measurement is either i) minimally informative (the measurement confines the state to an eigenstate of $\hat{X}$), or ii) maximally informative (the measurement is maximally disturbing, relative to the state, which is good for entanglement generation and purification). 
The latter typically correspond to the $\mathsf{s} = 0$ subspaces, provided $\hat{X}$ has been normalized to be traceless  (recall the discussions of \secref{sec_Entangle_sub-discuss} and \secref{sec_Purify}).
We do \emph{not} rule out the possibility of other ways to satisfy \eqref{mixed-state-NCC-exist}.

In practice, one may apply a bound on the magnitude of the feedback $||\hat{\omega}||$ (or elements thereof) that one can implement, and then find the corresponding bounds on $|X_{mn}|$ and $|\lambda_m - \lambda_n|$ for which the bounded controls admit a solution to \eqref{nc_rho_space}. 
We reiterate that the diagonal elements and some of the off-diagonal elements of $\hat{A}$ (specifically those within blocks of degenerate $\lambda$), and \emph{all} of the elements of $\hat{B}$, are free for us to choose. 
Thus, for a particular continuously monitored system in a specific state, either no noise--canceling feedback solutions exist, or an infinite number of solutions exist. 
We remark that  the examples developed in the main text suggest that the free parameters can often be chosen in helpful ways, e.g.~to simplify a noise--canceling feedback to use local operations, speed up the convergence to desired state, etc.
See Appendix \ref{app_Free_params} for further discussion of this point.

\subsection{Dynamics under Noise--Canceling Feedback \label{sec-nHH_stability}}

Let us investigate the noise--canceled dynamics in more detail. 
In the eigenbasis of $\hat{\rho}$ (recall \eqref{rho_diagonal_decomp}), we may write
\be 
    \lambda_m=\bra{m}\hat{\rho}\ket{m} \quad\text{such that}
\ee \be 
    \dot{\lambda}_m=\left(\partial_t\bra{m}\right)\hat{\rho}\ket{m}+\bra{m}\dot{\rho}\ket{m}+\bra{m}\hat{\rho}\left(\partial_t\ket{m}\right).
\ee
Using \eqref{rho_diagonal_decomp} for the first and the third terms on the right hand side, we get
\be \label{lambda_dot}
    \begin{split}\dot{\lambda}_m &=\bra{m}\dot{\rho}\ket{m}+\lambda_m\left(\partial_t\bra{m}\right)\ket{m}+\lambda_m\bra{m}\left(\partial_t\ket{m}\right)\\&=\bra{m}\dot{\rho}\ket{m}+\lambda_m\partial_t\left(\langle m|m\rangle\right) \\
    &= \bra{m}\dot{\rho}\ket{m}. \end{split}
\ee

\subsubsection{Manipulating the Noise--Canceled Dynamics} 

We next manipulate $\dot{\rho}$ \eqref{CME_ANC} to re-write the dynamics in a more convenient form (assuming $\eta = 1$). 
Note that the noise cancellation condition \eqref{NCC_eta_Xi} can be written in three equivalent forms
\begin{subequations} \label{noise_sub-forms}
\be \label{sub_no_1}
\widehat{\mathsf{b}}_L[\hat{\rho}]-i\left[\hat{\omega},\hat{\rho}\right]=0,
\ee \be \label{sub_no_2}
\hat{X}\hat{\rho}+\hat{\rho}\,\hat{X}-i\left[\tilde{\omega},\hat{\rho}\right]=\mathsf{s}\,\hat{\rho},
\ee \be \label{sub_no_3}
\hat{\Xi}\,\hat{\rho}+\hat{\rho}\,\hat{\Xi}^\dagger=\mathsf{s}\,\hat{\rho}.
\ee 
This last expression allows us to move $\hat{\rho}$ and $\hat{\Xi}$ through each other, because 
\be \label{sub_no_3b}
\hat{\Xi}\,\hat{\rho} = \hat{\rho}\,(\mathsf{s} - \hat{\Xi}^\dag) \quad\&\quad 
\hat{\rho}\,\hat{\Xi}^\dag = (\mathsf{s} - \hat{\Xi})\,\hat{\rho}.
\ee
Additionally, from these expressions we may obtain
\be  \label{sub_no_4}
\tilde{\omega} = i(\hat{\Xi} - \hat{X}) = i(\hat{X}-\hat{\Xi}^\dag) = \tfrac{i}{2}(\hat{\Xi}-\hat{\Xi}^\dag),
\ee and similarly \be \label{sub_no_5}
\hat{X} = \tfrac{1}{2}(\hat{\Xi} + \hat{\Xi}^\dag). 
\ee
\end{subequations}
Recall that $\hat{L}=\hat{X}+\hat{Y}$ and $\hat{\omega}=\tilde{\omega}-i\,\hat{Y}$, $\hat{\Xi}=\hat{X}-i\,\tilde{\omega}$ are implicit in all of the above definitions. 
Note that \eqref{sub_no_4} and \eqref{sub_no_5} are adequate to derive \eqref{HNC_Xi} from \eqref{NC_NHH_single}.

Now we re-arrange \eqref{CME_ANC}. 
Using \eqref{sub_no_1} and the fact that $\widehat{\mathsf{a}}_\omega[\hat{\rho}]=-\frac{1}{2}\left[\hat{\omega},\left[\hat{\omega},\hat{\rho}\right]\right]$, we can write the noise--canceled dynamics as
as 
\begin{equation}
    \dot{\rho}=-i[\hat{\Omega},\hat{\rho}]+\widehat{\mathsf{a}}_L[\hat{\rho}]-\widehat{\mathsf{a}}_\omega[\hat{\rho}].
\end{equation}
Next, we may decompose $\hat{L}$ into its Hermitian and anti-Hermitian parts to obtain 
\begin{equation}
  \begin{split}
      \dot{\rho}=& i[\hat{\rho},\hat{\Omega}]+\widehat{\mathsf{a}}_X[\hat{\rho}]-\widehat{\mathsf{a}}_{\tilde{\omega}}[\hat{\rho}]+\hat{Y}\,\hat{\rho}\,\hat{\Xi}^\dagger-\hat{\Xi}\,\hat{\rho}\,\hat{Y}\\&-\frac{1}{2}\left\lbrace\hat{\Xi}^\dagger\,\hat{Y}-\hat{Y}\,\hat{\Xi},\hat{\rho}\right\rbrace.
  \end{split}  
\end{equation}
Repeated use of \eqref{sub_no_3} and \eqref{sub_no_3b} leads us to
\begin{equation}
\dot{\rho}= i\left[\hat{\rho},\hat{\Omega}+i\,\mathsf{s}\,\hat{Y}-\tfrac{i}{2}\left(\hat{Y}\hat{\Xi}+\hat{\Xi}^\dagger\hat{Y}\right)\right]+\widehat{\mathsf{a}}_X[\hat{\rho}]-\widehat{\mathsf{a}}_{\tilde{\omega}}[\hat{\rho}].
\end{equation}
Let's consider the last two (dissipative) terms. 
As $\widehat{\mathsf{a}}_{\tilde{\omega}}[\hat{\rho}]=\frac{1}{2}\left[\tilde{\omega},\left[\hat{\rho},\tilde{\omega}\right]\right]$, we can substitute the inner commutator using \eqref{sub_no_2} to get
\begin{equation}
\begin{split}
    \dot{\rho}=&i\left[\hat{\rho},\hat{\Omega}+i\,\mathsf{s}\,\hat{Y}-\tfrac{i}{2}\left(\hat{Y}\,\hat{\Xi}+\hat{\Xi}^\dagger\,\hat{Y}\right)-\tfrac{1}{2}\,\mathsf{s}\,\tilde{\omega}\right]\\&+\tfrac{1}{2}\,\hat{X}\,\hat{\rho}\,\hat{\Xi}^\dagger+\tfrac{1}{2}\,\hat{\Xi}\,\hat{\rho}\,\hat{X}-\tfrac{1}{2}\,\hat{\Xi}^\dagger\,\hat{X}\,\hat{\rho}-\tfrac{1}{2}\,\hat{\rho}\,\hat{X}\,\hat{\Xi}.
\end{split}
\end{equation} 
Then finally, using \eqref{sub_no_2} and \eqref{sub_no_3}, 
the noise--canceled density matrix evolution \eqref{CME_ANC} can be simplified (for one measurement channel and $\eta = 1$) to
\begin{equation}
\begin{split}
    \dot{\rho}=&-i\left[\hat{\Omega}+i\,\mathsf{s}\,\hat{Y}-\frac{i}{2}\left(\hat{Y}\hat{\Xi}+\hat{\Xi}^\dagger\hat{Y}\right)-\mathsf{s}\,\tilde{\omega},\hat{\rho}\right]\\&+\left\lbrace\frac{\mathsf{s}^2}{4}-\frac{1}{2}\left(\hat{X}\hat{\Xi}+\hat{\Xi}^\dagger\hat{X}\right),\hat{\rho}\right\rbrace.
\end{split}
    \label{rho_evol_XY}
\end{equation}
We now shorthand this form, redefining
\begin{subequations} \label{rho_evol_PQ_compact} \be
\tilde{\Omega}= \hat{\Omega}+i\,\mathsf{s}\,\hat{Y}-\frac{i}{2}\left(\hat{Y}\hat{\Xi}+\hat{\Xi}^\dagger\hat{Y}\right)-\mathsf{s}\,\tilde{\omega}, \quad\&
\ee \be \label{def_G_nH}
\hat{G}=\frac{\mathsf{s}^2}{4}-\frac{1}{2}\left(\hat{X}\,\hat{\Xi}+\hat{\Xi}^\dagger\hat{X}\right) = \frac{\mathsf{s}^2}{4}-\hat{X}^2 + \tfrac{i}{2}[\hat{X},\tilde{\omega}],
\ee
so as to express \eqref{rho_evol_XY} as  
\begin{equation}
     \dot{\rho}= i\left[\hat{\rho},\tilde{\Omega}\right]+\left\lbrace \hat{G},\hat{\rho}\right\rbrace.
\end{equation}\end{subequations}
Notice that in this expression of the dynamics, we operate on $\hat{\rho}$ only from the left \emph{or} the right, but never both at once.


\subsubsection{\label{ncc_dynamics_app}Properties of the Noise--Canceled Dynamics} 

We now use \eqref{rho_evol_PQ_compact} to elucidate some general properties of the noise--canceled dynamics.  
Using \eqref{lambda_dot} and \eqref{rho_evol_PQ_compact}, we get
\begin{equation} \label{lambda_dot_m}
    \dot{\lambda}_m=2\,\lambda_m \bra{m}\hat{G}\ket{m} = 2\,\lambda_m\,\ensavg{\hat{G}}_m,
\end{equation}
where the index $m$ is not summed over. 
Using \eqref{nc_xmm}, \eqref{a_mnvals}, and \eqref{xmn_0}, we can show that
\begin{equation} \dot{\lambda}_m=\lambda_m\sum_{\substack{k\neq m\\ \lambda_k\neq\lambda_m}}\frac{4\lambda_k|X_{m,k}|^2}{\lambda_m-\lambda_k}.
\label{eq:lambda_m_evol}
\end{equation}
Note, if $\lambda_m(0)=0$, $\lambda_m(t)$ remains $0$ throughout. Thus, the evolution does not increase the rank of $\hat{\rho}$, and pure states remain pure, as expected.
Also for degenerate mixed states, $\dot{\lambda}_m=0$. Therefore, degenerate mixed states remain degenerate. 
Also note, for the largest $\lambda_m$, $\dot{\lambda}_m\geq 0$, thus, the largest eigenvalue never decreases (even if as the state is it is associated with may evolve over the course of the evolution).

We may elaborate on these comments in the case of pure--state dynamics. The noise canceled dynamics can be expressed in terms of a non-Hermitian Hamiltonian given by \eqref{NC_NHH_single}. $\hat{\mathcal{H}}_{NC}$ can be simplified to \eqref{HNC_Xi}, paired with the noise--cancellation condition \eqref{NCC_Xi_main}, which we now use. 
When $\hat{\mathcal{H}}_{NC}$ acts on a eigenstate of the noise operator $\hat{\Xi}$ (i.e.~any state for which $\hat{\mathcal{H}}_{NC}$ is a valid generator of dynamics), we may write
\be
    \hat{\mathcal{H}}_{NC}\ket{\psi}= \left(\tilde{\Omega} + \mathsf{s}\,\tilde{\omega}\right)\ket{\psi}-\frac{i}{4}\left(\hat{\Xi}^\dagger-\frac{\mathsf{s}}{2}\right)^2\ket{\psi},
\ee 
Moving forward, we note that the operator manipulations 
\be 
\hat{\Xi}^\dagger-\tfrac{1}{2}\,\mathsf{s}=\Delta\hat{X}+i\,\tilde{\omega} \quad\&\quad \hat{\Xi}-\tfrac{1}{2}\,\mathsf{s}=\Delta\hat{X}-i\,\tilde{\omega}
\ee
are valid. 
Using the expression for noise canceling feedback in \eqref{NCQF_form_general} with $\hat{A}=0$, and using the projector relations from \eqref{projector_eq1} and \eqref{projector_eq2}, we can write $\hat{\mathcal{H}}_{NC}\ket{\psi}=$
\begin{equation}
    \begin{split}
    &\left(\tilde{\Omega} + \mathsf{s}\,\tilde{\omega}\right)\ket{\psi} \\& -\frac{i}{2}\bigg\{\left(\Delta\hat{X}^2-\langle\Delta\hat{X}^2\rangle\right)  +i\left(\hat{B}\,\Delta\hat{X}-\langle\hat{B}\,\Delta\hat{X}\rangle\right)\bigg\}\ket{\psi}.
    \end{split}\label{Hncpsi_simplified}
\end{equation}
The evolution of a pure state under noise-canceling feedback is then 
\begin{equation} \label{HNC_evol_B}
    \begin{split}\dot{\ket{\psi}}&=-i\left(\tilde{\Omega} + \mathsf{s}\,\tilde{\omega}\right) \ket{\psi}\\&-\frac{1}{2}\left\{\left(\Delta\hat{X}^2-\langle\Delta\hat{X}^2\rangle\right)+i\left(\hat{B}\Delta\hat{X}-\langle\hat{B}\Delta\hat{X}\rangle\right)\right\}\ket{\psi}.
    \end{split}
\end{equation}
It is easy to recognize that with $\tilde{\Omega} + \mathsf{s}\,\tilde{\omega}=0$, eigenkets of $\hat{X}$ are fixed points of the dynamics.
This is compatible with the discussion of \secref{sec-Meas-ES}.
We also remark that \eqref{HNC_evol_B} does however imply an additional and less-intuitive result: Even though $\hat{B}$ is completely free in the sense of being irrelevant to satisfying the NCC, it can still impact the dynamics under feedback.

In the next analysis, we assume $\tilde{\Omega} + \mathsf{s}\,\tilde{\omega}=0$ and $\hat{B}=0$, such that the dynamics are primarily driven by a relatively simple feedback. 
The assumption $\tilde{\Omega} + \mathsf{s}\,\tilde{\omega} = 0$ is most natural when we monitor a Hermitian observable, although it is clearly obtainable in general using the appropriate open--loop control $\hat{\Omega}$.
Then, under noise canceling feedback, the state evolution is given by
\begin{equation}
    \dot{\ket{\psi}}=-\frac{1}{2}\left(\Delta\hat{X}^2-\langle\Delta\hat{X}^2\rangle\right)\ket{\psi}=-\frac{1}{2}\left(\Delta\hat{X}^2-V_X\right)\ket{\psi},
    \label{psidot_simplified}
\end{equation}
where $V_X= \langle\Delta\hat{X}^2\rangle$ is the variance. For simplicity, we also assume that the observable $\hat{X}$ is time-independent. Using \eqref{psidot_simplified}, we can prove
\begin{equation}
    \dot{\mathsf{s}}=-2\langle\Delta\hat{X}^3\rangle, \quad \dot{V}_X=V_X^2-\langle\Delta\hat{X}^4\rangle.
\end{equation}
We can easily show that the right hand side of the second equation is nonpositive. 
Consider the positive operator 
\be
\hat{\nu} \equiv \left(\Delta\hat{X}^2-V_X\right)^2.
\ee
Then, we must have $\ensavg{\hat{\nu}}\geq 0$, which implies $\dot{V}_X\leq 0$. 
The variance is non-increasing, which implies that with unitary contributions to the dynamics either non-existent or canceled--out via open loop control, the remaining NCQF dynamics drive the system towards eigenspaces of $\hat{X}$.
We may appreciate how many of our examples with Hermitian observables in the main text (recall \ref{sec:half-par} and \ref{sec:GHZ} in particular) reflect the following behavior: without $\hat{\Omega}$, NCQF tends to drive the dynamics asymptotically towards $\hat{X}$ eigenspaces, which are stable fixed points of the dynamics, and which generate no measurement noise when reached (such that the feedback controller is required to intervene to a smaller degree as we approach the stable space).

We can write the following 
\begin{equation}
    \left\langle\Delta\hat{X}^3\right\rangle=\left\langle\hat{X}^3-\frac{3\mathsf{s}}{2}\hat{X}^2\right\rangle+\frac{\mathsf{s}^3}{4}.
\end{equation}
If $\hat{X}$ is a sum of two commuting Paulis, $\left\langle\hat{X}^3\right\rangle\propto \mathsf{s}$. In that case, if we start in the zero signal space, we remain $\mathsf{s}=0$ throughout the evolution.

\section{Free Parameters for Noise Cancellation or Minimization with Control Constraints \label{app_Free_params}}

We here elaborate on the ideas of \secref{sec_Restricted_Controls}. 
We first consider an alternative approach based on the use of free parameters derived in the previous section (free parameters can be used to modify a noise--canceling feedback operation $\hat{\omega}_0$ such as \eqref{basic_NCQF_solution} without impeding the noise--cancellation property). 
Then we extend the discussion of \secref{sec_Restricted_Controls} more directly, deriving several special cases and generalizations of the equations shown there. 

\subsection{Free Parameters and Feedback Simplifiability \label{sec-NCC_Free}}

Recalling \eqref{NCQF_form_general}, we have seen that free feedback parameters may be added by choosing non-zero $\hat{A}$ and $\hat{B}$ in our feedback. 
In the pure--state case, $\hat{A}$ may be chosen to add a component to the feedback along $\hat{\omega}_\mathrm{free} = \omega_\psi\,\ket{\psi}\bra{\psi}$, i.e.
\be \label{NCC_phase}
\left( \hat{L} - i\,(\hat{\omega}_0 + \hat{\omega}_\mathrm{free})) \right)\ket{\psi} = \left(\tfrac{1}{2}\,\mathsf{s} -i\,\omega_\psi \right)\ket{\psi},
\ee
where $\hat{B}$ (acting entirely within the nullspace) may contribute indirectly to the dynamics (recall \eqref{HNC_evol_B}), and $\hat{A}$ engages the global phase freedom of the NCC discussed briefly near \eqref{CSSE_NCcondition} and in Ref.~\cite{zhang2020locally}. 

\begin{figure}
\centering
\includegraphics[width = .75\columnwidth]{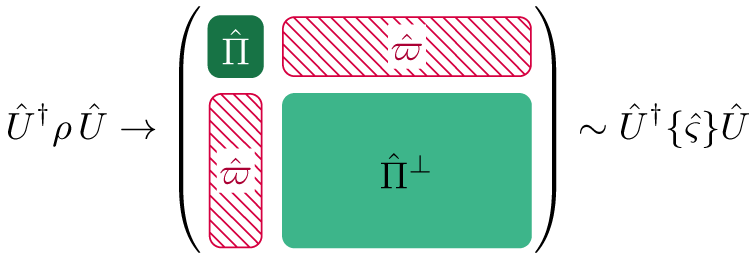}
\caption{
We illustrate the operator spaces $\hat{\Pi}$ (the current state), $\hat{\Pi}^\perp$ (the nullspace of $\hat{\rho}$) \eqref{diagonalize_NCC_basis}, and $\hat{\varpi}$ (transition elements) \eqref{transition_elems}, in the $\hat{\rho}$--eigenbasis. 
Hermitian operators acting on the $N\times N$ state, including the feedback $\hat{\omega}$, live in this space that may be expressed in terms of a basis of $N^2-1$ traceless and Hermitian operators $\lbrace\hat{\varsigma}\rbrace$, which are orthogonal with respect to an inner product $\tr{\hat{\varsigma}_m\,\hat{\varsigma}_n} \propto \delta_{nm}$ \cite{Bertlmann_2008}. 
NCQF operations with support on the green spaces $\hat{\Pi}$ and $\hat{\Pi}^\perp$ are free operations (for pure $\hat{\rho}$ and $\eta = 1$), while the noise cancellation condition \eqref{NCC_projection} constrains operations in the transition elements $\hat{\varpi}$.
This implies that if we have a noise--canceling operation $\hat{\omega}_0$ \eqref{pure_NC_single}, we are free to modify it only by adding any $\hat{\omega}_\mathrm{free}  = \sum_n c_n\,\hat{\varsigma}_n$ for $\hat{\varsigma}_n \notin \hat{\varpi}$. 
Any modifications to a NCQF protocol satisfying \eqref{NCC_projection} that use transition elements $\hat{\varpi}$, however, should be expected to break the protocol's noise cancellation property. 
This is the main restriction governing the constraint of $\hat{\omega}$ to a specific subset of operations $\varsigma' \in \lbrace \hat{\varsigma} \rbrace$ while satisfying the NCC \eqref{NCC_projection}. 
}\label{fig:transition_elems}
\end{figure}

For the sake of this discussion, let us suppose that we have chosen a basis for traceless Hermitian matrices $\lbrace \hat{\varsigma} \rbrace$ (see e.g.~\cite{Bertlmann_2008}) in which to represent our controls $\hat{\omega}$. 
We have some $\hat{\omega}_0$ obtained via \eqref{basic_NCQF_solution} or \eqref{pure_NC_single}, which is guaranteed to cancel the noise under ideal circumstances, but we wish to know if we can modify this feedback to write $\hat{\omega} = \hat{\omega}_0 + \hat{\omega}_\mathrm{free}$, such that $\hat{\omega}$ is still noise canceling, and \emph{only has support on a specific subset of} $\lbrace \hat{\varsigma} \rbrace$. 
Let us split our operator space into a desired and undesired space, i.e.~$\lbrace \hat{\varsigma} \rbrace = \varsigma' \oplus \varsigma^\perp$ where we wish to write a noise--canceling $\hat{\omega}$ that has support only on $\varsigma'$ but not on $\varsigma^\perp$. 
Note from \eqref{NCQF_form_general} that $\hat{A}$ and $\hat{B}$ are free matrices for $\eta = 1$ and $\hat{\Pi} = \op{\psi}{\psi} = \hat{\rho}$ is a pure state, i.e.~$\hat{\omega}_{\mathrm{free}}$ may include any terms that are strictly supported on the state $\hat{\Pi}$ or the nullspace $\lbrace\hat{\Pi}^\perp\rbrace$ of $\hat{\rho}$ (but \emph{not} the transition elements between those two spaces).
Such ``transition elements'' may be formally defined as being operations that can be written in the form 
\be \label{transition_elems}
\hat{\varpi} \propto \left( \sum_n c_n \ket{\psi_n^\perp}\right) \bra{\psi} + \ket{\psi}\left( \sum_n c_n^\ast \bra{\psi_n^\perp}\right)
\ee
using the notation of \eqref{diagonalize_NCC_basis}.
Then our feedback simplifiability question reduces thus: For every $\hat{\varsigma} \in \varsigma^\perp$ for which 
\be \label{simplify_condition}
\tr{\hat{\omega}_0\,\hat{\varsigma}} \neq 0,
\ee
is $\hat{\varsigma}$ in the space of free parameters? 
If yes (i.e.~all $\hat{\varsigma}$ described in \eqref{simplify_condition}, satisfy the condition $\tr{\hat{\varsigma}\,\hat{\varpi}} = 0$ for all $\hat{\varpi}$ of the form \eqref{transition_elems}) then we may choose $\hat{\omega}_\mathrm{free}$ so as to cancel out whatever support $\hat{\omega}_0$ had on $\varsigma^\perp$ \emph{without} jeopardizing the noise cancellation property of the feedback. 
If no (i.e.~$\hat{\omega}_0$ has support in $\varsigma^\perp$ as per \eqref{simplify_condition}, and some of that support is also in the space of transition elements where $\tr{\hat{\varsigma}\,\hat{\varpi}} \neq 0$), we conversely cannot have complete noise cancellation \emph{and} $\hat{\omega}$ only in $\varsigma'$. 
In short, we may impose a constraint on our feedback controls to avoid an operator subspace $\varsigma^\perp$, and still satisfy the NCC \eqref{NCC_both}, either if $\hat{\omega}_0$ has no support in $\varsigma^\perp$ to begin with, or if whatever support $\hat{\omega}_0$ does have in $\varsigma^\perp$ does not also overlap with transition elements \eqref{transition_elems} between the eigenspaces of $\hat{\rho}$. 
In the remaining case where $\hat{\omega}_0$ calls for an operation in part of the transition element space \eqref{transition_elems} that is in $\varsigma^\perp$, then our control constraint is incompatible with the situation: We must choose between noise cancellation or our control constraint, but we cannot obey both. 
See Fig.~\ref{fig:transition_elems} for a visualization of the various sub-spaces involved in this discussion. 

\subsection{Noise Minimization for Restricted Pure--State Controls}

We now elaborate on the alternative approach to using free parameters, searching for noise--minimizing solutions under explicit contraints to the feedback control configuration.
Let us consider the derivations of \secref{sec_Restricted_Controls} in the special case of pure states and perfect measurement efficiency. 
We again impose $\hat{\omega} = f\,\hat{\Theta}$, where $\hat{\Theta}$ is provided by the user, and then seek $f$ such that the noise is minimized to the extent possible with the given form of the feedback controller. 
The un-normalized ``noise state'' analogous to $\hat{\mathcal{B}}$ can be expressed as
\begin{equation} \label{noise-state}
   \ket{\tilde{\psi}}= \Pi^\perp \left(\hat{\omega}+i\hat{L}\right)\ket{\psi},
\end{equation}
where this expression should vanish for perfect noise cancellation. 
Complete noise cancellation is possible iff $\mathcal{N} = \ip{\tilde{\psi}}{\tilde{\psi}}= 0$. This leads to 
\begin{equation} \label{ft_expect}
  \mathcal{N} = \left \langle\left(f\,\hat{\Theta} -i\,\hat{L}^\dag\right)  \hat{\Pi}^\perp \left(f\,\hat{\Theta} + i \,\hat{L}\right)\right\rangle \geq 0 
\end{equation}
which is effectively a quadratic equation in $f$ at any given instant in time, such that
\begin{subequations} \label{quadratic_solve} \begin{align}
\mathcal{N} &= a\,f^2 + b\,f + c \quad\text{for}\\
a &\equiv  \ensavg{\hat{\Theta}\,\hat{\Pi}^\perp\,\hat{\Theta}} = \mathrm{var}\,\hat{\Theta}, \\
b &= i\ensavg{\hat{\Theta}\,\hat{\Pi}^\perp\,\hat{L} - \hat{L}^\dag\,\hat{\Pi}^\perp\,\hat{\Theta}}, \\
c &= \ensavg{\hat{L}^\dag\,\hat{\Pi}^\perp\,\hat{L}},
\end{align} \end{subequations}
in direct analogy with \eqref{noise-quad} and \eqref{f-quad-single-general}. 
We have use the variance for shorthand, defined as
\begin{equation}
    \mathrm{var}\,O =\left\langle\hat{O}^2\right\rangle-\left\langle\hat{O}\right\rangle^2,
\end{equation}
for any operator $\hat{O}$. 

If $\ket{\psi}$ is an eigenstate of $\hat{L}$, we have a special case: All of $a,b,c$ vanish, and any feedback operation is allowed (and moreover, none is needed, as measurement eigenstates are always in the nullspace of the diffusive noise operator $\widehat{\mathsf{b}}$ \eqref{eq:SME_wfb}).

If we wish to solve for constraints on $\tilde{\omega} = g\,\hat{M} = \hat{\omega} + i\,\hat{Y}$ instead, we obtain an expression
\be \label{gm_expect}
\ensavg{\left(g\,\hat{M} -i\,\hat{X}\right)  \hat{\Pi}^\perp \left(g\,\hat{M} + i \,\hat{X}\right)}=0,
\ee 
directly analogous to \eqref{ft_expect}. 
The remainder of the derivation follows, now with quadratic coefficients
\begin{subequations} \label{poly_hermitian} \begin{align}
a &= \ensavg{\hat{M}\,\hat{\Pi}^\perp\,\hat{M}} = \mathrm{var}\,M, \\
b &= \ensavg{\hat{M}\,\hat{\Pi}^\perp\,\hat{X} - \hat{X}\,\hat{\Pi}^\perp\,\hat{M}} = i\ensavg{[\hat{M},\hat{X}]}, \\
c &= \ensavg{\hat{X}\,\hat{\Pi}^\perp\,\hat{X}} = \mathrm{var}\,X, 
\end{align}\end{subequations}
such that $a\,g^2 + b\,g + c = 0$ may be solved for $g$ as in \eqref{quad_sol} to check for the existence of a noise canceling solution with the constraint $\tilde{\omega} \propto \hat{M}$. 
This reveals that the coefficients in our quadratic equation are substantially simplified when feedback is performed using a Hermitian measurement operator (i.e.~when $\hat{Y} = 0$). 
To be clear, these expressions are still valid when $\hat{Y}\neq 0$, but by using them we solve for a feedback $\hat{\omega} = g\,\hat{M} - i\,\hat{Y}$, not $\hat{\omega} \propto \hat{M}$.

All of the points of the discussion surrounding \eqref{quad_sol} still apply in the present case. 
In particular, while a completely noise--canceling feedback always exists for pure states, the particular controller form $\hat{\Theta}$ or $\hat{M}$ is \emph{not} guaranteed to allow this. 
The above expressions should still be understood as a means to test whether the choice $\hat{\omega} \propto \hat{\Theta}$ or $\hat{M}$ admits perfect noise cancellation, or merely an imperfect minimization. 

\subsection{Generalization to Multi--Dimensional Control Constraints and Mixed States}

We now extend the arguments of \secref{sec_Restricted_Controls} by considering a more flexible controller form:
\be \label{multi-om-form}
\hat{\omega} = \sum_j f_j\,\hat{\Theta}_j = \mathbf{f}\cdot\hat{\boldsymbol{\Theta}}.
\ee 
Our task is the same as above: We seek the controller weights $\mathbf{f}$ that minimize the noise in the dynamics given the constraint to the controller terms $\hat{\omega} \sim \sum_j \hat{\Theta}_j$. 

We may still use \eqref{noise-quad}, or any of its special cases detailed immediately above. 
Evaluation of \eqref{noise-quad} for the generalized controller form \eqref{multi-om-form} leads to 
\begin{subequations} \label{f-quad-multi-general} \begin{align}
\mathcal{N}(\mathbf{f}) &= \mathbf{f}^\top\cdot\mathbf{a}\cdot\mathbf{f} + \mathbf{f}^\top\cdot\mathbf{b} + c \quad\text{for} \\
a_{jk} & = \tr{[\hat{\Theta}_j,\hat{\rho}][\hat{\rho},\hat{\Theta}_k]}, \\
b_j &= 
2i\,\sqrt{\eta}\left[ \tr{\hat{\rho}\,[\hat{\Theta}_j,\hat{X}]\,\hat{\rho}} + \ensavg{\lbrace \hat{Y}, \lbrace\hat{\rho},\hat{\Theta}_j\rbrace \rbrace} \right], 
\end{align} \end{subequations}
in direct analogy with \eqref{f-quad-single-general}. 
The tensor $\mathbf{a}$ is real--valued, and is also symmetric provided that $[\hat{\Theta}_j,\hat{\Theta}_k] = 0$. 
Moreover, we still have
\be \label{amat_positivity}
\mathbf{f}^\top \cdot \mathbf{a} \cdot \mathbf{f} = \sum_{jk} f_j\,f_k\,a_{jk} = \tr{[\hat{\rho},\hat{\omega}]^\dag[\hat{\rho},\hat{\omega}]} \geq 0
\ee
for physically admissable controls. 
It is clear that $\mathbf{a}$ is positive semi-definite by construction, because $\nexists$ any $\mathbf{f}$ for which \eqref{amat_positivity} is negative.
For $\mathbf{a}$ to be positive--definite, we require that there $\nexists~\mathbf{f}$ such that \eqref{amat_positivity} evaluates to $0$. 
This stricter condition requires that we choose controls $\hat{\Theta}_j$ such that there does not exist any $\mathbf{f}$ for which $\hat{\omega} = \mathbf{f}\cdot\hat{\boldsymbol{\Theta}}$ leads to $[\hat{\rho},\hat{\omega}] = 0$. 
In the case of pure states $\hat{\rho} = \op{\psi}{\psi}$, we may say that $\mathbf{a}$ is positive definite if $\ket{\psi}$ is \emph{not} an eigenstate of any of the $\hat{\Theta}_j$ or any linear combination $\hat{\omega}$ of them.

When $\mathbf{a}$ is positive--definite, $\mathcal{N}(\mathbf{f})$ has a well--defined and unique minimum in the space of control weights $\mathbf{f}$, i.e.~there exists
\begin{subequations} \label{fmin-multi} \be 
\mathbf{f}^\star = (\mathbf{a} + \mathbf{a}^\top)^{-1}\cdot\mathbf{b} \quad\text{such that}
\ee \be 
\mathcal{N}^\star = \mathcal{N}(\mathbf{f}^\star) = c - \tfrac{1}{2}\,\mathbf{b}^\top\cdot(\mathbf{a} + \mathbf{a}^\top)^{-1}\cdot\mathbf{b},
\ee \end{subequations}
as a direct generalization of \eqref{fmin-single}. 
This solution minimizes the noise over the constraint to feedback controls $\hat{\omega} \sim \sum_j \hat{\Theta}_j$, and is a perfectly noise canceling solution in the event that $\mathcal{N}^\star = 0$. 
Thus, we find that the conditions for $\mathbf{a}$ to be positive--definite are indirectly conditions to establish a minimal and well--defined feedback control basis for noise minimization in a given physical situation. 
We emphasize the utility of this result: The tools above allow us to identify a minimal feedback controller configuration $\lbrace \hat{\Theta}_j \rbrace$ under constraints, and then find the relationships and coefficients between those constrained feedback channels that optimally cancel (minimize) stochasticity in the dynamics.

\section{Population based noise canceling feedback}

In this section, we present an alternative approach to noise canceling feedback for mixed states. Such an approach overcomes the issue of the existence of noise canceling feedback for mixed states, especially for state preparation problems.

We consider the case $\hat{Y}=0$ (i.e., Hermitian observable) and feedback given by \eqref{SSE_ito-op}. Consider a target state $\ket{\psi_0}$ with projector $\hat{\Pi}_0=\ket{\psi_0}\bra{\psi_0}$ that we  intend to reach. We assume the system's state is $\hat{\rho}$. The population of $\ket{\psi_0}$ at time $t$ can be expressed as 
\begin{equation}
    M(t)=\left\langle\psi_0|\hat{\rho}(t)|\psi_0\right\rangle=\textrm{Tr}\left(\hat{\Pi}_0\hat{\rho}(t)\right).
\end{equation}
Then at time $t+dt$, the population is
\begin{equation}
\begin{split}
    M(t+dt) &= \textrm{Tr}\left(\hat{\Pi}_0\frac{\hat{\mathcal{U}}\,\hat{\mathcal{M}}\,\hat{\rho}\,\hat{\mathcal{M}}^\dagger \,\hat{\mathcal{U}}^\dagger}{\left\langle\hat{\mathcal{M}}^\dagger \hat{\mathcal{M}}\right\rangle}\right)\\&= \textrm{tr}\left(\frac{\hat{\mathcal{M}}^\dagger \,\hat{\mathcal{U}}^\dagger\,\hat{\Pi}_0\hat{\mathcal{U}} \,\hat{\mathcal{M}}}{\left\langle\hat{\mathcal{M}}^\dagger \hat{\mathcal{M}}\right\rangle}\hat{\rho}\right)=\textrm{tr}\left(\hat{\Pi}^\prime_0\hat{\rho}\right),
    \end{split}\label{pop_rho_update}
\end{equation}
where 
\begin{equation}
    \hat{\Pi}^\prime_0=\frac{\hat{\mathcal{M}}^\dagger \,\hat{\mathcal{U}}^\dagger\,\hat{\Pi}_0\hat{\mathcal{U}} \,\hat{\mathcal{M}}}{\left\langle\hat{\mathcal{M}}^\dagger \hat{\mathcal{M}}\right\rangle}.
\end{equation}
We can express
\begin{equation}
    \begin{split}& \hat{\Pi}^\prime_0-\hat{\Pi}_0= 
    \left(\hat{\Pi}_0\hat{\Xi}+\hat{\Xi}^\dagger\hat{\Pi_0}-\mathsf{s}\,\hat{\Pi}_0\right)dW\\&+dt\left(\hat{\Xi}^\dagger\hat{\Pi}_0\hat{\Xi}-\hat{\Pi}_0\left(\hat{J}+\mathsf{s}\,\hat{\Xi}\right)-\left(\hat{J}^\dagger+\mathsf{s}\,\hat{\Xi}^\dagger\right)\hat{\Pi}_0\right),
    \end{split}
\end{equation}
with $\hat{J}=i\hat{\Omega}+\frac{1}{2}\left(\hat{\omega}^2+\hat{X}^2\right)+i\hat{\omega}\hat{X}-\mathsf{s}\,\hat{X}$. Then, 
\begin{equation}
    \begin{split}dM&=\textrm{tr}\Bigg(\hat{\rho}(t)\Big(\left(\hat{\Pi}_0\hat{\Xi}+\hat{\Xi}^\dagger\hat{\Pi_0}-\mathsf{s}\,\hat{\Pi}_0\right)dW\\&+dt\left(\hat{\Xi}^\dagger\hat{\Pi}_0\hat{\Xi}-\hat{\Pi}_0\left(\hat{J}+\mathsf{s}\,\hat{\Xi}\right)-\left(\hat{J}^\dagger+\mathsf{s}\,\hat{\Xi}^\dagger\right)\hat{\Pi}_0\right)\Big)\Bigg).
    \end{split}
\end{equation}
We can choose to cancel the noise in the above equation, i.e.,
\begin{equation}
    \hat{\Pi}_0\hat{\Xi}+\hat{\Xi}^\dagger\hat{\Pi_0}=2p_0\hat{\Pi}_0,
    \label{ncc_state_prep}
\end{equation}
where $p_0=\textrm{tr}\left(\hat{X}\hat{\Pi}_0\right)$. Note, in general $2p_0\neq \mathsf{s}$. However, it is always possible to satisfy \eqref{ncc_state_prep} since $\hat{\Pi}_0$ is a pure state. We can choose 
\begin{equation}
   \hat{ \omega }= \omega_{\psi}\hat{\Pi}_0-i\left[\hat{\Pi}_0,\hat{X}\right]+\hat{\Pi}^{\perp}_0\hat{B}\hat{\Pi}^{\perp}_0,
    \label{state_prep_w}
\end{equation}
where $\hat{\Pi}^{\perp}_0= \hat{\mathds{1}}-\hat{\Pi}_0$, $\omega_\psi$ is a free parameter and $\hat{B}$ is an arbitrary operator. Thus, the concept of noise-canceling feedback is applied to the population instead of the dynamics of the state. That way (near) deterministically population increase might be possible. Another benefit is that the feedback can be largely time-independent. 
If $\left[\hat{X},\hat{\Pi}_0\right]=0$, the population evolution is given by
\begin{equation}
    dM = 2(p_0-\mathsf{s})MdW+\textrm{Tr}\left(i\hat{\Omega}\left[\hat{\Pi}_0,\hat{\rho}\right]\right)dt.
    \label{pop_SDE_5q}
\end{equation}
Thus, in this case, the free parameters of the NCQF in \eqref{state_prep_w} do not contribute to the dynamics of the population of interest.

\vspace{30pt}

\bibliography{refs}
\end{document}